\documentclass[twocolumn]{aastex7}
\usepackage{soul}
\usepackage{hyperref}
\usepackage{makecell}
\usepackage{amsmath}
\usepackage{natbib}
\defcitealias{2018MNRAS.475.3369C}{CO18}
\defcitealias{2015A&A...576A.128D}{DO15}
\defcitealias{2004A&A...416.1117C}{CA04}

\usepackage{multirow}
\usepackage{CJKutf8}

\tabletypesize{\scriptsize}
\usepackage{lipsum}  

\begin{document}

\title{Revealing $\alpha$-Element's Past with Subaru/IRD: \\ Oxygen Abundance of 35 Very Metal-Poor Stars from Near-IR OH lines}

\author[0009-0002-4853-9593]{Bakuh Danang Setyo Budi} 
\email{danang.budi@grad.nao.ac.jp}
\affiliation{Astronomical Science Program, The Graduate University for Advanced Studies, SOKENDAI, 2-21-1 Osawa, Mitaka, Tokyo, 181-8588, Japan}
\affiliation{National Astronomical Observatory of Japan, 2-21-1 Osawa, Mitaka, Tokyo, 181-8588, Japan}

\author[0000-0002-8975-6829]{Wako Aoki (\begin{CJK*}{UTF8}{min}青木和光\end{CJK*})} 
\email{aoki.wako@nao.ac.jp}
\affiliation{Astronomical Science Program, The Graduate University for Advanced Studies, SOKENDAI, 2-21-1 Osawa, Mitaka, Tokyo, 181-8588, Japan}
\affiliation{National Astronomical Observatory of Japan, 2-21-1 Osawa, Mitaka, Tokyo, 181-8588, Japan}

\author[0000-0002-5259-3974]{Nicholas Storm} 
\email{storm@mpia.de}
\affiliation{Max-Planck-Institut für Astronomie, Königstuhl 17, D-69117 Heidelberg, Germany}
\affiliation{Heidelberg University, Grabengasse 1, D-69117 Heidelberg, Germany}

\author[0000-0003-4656-0241]{Miho Ishigaki (\begin{CJK*}{UTF8}{min}石垣美歩\end{CJK*})} 
\email{miho.ishigaki@nao.ac.jp}
\affiliation{National Astronomical Observatory of Japan, 2-21-1 Osawa, Mitaka, Tokyo, 181-8588, Japan}

\author[0000-0002-8077-4617]{Tadafumi Matsuno (\begin{CJK*}{UTF8}{min}松野允郁\end{CJK*})} 
\email{matsuno@uni-heideberg.de}
\affiliation{Astronomisches Rechen-Institut, Zentrum für Astronomie der Universität Heidelberg, Mönchhofstraße 12–14, 69120 Heidelberg,
Germany}

\author[0000-0003-3618-7535]{Teruyuki Hirano (\begin{CJK*}{UTF8}{min}平野照幸\end{CJK*})}
\email{teruyuki.hirano@nao.ac.jp}
\affiliation{Astronomical Science Program, The Graduate University for Advanced Studies, SOKENDAI, 2-21-1 Osawa, Mitaka, Tokyo, 181-8588, Japan}
\affiliation{National Astronomical Observatory of Japan, 2-21-1 Osawa, Mitaka, Tokyo, 181-8588, Japan}
\affiliation{Astrobiology Center, National Institutes of Natural Sciences, 2-21-1 Osawa, Mitaka, Tokyo 181-8588, Japan}

\author[0000-0002-4677-9182]{Masayuki Kuzuhara (\begin{CJK*}{UTF8}{min}葛原 昌幸\end{CJK*})}
\email{m.kuzuhara@nao.ac.jp}
\affiliation{National Astronomical Observatory of Japan, 2-21-1 Osawa, Mitaka, Tokyo, 181-8588, Japan}
\affiliation{Astrobiology Center, National Institutes of Natural Sciences, 2-21-1 Osawa, Mitaka, Tokyo 181-8588, Japan}

\author[0000-0001-9326-8134]{Jun Nishikawa (\begin{CJK*}{UTF8}{min}西川淳\end{CJK*})}
\email{jun.nishikawa@nao.ac.jp}
\affiliation{National Astronomical Observatory of Japan, 2-21-1 Osawa, Mitaka, Tokyo, 181-8588, Japan}
\affiliation{Astrobiology Center, National Institutes of Natural Sciences, 2-21-1 Osawa, Mitaka, Tokyo 181-8588, Japan}

\author[0000-0002-5051-6027]{Masashi Omiya (\begin{CJK*}{UTF8}{min}大宮正士\end{CJK*})}
\email{omiya.masashi@nao.ac.jp}
\affiliation{National Astronomical Observatory of Japan, 2-21-1 Osawa, Mitaka, Tokyo, 181-8588, Japan}
\affiliation{Astrobiology Center, National Institutes of Natural Sciences, 2-21-1 Osawa, Mitaka, Tokyo 181-8588, Japan}

\author[0000-0001-6181-3142]{Takayuki Kotani (\begin{CJK*}{UTF8}{min}小谷隆行\end{CJK*})}
\email{t.kotani@nao.ac.jp}
\affiliation{Astronomical Science Program, The Graduate University for Advanced Studies, SOKENDAI, 2-21-1 Osawa, Mitaka, Tokyo, 181-8588, Japan}
\affiliation{National Astronomical Observatory of Japan, 2-21-1 Osawa, Mitaka, Tokyo, 181-8588, Japan}
\affiliation{Astrobiology Center, National Institutes of Natural Sciences, 2-21-1 Osawa, Mitaka, Tokyo 181-8588, Japan}

\author[0000-0002-9294-1793]{Tomoyuki Kudo (\begin{CJK*}{UTF8}{min}工藤 智幸\end{CJK*})}
\email{kudotm@naoj.org}
\affiliation{Subaru Telescope, 650 N. Aohoku Place, Hilo, HI 96720, USA}

\author[0000-0003-4018-2569]{Sebastien Vievard}
\email{vievard@naoj.org}
\affiliation{Subaru Telescope, 650 N. Aohoku Place, Hilo, HI 96720, USA}
\affiliation{Space Science and Engineering Initiative, College of Engineering, University of Hawai‘i, Hilo, HI 96720, USA}
\affiliation{Institute for Astronomy, University of Hawaii, Hilo, HI 96720, USA}

\author[0000-0002-6510-0681]{Motohide Tamura (\begin{CJK*}{UTF8}{min}田村元秀\end{CJK*})}
\email{motohide.tamura@abc-National Institutes of Natural Sciences.jp}
\affiliation{National Astronomical Observatory of Japan, 2-21-1 Osawa, Mitaka, Tokyo, 181-8588, Japan}
\affiliation{Astrobiology Center, National Institutes of Natural Sciences, 2-21-1 Osawa, Mitaka, Tokyo 181-8588, Japan}
\affiliation{Department of Astronomy, Graduate School of Science, The University of Tokyo, 7-3-1 Hongo, Bunkyo-ku, Tokyo 113-0033, Japan}

\begin{abstract}
Oxygen abundances in very and extremely metal-poor (V/EMP) stars provide critical constraints on early massive stars' nucleosynthesis. An Oxygen abundance analysis is presented for 35 V/EMP stars ($-4.0 < \text{[Fe/H]} < -1.5$) using near-infrared $H$-band OH vibro-rotational lines from high-resolution Subaru/IRD spectra. To examine the reliability of these {near infrared} (NIR) OH lines, the results are compared with the abundances obtained from the 3D/NLTE-insensitive forbidden $\text{[OI]}$ 6300 \AA\ line using archival high-resolution optical spectra. After homogeneously rederiving stellar parameters and 1D/NLTE Fe abundances using \textit{Gaia} photo-astrometry and literature optical Fe equivalent width data, oxygen abundance from OH and [OI] lines is determined through 1D/LTE spectral synthesis. A sensitivity analysis confirms that {NIR} OH lines are highly sensitive to the adopted temperature ($\Delta\log\epsilon(\text{O})/\Delta T\sim\pm0.25\text{ dex}/\pm100\text{ K}$) compared to the forbidden line. A temperature-dependent discrepancy between the tracers is identified: in cool red giants ($T_\text{eff} \lesssim 4600$ K), OH-based abundances are systematically lower than $\text{[OI]}$-based abundance by 0.05 to 0.25 dex, while warmer red giants show higher OH-based abundances as expected from 3D effects. Despite this systematic offset, the numerous measurable NIR OH lines yield significantly smaller random abundance errors than that of the single, weak $\text{[OI]}$ line. Leveraging this statistical precision, an empirical calibration as a function of $T_\text{eff}$, $\log g$, $\text{[Fe/H]}$, and $\text{[C/Fe]}$ is derived to align the 1D/LTE OH abundances onto the $\text{[OI]}$ scale. Applying this correction substantially reduces the scatter and temperature dependence in the $\text{[O/Fe]}$ versus $\text{[Fe/H]}$ plane and flattens the trend, bringing the results into fairly good agreement with Galactic chemical evolution models.

\end{abstract}



\section{Introduction} 
\label{sec:Introduction}

Oxygen (O) is the third most abundant chemical element in the Universe, after hydrogen and helium. Its chemical evolution, as well as that of other $\alpha$-elements, is closely linked to the life cycle of massive stars ($M > 10 M_{\odot}$), as it is synthesized almost exclusively during hydrostatic burning phases and later ejected into the interstellar medium (ISM) by Type II core-collapse supernovae (CCSNe) \citep{1995ApJS..101..181W,2009A&A...Fabian}. In contrast, iron (Fe) is produced both by CCSNe of massive stars and by Type Ia supernovae, which are white-dwarf binary explosions. 

The oxygen-to-iron abundance ratio ([O/Fe]\footnote{$\text{[A/B]} = \log(N_\text{A}/N_\text{B}) - \log(N_\text{A}/N_\text{B})_\odot$, and $A(\text{X}) = \log\epsilon_\text{X} = \log(N_\text{X}/N_\text{H}) + 12$, where A, B, and X are elements.}) serves as useful indicator of relative contributions of these two types of supernovae over time. To trace the evolution of this ratio from the earliest epochs, chemical abundances are derived in very metal-poor (VMP) stars with $\text{[Fe/H]} < -2$ and extremely metal-poor (EMP) stars with $\text{[Fe/H]} < -3$ \citep{2005ARA&A..43..531B}. These ancient stars, particularly in the Galactic halo, preserve the primordial abundance patterns in their photospheres, directly reflecting the nucleosynthesis yields of their progenitors—the first generation of massive stars, or Population III stars \citep{2012ApJ...Norris}. At such extremely low metallicity, the contributions of Fe production from Type Ia supernovae are not expected, due to their long timescale. Therefore, the [O/Fe] evolution over metallicity is expected to directly record the CCSN yield.

Recent simulations of galactic chemical evolution indicate that typical CCSN models incorporating stellar rotation yield a high average [O/Fe] plateau ($\sim+0.8$ dex) with large scatter ($\sim0.3$ to $0.6$ dex as metallicity decreases) in the extremely metal-poor regime under multi-explosion scenarios \citep{2022MNRAS.516.6075S}. The large scatter is possibly due to variations in O/Fe ratios produced by supernovae, which depend on the progenitor mass, as well as incomplete mixing of the gas clouds from which metal-poor stars formed. 

When the models include failed-SNe, supernovae that collapse their entire CO core into black holes, it lowers the average [O/Fe] at extremely metal-poor regime. This trend heavily depends on the adopted lower mass limit for failed-SNe within the initial mass function (IMF), for which current models assume failed-SNe occur at masses $\geq30\text{ M}_\odot$ \citep{2020ApJ...900..179K}. In addition, including hypernovae (HNe), or high-energy explosions $(E\gtrsim10^{52}\text{ erg, or ten times the typical CCSN energy})$ from stars with masses $>20 M_{\odot}$, also lowers the [O/Fe] trend {with} smaller scatter in the extremely metal-poor regime. This is due to a tight correlation between mass-explosion energy and the amount of Fe ejected in HNe simulation \citep{2007ApJ...660..516T,2013ARA&A..51..457N}. Although the exact fraction of HNe among all SNe is currently still uncertain, \cite{2020ApJ...900..179K} adopt a value of $\epsilon_\text{HN}=0.5$ for stars with $Z=0$ in their calculations. Therefore, observational [O/Fe] trends and the scatter size of V/EMP stars as a function of [Fe/H] provide constraints on the contributions of CCSN, HNe, and failed-SNe in the earliest era.

On the other hand, measuring oxygen abundance \textit{reliably} in this metallicity regime is a longstanding observational challenge. To derive oxygen abundance in metal-poor stars, four tracers are typically used across different wavelength regimes. The derivation from these tracers for the same stars sometimes result in large discrepancies. One of the most reliable tracers is the forbidden [OI] 6300 \AA\ line, which forms {in the condition }very close to local thermodynamic equilibrium (LTE) and is relatively insensitive to 3D atmospheric effects \citep[e.g.,][]{2003AA402..343T,2016MNRAS.455.3735A,2019A&A...622L...4A, 2019A&A...630A.104A}. The difficulties in measuring this line arise from its weakness in V/EMP stars and potential contamination of telluric lines. For FGK dwarf stars, the permitted OI triplet 7772-7775 \AA\ lines are frequently used. These lines are, however, known to depart from LTE, resulting in significantly higher O abundance th{a}n to the one derived with the [OI] line. Large efforts have been made to develop 1D/NLTE and 3D/NLTE models for these lines \citep[e.g., ][]{2003AA402..343T,2015A&A...583A..57S,2019A&A...630A.104A}. In addition, these lines remain extremely weak and inaccessible for V/EMP stars. Another tracer is the near-ultraviolet (NUV) electronic transitions of the OH molecule in 3100-3300 \AA, which are relatively strong in EMP stars with [Fe/H] $<-4$ or lower \citep{2015ApJ...811...64A}. The main drawbacks of NUV OH lines are that they are strongly affected by 3D hydrodynamical effects and depend on the adopted carbon abundance; hence, simultaneous derivation of C and O abundances is needed \citep{2017A&A...599A.128P}. In addition, these lines are only observable by a limited number of ground-based spectrographs due to low efficiency and low atmospheric transmission at the blue end of optical spectra.

In this work, the other tracer of O abundance, the near-infrared (NIR) vibro-rotational transition of the OH molecule located in the \textit{H}-band between 1.5–1.7 $\mu$m, is investigated. The major advantages of this tracer compared to others are as follows: (1) there are many measurable lines \citep{2016ApJ...819..103A}, with line strengths at least twice stronger than the [OI] line \citep{2015A&A...576A.128D}; (2) the O abundance derived from the NIR OH line is relatively independent of the adopted C abundance, even in 3D/LTE modelling, compared to the NUV OH lines \citep[as demonstrated by][]{2017A&A...599A.128P}; and (3) from an observational perspective, since most of the targets are cool red giants, they are relatively bright in the \textit{H}-band, making it easier to obtain high quality spectra in a relatively short observing time, large dataset can be obtained with NIR spectrometers that recently available.

Despite the absence of full 3D hydrodynamical atmospheric modelling with complete non-LTE treatment for the NIR OH lines, with careful temperature calibration, a larger dataset than in previous studies, and calibration using the [OI] line, we demonstrate in this work that the agreement in 1D/LTE O abundance between the NIR OH and [OI] tracers can be brought closer, showing its potential application in V/EMP stars.

The sample selection, observations, and data used in this work are reported in Section \ref{sec:Observation}. 
In Sections \ref{sec:StellarParameters} and \ref{sec:AbundanceAnalysis}, the details of the stellar parameter rederivation and chemical abundance determination are explained. The results, sensitivity analysis on OH and [OI] lines, the abundance discrepancy between the NIR OH and [OI] tracers, and its effect on the [O/Fe] trends of the samples are presented and explained in Section \ref{sec:Discussions}. In the final part, Section \ref{sec:conclusion}, the findings of this study are summarized.

\section{Observations and Target Selection} \label{sec:Observation}
\subsection{Subaru/IRD observations}
To derive infrared OH lines located between $\lambda\lambda1500-1700$ nm in V/EMP giants, this study utilizes the high-quality, high-resolution spectra obtained with the \textbf{I}nfra\textbf{R}ed \textbf{D}oppler (IRD) NIR spectrograph \citep{2012SPIE.8446E..1TT,2018SPIE10702E..11K} installed on the Subaru Telescope, which cover the \textit{YJH}-bands. All of the spectra used in this paper are taken from the dataset of 46 metal-poor stars with a metallicity range of $-4.0<$ \text{[Fe/H]} $<-1.5$ that has been published before \citep{2022PASJ...74..273A,aoki2025elementalabundances44metalpoor}. The details of the observations and the data reduction process were reported by these papers.

After {a} radial velocity correction, a visual inspection is performed on the suspected locations of OH lines based on the \texttt{APOGEE} linelist \citep{2021AJ....161..254S}. Spectra of rapidly rotating stars{, HR2783 \& HD188651 ($\text{S/N}\sim100$ at \textit{H}-band),} are used to identify telluric lines. OH lines with clear absorption features not affected by either residual atmospheric OH glow-lines or telluric absorptions are selected. The final sample consists of 34 red giants and {one} dwarf star. The photometry and parallax parameters from \textit{Gaia} and \textit{2MASS} catalogues, signal-to-noise ratio (S/N) at {the} \textit{H}-band, and radial velocity {derived} from the Subaru/IRD spectra for the sample are given in Table \ref{tab:obs_phot_rv}.

\startlongtable
\begin{deluxetable*}{lcccccccc}
\centering
\tabletypesize{\small}
\setlength{\tabcolsep}{2.5pt}
\digitalasset
\tablewidth{0pt}
\tablecaption{Observational parameters, spectral quality \& radial velocities of our samples\label{tab:obs_phot_rv}}
\tablehead{
\colhead{\multirow{2}{*}{Target}} & \colhead{$V^\dagger$} & \colhead{$G^\ddagger$} & \colhead{$BP^\ddagger$} & \colhead{$RP^\ddagger$} & \colhead{$K_s^\dagger$} & \colhead{$\varpi^\ddagger$} & \colhead{$\text{RV}_\text{IRD}$} & \colhead{S/N} \\
\colhead{} & \colhead{(mag)} & \colhead{(mag)} & \colhead{(mag)} & \colhead{(mag)} & \colhead{(mag)} & \colhead{(mas)} & \colhead{({km s$^{-1}$})} & \colhead{@1600 nm}
}
\startdata
BD-02$^\circ$5957 & 11.38(4) & 10.761(3) & 11.392(3) & 9.994(4) & 8.31(2) & 0.204(17) & -111.3 &  86 \\
BD-07$^\circ$2674 & 10.26(1) & 9.982(3) & 10.532(3) & 9.271(4) & 7.74(3) & 0.433(18) & 267.0 &  149 \\
BD-14$^\circ$5890 & 10.21(3) & 9.949(3) & 10.404(3) & 9.313(4) & 7.94(4) & 0.971(15) & 232.0 &  124 \\
BD-15$^\circ$5781 & 10.72(3) & 10.435(3) & 10.931(3) & 9.766(4) & 8.29(2) & 0.569(16) & 36.9 &  67 \\
BD-18$^\circ$271 & 9.81(3) & 9.348(3) & 10.038(5) & 8.547(5) & 6.79(2) & 0.394(18) & -114.2 &  148 \\
BD-18$^\circ$5550 & 9.28(3) & 8.927(3) & 9.486(3) & 8.197(4) & 6.56(2) & 1.972(20) & -13.8 &  188 \\
BD-20$^\circ$6008 & 9.84(3) & 9.552(3) & 10.055(3) & 8.874(4) & 7.40(3) & 0.721(18) & 135.6 &  115 \\
BD+03$^\circ$2782 & 9.70(3) & 9.349(3) & 9.894(3) & 8.641(4) & 7.08(3) & 0.680(16) & 101.2 &  165 \\
BD+30$^\circ$2611 & 9.13(3) & 8.728(3) & 9.395(3) & 7.937(4) & 6.09(2) & 0.655(15) & -214.4 &  173 \\
CS 29502-092 & 12.01(8) & 11.607(3) & 12.060(3) & 10.973(4) & 9.60(2) & 0.757(17) & 43.2 &  91 \\
CS 30314-067 & 11.85(3) & 11.461(3) & 12.088(3) & 10.700(4) & 9.03(2) & 0.200(20) & 256.2 &  130 \\
HD 107752 & 10.01(3) & 9.704(3) & 10.179(3) & 9.053(4) & 7.66(2) & 0.730(21) & 298.9 &  154 \\
HD 110184 & 8.27(3) & 7.874(3) & 8.534(3) & 7.086(4) & 5.35(3) & 0.623(24) & 216.0 &  155 \\
HD 115444 & 8.96(3) & 8.686(3) & 9.153(3) & 8.035(4) & 6.61(2) & 1.179(18) & 52.4 &  175 \\
HD 118055 & 8.86(3) & 8.430(3) & 9.133(3) & 7.614(4) & 5.72(2) & 0.691(27) & -36.5 &  257 \\
HD 122563 & 6.20(2) & 5.875(3) & 6.397(3) & 5.179(4) & 3.73(4) & 3.099(33) & 41.6 &  297 \\
HD 187111 & 7.71(3) & 7.279(3) & 7.981(3) & 6.460(4) & 4.55(2) & 1.791(20) & -75.8 &  186 \\
HD 204543 & 8.28(3) & 7.994(3) & 8.511(3) & 7.303(4) & 5.78(2) & 1.414(23) & 12.2 &  214 \\
HD 221170 & 7.67(3) & 7.300(3) & 7.915(3) & 6.536(4) & 4.84(2) & 1.807(23) & 62.1 &  230 \\
HD 25329 & 8.51(3) & 8.232(3) & 8.697(3) & 7.590(4) & 6.20(2) & 54.042(26) & 32.1 &  207 \\
HD 4306 & 9.02(3) & 8.771(3) & 9.194(3) & 8.160(4) & 6.82(2) & 2.047(33) & -12.8 &  78 \\
HD 6268 & 8.09(3) & 7.820(3) & 8.302(3) & 7.154(4) & 5.71(2) & 1.472(25) & 140.0 &  115 \\
HD 85773 & 9.42(3) & 8.986(3) & 9.630(3) & 8.208(4) & 6.51(2) & 0.467(15) & 237.1 &  182 \\
HD 8724 & 8.30(1) & 7.969(3) & 8.533(3) & 7.241(4) & 5.64(2) & 2.310(26) & -20.0 &  116 \\
HD 88609 & 8.59(3) & 8.270(3) & 8.815(3) & 7.558(4) & 6.01(2) & 0.857(21) & 60.0 &  165 \\
HE 1116-0634 & 11.84(3) & 11.273(3) & 11.825(3) & 10.564(4) & 9.01(2) & 0.203(21) & 198.0 &  137 \\
HE 1523-0901 & 11.50(3) & 10.746(3) & 11.372(3) & 9.982(4) & 8.35(3) & 0.328(20) & -105.6 &  101 \\
LAMOST J0040+2729 & 11.13(2) & 10.810(3) & 11.364(3) & 10.097(4) & 8.52(2) & 0.346(16) & 3.5 &  121 \\
LAMOST J2114-0616 & 11.41(2) & 10.862(3) & 11.392(3) & 10.171(4) & 8.66(2) & 0.439(17) & -50.2 &  117 \\
LAMOST J2217+2104 & 13.25(9) & 13.020(3) & 13.630(3) & 12.267(4) & 10.64(2) & 0.065(17) & -51.4 &  110 \\
LAMOST J2347+2851 & 11.13(1) & 10.867(3) & 11.351(3) & 10.205(4) & 8.77(2) & 0.473(17) & -147.1 &  94 \\
TYC 3407-1352-1 & 11.10(2) & 10.545(3) & 11.076(3) & 9.849(4) & 8.33(2) & 0.475(15) & 122.4 &  89 \\
TYC 3814-1598-1 & 10.10(1) & 9.722(3) & 10.355(3) & 8.954(4) & 7.30(2) & 0.318(12) & 30.1 &  150 \\
UCAC4 425-121652 & 12.78(8) & 12.342(3) & 12.959(3) & 11.580(4) & 9.88(2) & 0.185(14) & 114.4 &  85 \\
UCAC4 515-137892 & 12.52(7) & 12.143(3) & 12.772(3) & 11.375(4) & 9.69(2) & 0.141(14) & 19.2 & 96 \\
\enddata
\tablenotetext{}{Note: Values in parentheses indicate the uncertainty in the last digit (e.g., 1.00(9) represents $1.00\pm0.09$).}
\tablenotetext{\dagger}{Photometric data references: $V$: \cite{2000AATycho2} and \cite{2012yUSSNaval}; $K_s$: \cite{20032MASS}.}
\tablenotetext{\ddagger}{Taken from \cite{2023AAGaiaDR3}.}
\end{deluxetable*}

\subsection{Archival Optical Spectra}
In order to check the consistency of the derived oxygen abundance, an attempt is made to rederive the O abundance from the neutral forbidden oxygen line ([OI]) at $\lambda630$ nm, which is known to be insensitive to the non-LTE effect \citep{2003AA402..343T,2016MNRAS.455.3735A}. All available high-resolution archival spectra are collected from the \textbf{H}igh \textbf{D}ispersion \textbf{S}pectrograph \citep[HDS:][]{2002PASJ...54..855N}, the \textbf{U}ltraviolet \& \textbf{V}isual \textbf{E}chelle \textbf{S}pectrograph \citep[UVES:][]{2000SPIE.4008..534D}, and the \textbf{HI}gh \textbf{R}esolution \textbf{E}chelle \textbf{S}pectrometer \citep[HIRES:][]{1994SPIE.2198..362V} installed at the Subaru, VLT-Kueyen, and Keck-I telescopes, respectively. In addition, for the benchmark very metal-poor star, HD 122563, additional ultra-high resolution and high-quality spectra from the \textbf{H}igh \textbf{A}ccuracy \textbf{R}adial velocity \textbf{P}lanet \textbf{S}earcher \citep[HARPS:][]{2003Msngr.114...20M} spectrograph installed at the ESO La Silla 3.6m telescope and the \textbf{E}chelle \textbf{SP}ectrograph for \textbf{R}ocky \textbf{E}xoplanets and \textbf{S}table \textbf{S}pectroscopic \textbf{O}bservations \citep[ESPRESSO:][]{2021A&A...645A..96P} spectrograph installed at the VLT are also obtained.

The reduced spectra from each telescope are accessed and queried through the online interfaces of the Japan Virtual Observatory (JVO)\footnote{\url{https://jvo.nao.ac.jp/index-e.html}}, the ESO Archive Science Portal\footnote{\url{https://archive.eso.org/scienceportal/home}}, and the Keck Observatory Archival (KOA) Portal\footnote{\url{https://koa.ipac.caltech.edu/}}. The JVO {provides reduced spectra of} HDS data from the SMOKA database\footnote{\url{https://smoka.nao.ac.jp}}{. T}he normalized spectra are carefully inspected and found to {have} sufficient quality for the purpose of this study. For all ESO facility spectrographs, the reduction process is done using dedicated pipelines for each instrument\footnote{\url{https://www.eso.org/sci/software/pipe_aem_table.html}}. For the HIRES spectra, the reduction process is done by the \texttt{MAKEE} software \citep{2024ascl.soft07001B}. All spectra are renormalized through the \texttt{SUPPNet}\footnote{\url{https://rozanskit.com/suppnet/}} package \citep{2022AA_suppnet} for a wider wavelength range and through high-order polynomial fitting for a local continuum near 6300 \AA.

The archival spectra of rapidly rotating B-type stars from each instrument are also taken to check the possibility of telluric line contamination around 6300 \AA\;by shifting and matching telluric lines in both spectra. If there is possible contamination at either end of the wing, the telluric line is subtracted/removed through Gaussian fitting. The O {abundances derived from} non-subtracted and telluric-subtracted spectra{show negligible difference.} Some of the archival spectra are discarded due to {severe blending of} telluric lines. In total, 42 optical archival spectra for 23 stars are obtained with more than one instrument for some objects, which enables an assessment of the consistency of the derived [OI]-based abundance. The observation log for the archive data, derived radial velocity, and spectral quality used in this paper are described in Table~\ref{tab:logobs}.

\startlongtable
\begin{deluxetable*}{lcccccc}
\centering
\digitalasset
\tablewidth{0pt}
\tablecaption{Log of observations for archival Optical spectra\label{tab:logobs}}
\tablehead{
\colhead{Target} & \colhead{Proposal ID} & \colhead{Telescope/Instrument}  & \colhead{DATE-OBS} & \colhead{$R$} & \colhead{$\text{RV}^\dagger_\text{opt}$} & \colhead{S/N} \\
\colhead{}  & \colhead{} & \colhead{} & \colhead{} & \colhead{$(\lambda/\Delta \lambda)$} & \colhead{(km s$^{-1}$)}  & \colhead{@636 nm} 
}
\startdata
BD -07$^\circ$2674 & 111.251N.001 & VLT-Kueyen/UVES & 2023/04/29 & 34540 & 195.4 & 148 \\
BD -18$^\circ$271 & U14H & Keck/HIRES & 1997/08/10 & 35800 & -126.1 & 181 \\
BD -18$^\circ$271 & o08137 & Subaru/HDS & 2008/07/28 & 51428 & -233.9 & 125 \\
BD -18$^\circ$5550 & o05114 & Subaru/HDS & 2005/06/20 & 60000 & -139.1 & 75 \\
BD +03$^\circ$2782 & o11140 & Subaru/HDS & 2011/02/16 & 60000 & 31.1 & 183 \\
BD +30$^\circ$2611 & o14185 & Subaru/HDS & 2014/08/19 & 32727 & -282.6 & 116 \\
CS 30314-067 & 165.N-0276(A) & VLT-Kueyen/UVES & 2001/10/21 & 42310 & 174.0 & 164 \\
HD 107752 & o08106 & Subaru/HDS & 2008/05/01 & 60000 & 219.1 & 209 \\
HD 107752 & 0104.D-0059(A) & VLT-Kueyen/UVES & 2020/03/03 & 42310 & 210.7 & 153 \\
HD 110184 & o08106 & Subaru/HDS & 2008/05/02 & 60000 & 138.9 & 329 \\
HD 115444 & C03H & Keck/HIRES & 2005/06/14 & 102700 & 55.8 & 245 \\
HD 115444 & C01H & Keck/HIRES & 2005/06/17 & 102700 & 54.6 & 214 \\
HD 115444 & U100Hb & Keck/HIRES & 2007/06/05 & 102700 & 55.3 & 338 \\
HD 115444 & U100Hb & Keck/HIRES & 2007/06/07 & 102700 & 55.3 & 323 \\
HD 118055 & 090.B-0605(A) & VLT-Kueyen/UVES & 2013/01/04 & 66320 & -131.6 & 113 \\
HD 122563 & 080.D-0347(A) & ESO3.6m/HARPS & 2008/02/24 & 115000 & -25.9 & 319 \\
HD 122563 & U084Hr & Keck/HIRES & 2007/06/06 & 47700 & 57.0 & 233 \\
HD 122563 & Y267Hr & Keck/HIRES & 2009/05/11 & 35800 & 54.2 & 205 \\
HD 122563 & U079Hr & Keck/HIRES & 2014/05/24 & 95500 & 56.6 & 250 \\
HD 122563 & Z079Hr & Keck/HIRES & 2016/06/28 & 47700 & 56.0 & 197 \\
HD 122563 & o05114 & Subaru/HDS & 2005/06/19 & 60000 & -1.5 & 200 \\
HD 122563 & 0103.D-0118(A) & VLT/ESPRESSO & 2019/04/06 & 190000 & 56.06 & 219 \\
HD 122563 & 0103.D-0118(A) & VLT/ESPRESSO & 2019/04/11 & 190000 & 56.55 & 661 \\
HD 187111 & 71.B-0529(A) & VLT-Kueyen/UVES & 2003/08/02 & 45254 & -179.0 & 179 \\
HD 204543 & o01117 & Subaru/HDS & 2001/06/04 & 50000 & -99.4 & 273 \\
HD 204543 & 71.B-0529(A) & VLT-Kueyen/UVES & 2003/06/07 & 45254 & -124.7 & 224 \\
HD 221170 & U100Hb & Keck/HIRES & 2007/06/05 & 102700 & -39.0 & 336 \\
HD 221170 & U100Hb & Keck/HIRES & 2007/06/07 & 102700 & -39.1 & 241 \\
HD 221170 & o05114 & Subaru/HDS & 2005/06/22 & 60000 & -122.3 & 368 \\
HD 4306 & 68.D-0546(A) & VLT-Kueyen/UVES & 2001/10/09 & 56990 & -64.3 & 290 \\
HD 6268 & N026Hr & Keck/HIRES & 2006/09/04 & 47700 & 120.9 & 289 \\
HD 6268 & o06138 & Subaru/HDS & 2007/01/26 & 36000 & 39.7 & 74 \\
HD 6268 & 165.N-0276(A) & VLT-Kueyen/UVES & 2001/11/07 & 66320 & 57.2 & 407 \\
HD 85773 & o07229 & Subaru/HDS & 2008/01/05 & 90000 & 147.1 & 183 \\
HD 85773 & 165.N-0276(A) & VLT-Kueyen/UVES & 2001/11/06 & 107200 & 124.7 & 365 \\
HD 8724 & 71.B-0529(A) & VLT-Kueyen/UVES & 2003/08/07 & 45254 & -140.6 & 244 \\
HD 88609 & U084Hr & Keck/HIRES & 2007/06/06 & 47700 & 45.4 & 223 \\
HD 88609 & o06138 & Subaru/HDS & 2007/01/27 & 36000 & -40.2 & 400 \\
LAMOST J0040+2729 & o16123 & Subaru/HDS & 2016/11/17 & 45000 & -92.0 & 94 \\
LAMOST J2114-0616 & o16123 & Subaru/HDS & 2016/05/24 & 45000 & -162.9 & 50 \\
LAMOST J2217+2104 & o16123 & Subaru/HDS & 2017/08/05 & 60000 & -116.4 & 33 \\
LAMOST J2347+2851 & o16123 & Subaru/HDS & 2017/08/05 & 45000 & -245.7 & 125 \\
Arcturus & 073.D-0590(A) & ESO3.6m/HARPS & 2004/07/08 & 115000 & -5.0 & 351 \\
Arcturus & 273.D-5032(A) & VLT-Kueyen/UVES & 2004/08/03 & 107200 & 19.0 & 352 \\
Arcturus & 0103.D-0118(A) & VLT/ESPRESSO & 2019/06/05 & 190000 & 77.8 & 424 \\
\enddata
\tablenotetext{\dagger}{This radial velocity have not been corrected for Barrycentric motion of Earth.}
\end{deluxetable*}

\section{Stellar Parameters} \label{sec:StellarParameters}
To ensure consistency and uniformity among the analyses, all of the input parameters are recalibrated: effective temperature, surface gravity, microturbulence velocity, and iron abundance or metallicity. In this section, the method to derive the first three parameters is described; Fe abundance is determined in the next section.

\subsection{Effective Temperature}

A similar approach to \cite{2025AJ....169..172M} is followed to rederive the effective temperature using a combination of colours from the \textit{2MASS} $K_s$ band and the \textit{Gaia} \textit{G}, \textit{BP}, and \textit{RP} bands. 

First, the intrinsic colours are derived through a dereddening process across all photometric bands. For the $K_s$ band, the reddening parameter $E(B-V)$ is adopted from the 2D dust maps of \cite{Schlafly_2011} and corrected for the object's distance and galactic latitude following the formulae provided by \cite{2000AJ....120.2065Bonifacio}. The corrected reddening values are presented in Table \ref{tab:obs_color_teff}. The $K_s$-band extinction is calculated using the relation $A_{K_s} = 0.310 \cdot E(B-V)$ \citep{Schlafly_2011}. For the \textit{Gaia} bands, each band magnitude is iteratively dereddened using colour-dependent coefficients derived from the formulae of \cite{2018AA616A..10G}.

Using the dereddened colours, the effective temperature for each colour is determined based on the empirical colour–$T_\text{eff}$–[Fe/H] relation provided in \cite{2021AA653A..90M}. Monte Carlo simulations are conducted, drawing $10^6$ samples for each input parameter (magnitude, metallicity, reddening, distance, and all of the given polynomial coefficients) considering their uncertainties. The temperature and its uncertainty for each colour are determined by calculating the median and standard deviation of the resulting distribution. The weighted average of $T_\text{eff}$ is derived from six bands, with the uncertainty calculated using the following relation:
\begin{align}
 \sigma_{T_\text{eff}}=\sqrt{\frac{1}{\sum(1/\sigma_C^2)}+\frac{\sigma_\text{scat}^2}{6}},
\end{align}
where $\sigma_C$ and $\sigma_\text{scat}$ represent the $T_\text{eff}$ uncertainty {resulting from each colour} and scatter size , respectively.

The initial Fe abundance is assumed from the literature, and the process is iterated after re-deriving the Fe abundance from Section \ref{subsec:OptFelines} until a consistent value within the uncertainty is found. The derived temperature{s} with their uncertaint{ies} for each colour are given in Table \ref{tab:obs_color_teff}, while the adopted values are given in Table \ref{tab:adopted_params}.

\subsection{Surface Gravity}
Surface gravity, $\log g$, for each star is derived using the fundamental relation:
\begin{align}
\log g = &4\log T_\text{eff}+\log(M/M_\odot)-10.61+\\&0.4(\text{BC}_V+V+5\log \varpi+5-\notag\\&R_V\cdot E(B-V)_\text{cor}-M_{\text{bol},\odot}\notag),
\end{align}
with $T_\text{eff}$ derived from the previous step and stellar mass $(M)$ { of $0.80\pm0.25M_\odot$, which is the typical mass for metal-poor Red Giants obtained from asteroseismology \citep{Huber_2024,Marasco_2025}}. The bolometric correction in the \textit{V}-band ($\text{BC}_V$) is calculated by interpolation from the bolometric correction grids provided by the \textbf{M}ESA \textbf{I}sochrones and \textbf{S}tellar \textbf{T}racks \citep[\texttt{MIST}:][]{2016ApJS..222....8D,2016ApJ...823..102C} project via the \texttt{isochrones}\footnote{\url{https://isochrones.readthedocs.io/en/latest/}} package \citep{2015ascl.soft03010M}, which depends on $T_\text{eff}$, $\log g$, $\text{[Fe/H]}$, and $A_V$. The $V$-band magnitudes with their uncertainties are taken from the catalogue \citep{2000AATycho2}. The parallaxes and their uncertainties are taken from the \textit{Gaia} DR3 catalogue \citep{2023AAGaiaDR3}. $R_V$ is assumed to follow the Galactic extinction law, $R_V=3.1$ \citep{1989ApJ_cardelli}, along with the solar constants of $\log T_{\text{eff},\odot}=3.7617$, $\log g_\odot=4.438$, and $M_{\text{bol,}\odot}=4.74$ \citep{2016AJ....152...41P}. Using the same method as for $T_\text{eff}$, $10^6$ samples are drawn for each input parameter; $\log g$ and its uncertainty are determined from the median and standard deviation of the distribution. Again, because of the dependency between $\text{BC}_V$ and $\log g$, this analysis is iterated several times until convergence for $\log g$ is achieved. The input parameters ($V$-band magnitude, parallax, reddening) are given in Table \ref{tab:obs_color_teff}, while the final adopted $\log g$ values are given in Table \ref{tab:adopted_params}.

\begin{deluxetable}{lccccccc}
\centering
\tabletypesize{\small}
\setlength{\tabcolsep}{2.5pt}
\digitalasset
\tablewidth{0pt}
\tablecaption{Derived Color Temperatures\label{tab:obs_color_teff}}
\tablehead{
\colhead{\multirow{2}{*}{Target}} & \colhead{$\text{E(B-V)}_\text{cor}$} & \colhead{$T_{BP-RP}$} & \colhead{$T_{BP-G}$} & \colhead{$T_{G-RP}$} & \colhead{$T_{BP-K_s}$} & \colhead{$T_{RP-K_s}$} & \colhead{$T_{G-K_s}$} \\
\colhead{} & \colhead{(mag)} & \colhead{(K)} & \colhead{(K)} & \colhead{(K)} & \colhead{(K)} & \colhead{(K)} & \colhead{(K)}
}
\startdata
BD-02$^\circ$5957 & 0.046(2) & 4472(20) & 4472(21) & 4479(25) & 4485(20) & 4509(35) & 4506(25) \\
BD-07$^\circ$2674 & 0.027(1) & 4653(22) & 4661(24) & 4650(26) & 4675(27) & 4697(49) & 4686(34) \\
BD-14$^\circ$5890 & 0.030(2) & 4981(16) & 4984(20) & 4975(23) & 4980(35) & 4976(63) & 4978(45) \\
BD-15$^\circ$5781 & 0.036(1) & 4849(18) & 4848(21) & 4850(24) & 4833(21) & 4818(37) & 4830(26) \\
BD-18$^\circ$271 & 0.014(1) & 4277(17) & 4267(19) & 4291(24) & 4306(19) & 4365(32) & 4350(24) \\
BD-18$^\circ$5550 & 0.137(17) & 4877(50) & 4897(51) & 4857(53) & 4829(47) & 4785(53) & 4808(48) \\
BD-20$^\circ$6008 & 0.037(1) & 4820(18) & 4828(21) & 4811(24) & 4819(26) & 4818(47) & 4819(33) \\
BD+03$^\circ$2782 & 0.024(1) & 4668(13) & 4663(16) & 4666(21) & 4659(28) & 4660(50) & 4664(36) \\
BD+30$^\circ$2611 & 0.017(1) & 4343(14) & 4336(15) & 4342(20) & 4295(16) & 4297(26) & 4305(20) \\
CS 29502-092 & 0.084(4) & 5144(27) & 5147(32) & 5143(30) & 5142(26) & 5138(42) & 5140(30) \\
CS 30314-067 & 0.059(1) & 4515(22) & 4512(23) & 4529(26) & 4532(19) & 4555(33) & 4552(24) \\
HD 107752 & 0.026(2) & 4895(18) & 4893(22) & 4897(24) & 4919(18) & 4938(31) & 4928(22) \\
HD 110184 & 0.019(1) & 4346(15) & 4342(16) & 4349(21) & 4361(20) & 4400(35) & 4391(26) \\
HD 115444 & 0.012(1) & 4873(17) & 4885(21) & 4858(23) & 4857(20) & 4841(36) & 4848(25) \\
HD 118055 & 0.080(1) & 4369(14) & 4361(15) & 4369(21) & 4332(16) & 4335(26) & 4345(19) \\
HD 122563 & 0.020(1) & 4709(16) & 4718(18) & 4695(23) & 4770(32) & 4827(60) & 4792(42) \\
HD 187111 & 0.118(7) & 4442(20) & 4439(21) & 4434(26) & 4392(21) & 4380(27) & 4394(23) \\
HD 204543 & 0.035(2) & 4776(14) & 4780(17) & 4764(22) & 4754(16) & 4739(26) & 4748(19) \\
HD 221170 & 0.107(5) & 4642(18) & 4641(19) & 4635(24) & 4633(18) & 4634(27) & 4637(21) \\
HD 25329 & 0.000(1) & 4798(13) & 4802(19) & 4794(24) & 4777(16) & 4792(28) & 4795(20) \\
HD 4306 & 0.030(1) & 5096(21) & 5111(26) & 5081(26) & 5074(26) & 5052(45) & 5061(32) \\
HD 6268 & 0.015(1) & 4826(15) & 4839(18) & 4807(22) & 4827(18) & 4828(31) & 4825(22) \\
HD 85773 & 0.040(1) & 4428(15) & 4423(16) & 4428(21) & 4450(17) & 4491(30) & 4478(22) \\
HD 8724 & 0.092(5) & 4770(19) & 4774(21) & 4758(25) & 4756(22) & 4747(33) & 4751(26) \\
HD 88609 & 0.008(1) & 4616(16) & 4622(18) & 4607(23) & 4622(17) & 4633(30) & 4630(21) \\
HE 1116-0634 & 0.047(1) & 4701(26) & 4704(27) & 4707(29) & 4702(22) & 4702(38) & 4706(27) \\
HE 1523-0901 & 0.138(2) & 4692(16) & 4681(19) & 4702(23) & 4752(25) & 4809(46) & 4781(32) \\
LAMOST J0040+2729 & 0.046(3) & 4686(18) & 4682(20) & 4690(24) & 4683(19) & 4683(32) & 4689(23) \\
LAMOST J2114-0616 & 0.094(4) & 4897(19) & 4880(22) & 4913(26) & 4900(24) & 4902(40) & 4907(29) \\
LAMOST J2217+2104 & 0.054(3) & 4553(38) & 4577(39) & 4553(37) & 4579(28) & 4603(46) & 4589(33) \\
LAMOST J2347+2851 & 0.056(2) & 4941(16) & 4944(20) & 4936(23) & 4940(19) & 4936(33) & 4939(23) \\
TYC 3407-1352-1 & 0.056(4) & 4783(23) & 4783(25) & 4785(27) & 4785(21) & 4787(32) & 4789(24) \\
TYC 3814-1598-1 & 0.007(1) & 4388(19) & 4386(20) & 4398(24) & 4429(17) & 4483(29) & 4465(21) \\
UCAC4 425-121652 & 0.100(6) & 4620(22) & 4620(24) & 4619(28) & 4611(24) & 4609(36) & 4616(28) \\
UCAC4 515-137892 & 0.102(6) & 4597(26) & 4600(27) & 4600(30) & 4613(23) & 4632(33) & 4626(25) \\
\enddata
\end{deluxetable}

\subsection{Microturbulence velocity \& Carbon Abundance}\label{ssec:vmic_cabund}
The microturbulence velocity is determined in this work. The empirical relation among $T_\text{eff}$, $\log g$, and $\xi_t$ is adopted using Eq. (2) in \cite{2016AA_3Dmicro}, based on 3D atmospheric modelling of Fe lines in member stars of the Hyades open cluster. To derive $\xi_t$, a Monte Carlo simulation with $10^6$ samples is employed, using parameter inputs from previous steps, and the median and standard deviation of the resulting distribution are adopted.

Since carbon and oxygen are interrelated through CO molecules \citep{2007A&A...469..687C}, {the derived oxygen abundances could depend on the carbon abundances assumed in the analysis}. However, a previous study for HD 122563, which involves simultaneous abundance analysis of CNO elements, shows that the derived O abundance from NIR OH lines is much less sensitive to the adopted carbon abundance \citep{2018MNRAS.475.3369C}. In addition, the targets have low [C/O] ratio{s}, which, as confirmed by sensitivity analysis (Section \ref{ssec:sensi_Oxy}), resultin a low dependency between the C and O abundances, at least in the 1D/LTE framework. Therefore, the carbon abundance values are adopted from the literature with no correction, mostly derived from the optical CH-band.

The adopted microturbulence velocity and carbon abundance $\text{A(C)}$ for each star are provided in Table \ref{tab:adopted_params}.
\begin{deluxetable}{lcccccccc}
\centering
\digitalasset
\tablewidth{0pt}
\tablecaption{Adopted stellar  parameters\label{tab:adopted_params}}
\tablehead{
\colhead{\multirow{2}{*}{Target}} & \colhead{$T_\text{eff}$} & \colhead{$\sigma_{T_\text{eff}}$} & \colhead{$\log g$} & \colhead{$\sigma_{\log g}$} & \colhead{$\xi_t$} & \colhead{$\sigma_{\xi_t}$} & \colhead{$\text{A(C)}_\text{lit}$} & \colhead{\multirow{2}{*}{Ref}} \\
\colhead{} & \colhead{(K)} & \colhead{(K)} & \colhead{(cgs)} & \colhead{(cgs)} & \colhead{($\text{km s}^{-1}$)} & \colhead{($\text{km s}^{-1}$)} & \colhead{(dex)} & \colhead{}
}
\startdata
BD-02$^\circ$5957 & 4483 & 11 & 0.903 & 0.142 & 2.14 & 0.14 & 4.67 & [4] \\
BD-07$^\circ$2674 & 4664 & 13 & 1.236 & 0.127 & 2.03 & 0.12 & 6.33 & [3] \\
BD-14$^\circ$5890 & 4980 & 10 & 2.090 & 0.121 & 1.64 & 0.09 & 6.53 & [9] \\
BD-15$^\circ$5781 & 4842 & 10 & 1.749 & 0.123 & 1.77 & 0.10 & 5.22 & [6] \\
BD-18$^\circ$271 & 4299 & 17 & 0.782 & 0.129 & 2.04 & 0.13 & 5.47 & [11] \\
BD-18$^\circ$5550 & 4842 & 26 & 2.124 & 0.124 & 1.49 & 0.09 & 5.28 & [6] \\
BD-20$^\circ$6008 & 4820 & 10 & 1.591 & 0.123 & 1.88 & 0.11 & 4.90 & [6] \\
BD+03$^\circ$2782 & 4665 & 8 & 1.426 & 0.123 & 1.86 & 0.11 & 5.56 & [12] \\
BD+30$^\circ$2611 & 4324 & 11 & 0.966 & 0.123 & 1.89 & 0.11 & 6.27 & [13] \\
CS29502-092 & 5143 & 12 & 2.582 & 0.127 & 1.43 & 0.08 & 6.59 & [7] \\
CS30314-067 & 4530 & 12 & 1.079 & 0.149 & 2.03 & 0.14 & 6.27 & [6] \\
HD 107752 & 4909 & 11 & 1.727 & 0.124 & 1.85 & 0.10 & 5.28 & [8] \\
HD 110184 & 4355 & 12 & 0.592 & 0.126 & 2.30 & 0.13 & 5.36 & [10] \\
HD 115444 & 4864 & 11 & 1.720 & 0.122 & 1.82 & 0.10 & 5.38 & [15] \\
HD 118055 & 4354 & 9 & 0.846 & 0.127 & 2.04 & 0.12 & 5.77 & [16] \\
HD 122563 & 4722 & 22 & 1.374 & 0.121 & 1.97 & 0.11 & 5.06 & [6] \\
HD 187111 & 4416 & 14 & 1.208 & 0.123 & 1.78 & 0.11 & 6.24 & [8] \\
HD 204543 & 4764 & 9 & 1.538 & 0.122 & 1.87 & 0.10 & 6.02 & [13] \\
HD 221170 & 4638 & 9 & 1.342 & 0.121 & 1.91 & 0.11 & 5.35 & [17] \\
HD 25329 & 4793 & 8 & 4.843 & 0.149 & 0.85 & 0.03 & 7.29 & [8] \\
HD 4306 & 5085 & 14 & 2.298 & 0.121 & 1.57 & 0.08 & 5.78 & [14] \\
HD 6268 & 4826 & 9 & 1.547 & 0.122 & 1.93 & 0.11 & 5.05 & [6] \\
HD 85773 & 4441 & 13 & 0.830 & 0.125 & 2.16 & 0.13 & 5.54 & [8] \\
HD 8724 & 4762 & 10 & 1.902 & 0.121 & 1.58 & 0.09 & 6.45 & [8] \\
HD 88609 & 4621 & 9 & 1.169 & 0.123 & 2.05 & 0.12 & 4.92 & [6] \\
HE 1116-0634 & 4704 & 11 & 1.202 & 0.152 & 2.11 & 0.14 & 4.49 & [18] \\
HE 1523-0901 & 4711 & 22 & 1.391 & 0.132 & 1.94 & 0.12 & 4.99 & [19] \\
J0040+2729 & 4685 & 9 & 1.388 & 0.127 & 1.92 & 0.11 & 5.28 & [1] \\
J2114-0616 & 4898 & 11 & 1.758 & 0.125 & 1.82 & 0.10 & 6.52 & [1] \\
J2217+2104 & 4575 & 16 & 0.669 & 0.279 & 2.50 & 0.30 & 5.50 & [1] \\
J2347+2851 & 4940 & 9 & 1.780 & 0.125 & 1.84 & 0.10 & 5.80 & [1] \\
TYC 3407-1352-1 & 4785 & 10 & 1.687 & 0.125 & 1.77 & 0.10 & 5.34 & [1] \\
TYC 3814-1598-1 & 4420 & 18 & 0.788 & 0.126 & 2.18 & 0.13 & 4.94 & [5] \\
UCAC4 425-121652 & 4617 & 11 & 1.388 & 0.140 & 1.85 & 0.12 & 5.43 & [2] \\
UCAC4 515-137892 & 4611 & 12 & 1.035 & 0.152 & 2.16 & 0.15 & 6.11 & [3] \\
\enddata
\tablecomments{ References for $\mathrm{A}(\mathrm{C})$
 : [1] \cite{2022ApJ...Li}; [2] \cite{2019ApJ...Placco}; [3] \cite{2022ApJ...Shank}; [4] \cite{2018ApJ...hansen}; [5] \cite{2018ApJ...Holmbeck}; [6] \cite{Roederer_2014}; [7] \cite{2019AAarentsen}; [8] \cite{2004ApJ...simmerer}; [9] \cite{2022ApJ...zepeda}; [10] \cite{2004ApJ...Honda}; [11] \cite{2002ApJ...Melendez} ; [12] \cite{1982PASP...Kraft} ; [13] \cite{2008ApJ...Aok} ; [14] \cite{2011ApJ...Honda} ; [15] \cite{2007ApJ...Lai} ; [16] \cite{2013ApJ...Li} ; [17] \cite{2013PASJ...Takeda} ; [18] \cite{2011ApJ...Hollek} ; [19] \cite{2024AAMishenina}.}
\end{deluxetable}




\section{Chemical Abundance Analysis} \label{sec:AbundanceAnalysis}
\subsection{Model Atmosphere \& Radiative Transfer Code}
To derive Fe and O abundances, the commonly used 1D, LTE MARCS atmospheric model {grid} \citep{2008AAMARCS} is adopted, with spherical models for red giant stars ($\log g<3$) and plane-parallel models for dwarf stars. This has an advantage for the samples, which are mostly red giant (RG) stars. To generate synthetic spectra, the spectral synthesis code \texttt{turbospectrum} \citep{2012ascl.TS} is used{. This code} is capable of incorporating non-LTE departure coefficient grids for some elements while generating the spectra \citep[\texttt{TS-NLTE},][]{2023AAGerberTSFitPy}. All of the spectral synthesis from \texttt{TS-NLTE} and the spectral fitting process are done under a \texttt{PYTHON}-based wrapper \citep[\texttt{TSFitPy,}][]{2023AAGerberTSFitPy, 2023MNRAS_TSFitypy}.

In addition, 3D LTE synthetic spectra for the oxygen forbidden line at 6300 \AA~are generated to investigate 3D effects on the abundance determination from the atomic [OI] line. This line is insensitive to NLTE, simplifying the radiative transfer calculations. The synthetic spectra are computed using the \texttt{M3DIS} code \citep{2024A&A...688A..52E}. The 3D {radiative hydrodynamic} (RHD) model atmospheres are adopted from the updated \texttt{Stagger} grid \citep{RodriguezDiaz2024}, while the oxygen model atom is taken from \citet{2021MNRAS.508.2236B}. 

To determine the Fe abundance, standard equivalent width analysis is adopted, while for the O abundance spectral profile fitting around {[OI] and OH lines are} performed. While generating the 1D spectra, the line broadening is determined by the macroturbulence parameter, the rotational velocity of the star, and the instrumental profile. Since {determining each parameter is beyond the cope of this work,} a fixed $v\sin i\sim3\text{ $\text{km s}^{-1}$}$ is adopted, and macroturbulence is set as a freely adjustable parameter for the [OI] and OH lines analysis. For instrumental broadening, a constant $\mathcal{R}\sim70,000$ is adopted for the IRD spectra, and various $\mathcal{R}$ values corresponding to each instrumental setup are adopted for the archival [OI] spectra, which are given in Table \ref{tab:logobs}. All of the abundance analyses in this work adopt the solar abundance values from \cite{2022A&A...661A.140M}.

\subsection{Iron Abundance} \label{subsec:OptFelines}
The iron abundance is derived from optical line{s} , as the available NIR Fe lines are very limited \citep{aoki2025elementalabundances44metalpoor}. The EWs of Fe I and Fe II lines are gathered from the literature (see Table \ref{tab:Fe_abund} for details), except for UCAC4 515-137892, for which no available sources can be found. For this star, $\text{[Fe/H]} = -3.27$ is adopted \citep{2019ApJ...Placco}. For the remaining stars, line-by-line abundances are derived by matching the synthetic EWs to the observed values, using the adopted stellar parameters from the previous section as inputs. The stellar parameters of LAMOST J2217+2104 and TYC 3814-1598-1, which are $\text{[Fe/H]} \leqslant -3$ and $\log g \leqslant 1$, are not covered by the \texttt{MARCS} model grid. To address this issue, a fixed $\log g = 1$ is set, allowing a more flexible adjustment of the iron abundance. 
For the spectral line data for Fe, the line list from the Gaia-ESO Survey \citep[\texttt{GES} :][]{2021A&A...645A.106H} is adopted, which covers 475 to 685 nm and from 850 to 895 nm. For the rest of the Fe lines outside these regions, data from the \texttt{VALD} database \citep{1995A&ASvald1,2000BaltA.vald2,2015PhySvald3} are then adopted.

\begin{figure}[t!]
 \centering
 \includegraphics[width=\linewidth]{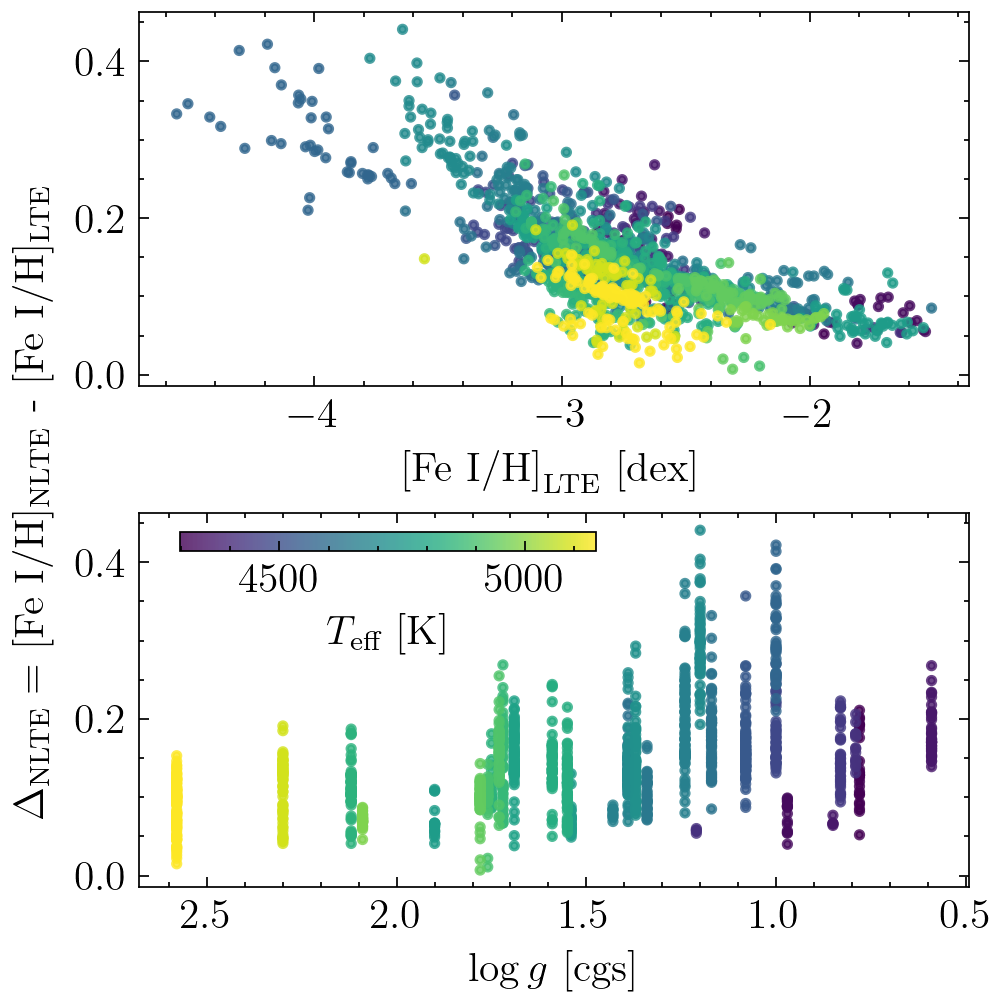}
 \caption{The non-LTE corrections of the Fe abundance from Fe I lines. Each point represents the result for an individual Fe I line for each star, with colour-codes according to the adopted effective temperature. The upper panel shows the correction as a function of the Fe I LTE abundance, while the lower panel shows the correction as a function of the adopted surface gravity.}
 \label{fig:nlteFeI}
\end{figure}

It is widely known that Fe I lines are significantly influenced by non-LTE effects. To enhance accuracy, non-LTE corrections are applied to Fe I lines using the online service \texttt{INASAN}\footnote{\url{https://spectrum.inasan.ru/nLTE2/}}, which provides a literature-based correction grid for many elements{, including} Fe I \citep{2011A&A...528A..87M}. A \texttt{PYTHON}-based program is developed that enables automatic querying of multiple lines in a single run, inspired by previous works \citep{2025AJ....169..172M}. The non-LTE corrections are displayed in Figure \ref{fig:nlteFeI}. 

\begin{figure}[t!]
 \centering
 \includegraphics[width=\linewidth]{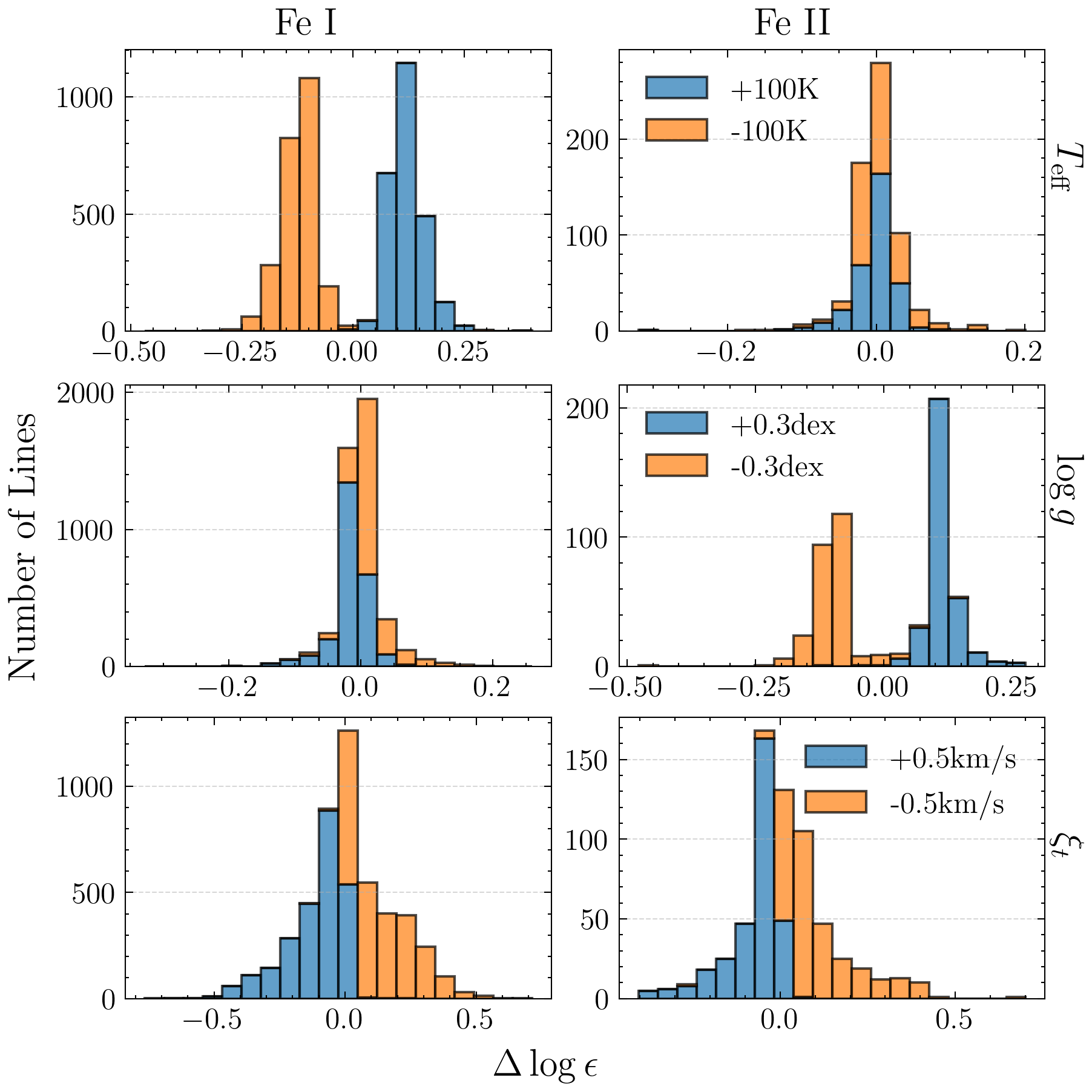}
 \caption{The histogram of Fe I (left panels) and Fe II (right panels) abundance changes corresponding to each stellar parameter: $T_\text{eff}$ on the first row, $\log g$ on the second row, and microturbulence velocity on the last row.}
 \label{fig:sens_Fe}
\end{figure}

To test the sensitivity of the derived Fe abundance, the input parameters $T_\text{eff}$, $\log g$, and $\xi_t$ are perturbed by $\pm100\text{ K}$, $\pm0.3\text{ dex}$, and $\pm0.5\text{ km s}^{-1}$, respectively. This procedure is carried out for each Fe line in all stars. The results of the sensitivity analysis for Fe I and Fe II lines are presented in Figure \ref{fig:sens_Fe}. In general, Fe I lines are sensitive to the changing adopted $T_\text{eff}$ {due to their minor population in cool star atmosphere,} where an increase of 100 K lead{s} to an Fe abundance change of $\sim+0.15$ dex. Fe II {lines, on the other hand, are sensitive to the change of surface gravity, as their line strength is affected by the change of continous opacity which depends on electron pressure, with a change of $\pm0.3$ dex in $\log g$, give an Fe abundance change of $\sim\pm0.1$ dex} The chang{e} of microturbulence gives a varied impact depending on the Fe line strength.

The weighted mean values of the Fe I abundance, which have been corrected for non-LTE effects, are calculated and adopted as $\text{[Fe/H]}${. The exception is} the dwarf star HD 25329, for which the NLTE corrections are not available by the above procedures{, h}ence, the Fe I abundance from the LTE analysis is adopted. The weight assigned to each Fe line is based on the line-by-line uncertainty arising from input atmospheric parameters, denoted as $\sigma_\text{atm}$. This parameter is calculated following Equation \ref{eq:sig_atm_line}:
\begin{align}
\sigma_\text{atm,line}^2=\sum_{i=1}^{N}\left(\frac{\partial\log\epsilon}{\partial X_i}\cdot \sigma_{X_i}\right)^2,\label{eq:sig_atm_line}
\end{align}
where $\partial\log\epsilon/\partial X_i$ represents the abundance change due to the change of input parameter $X_i$, and $\sigma_{X_i}$ denotes the uncertainty of each input parameter, as obtained in the previous section, except for temperature. Here, $N$ is 4, corresponding to $T_\text{eff},\; \log g,\; \xi_t$, and [Fe/H]. The {uncertainties of} temperature due to photometry errors are as small as $\sim10-20$ K, which could be underestimate as $T_\text{eff}$ error; hence, for a more reasonable and conservative approach, a constant $\sigma_{T_\text{eff}}=50$ K is assumed across the sample, which is supported by the typical scatter of the fit residual {in determining the color-$T_\text{eff}$ relation} \citep[$\sim49-80$ K for giant stars,][]{2021AA653A..90M}.

To compute the abundance uncertainty of Fe I (both LTE and NLTE) and Fe II (LTE), the following procedure is employed. First, the line-by-line scatter, $\sigma_\text{scat}$, is calculated as the standard deviation of Fe I abundances from both the LTE and NLTE analyses, and the Fe II abundance from the LTE analysis. To assess the impact of the atmospheric parameters, the mean of $\partial\log\epsilon/\partial X_i$ is derived for all lines with respect to each parameter, and the following equation is applied:
\begin{align}
\sigma_\text{atm,star}^2 = \sum_{i=1}^{N} \left( \biggl\langle \frac{\partial \log \epsilon}{\partial X_i} \biggr\rangle \cdot \sigma_{X_i} \right)^2, \label{eq:sig_atm_star}
\end{align}
which differs slightly from Equation \ref{eq:sig_atm_line}. Finally, the total uncertainty for the Fe lines is computed using the relation:
\begin{align}
\sigma_\text{tot} = \sqrt{\sigma_\text{atm,star}^2 + \frac{\sigma_\text{scat}^2}{N_\text{line}}}. \label{eq:sigma_tot}
\end{align}
All the values discussed in this section are provided in Table \ref{tab:Fe_abund}.



\startlongtable
\begin{longrotatetable}
\begin{deluxetable*}{lccccccccccccccc}
\centering
\digitalasset
\tablewidth{0pt}
\tablecaption{Derived Fe abundances of our samples\label{tab:Fe_abund}}
\tablehead{
\colhead{Designation} & \colhead{Ref} & \colhead{$\sigma_\text{atm,I}^\text{1L}$} & \colhead{$N_\text{lin,I}^\text{1L}$} & \colhead{$\sigma_\text{scat,I}^\text{1L}$} & \colhead{$\sigma_\text{tot,I}^\text{1L}$} & \colhead{$\langle\text{[Fe I/H]}\rangle^\text{1L}$} & \colhead{$N_\text{lin,I}^\text{1N}$} & \colhead{$\sigma_\text{scat,I}^\text{1N}$} & \colhead{$\sigma_\text{tot,I}^\text{1N}$} & \colhead{$\langle\text{[Fe I/H]}\rangle^\text{1N}$} & \colhead{$\sigma_\text{atm,II}^\text{1L}$} & \colhead{$N_\text{lin,II}^\text{1L}$} & \colhead{$\sigma_\text{scat,II}^\text{1L}$} & \colhead{$\sigma_\text{tot,II}^\text{1L}$} & \colhead{$\langle\text{[Fe II/H]}\rangle^\text{1L}$}
}
\startdata
BD-02$^\circ$5957 & [1] & 0.069 & 64 & 0.155 & 0.072 & -3.092 & 43 & 0.144 & 0.073 & -2.918 & 0.051 & 11 & 0.263 & 0.094 & -3.006 \\
BD-07$^\circ$2674 & [2] & 0.065 & 162 & 0.174 & 0.067 & -3.052 & 88 & 0.137 & 0.067 & -2.843 & 0.039 & 16 & 0.063 & 0.042 & -3.044 \\
BD-14$^\circ$5890 & [3] & 0.067 & 48 & 0.100 & 0.068 & -2.147 & 34 & 0.110 & 0.069 & -2.071 & 0.049 & 10 & 0.093 & 0.057 & -2.214 \\
BD-15$^\circ$5781 & [4] & 0.062 & 59 & 0.180 & 0.066 & -2.517 & 32 & 0.162 & 0.068 & -2.341 & 0.048 & 14 & 0.116 & 0.057 & -2.540 \\
BD-18$^\circ$271 & [5] & 0.083 & 43 & 0.128 & 0.085 & -2.624 & 27 & 0.146 & 0.088 & -2.468 & 0.067 & 4 & 0.065 & 0.074 & -2.378 \\
BD-18$^\circ$5550 & [6] & 0.063 & 83 & 0.094 & 0.064 & -2.968 & 70 & 0.076 & 0.064 & -2.857 & 0.048 & 8 & 0.077 & 0.055 & -2.806 \\
BD-20$^\circ$6008 & [6] & 0.066 & 64 & 0.129 & 0.068 & -2.718 & 53 & 0.108 & 0.068 & -2.579 & 0.045 & 6 & 0.046 & 0.048 & -2.803 \\
BD+03$^\circ$2782 & [7] & 0.073 & 23 & 0.114 & 0.077 & -2.074 & 10 & 0.087 & 0.078 & -1.967 & 0.053 & 9 & 0.174 & 0.079 & -2.021 \\
BD+30$^\circ$2611 & [5] & 0.089 & 23 & 0.103 & 0.092 & -1.654 & 15 & 0.100 & 0.093 & -1.604 & 0.092 & 3 & 0.045 & 0.095 & -1.481 \\
CS29502-092 & [8] & 0.063 & 169 & 0.202 & 0.065 & -2.745 & 107 & 0.159 & 0.065 & -2.682 & 0.048 & 11 & 0.097 & 0.056 & -2.669 \\
CS30314-067 & [8] & 0.087 & 148 & 0.260 & 0.089 & -2.916 & 96 & 0.178 & 0.088 & -2.778 & 0.055 & 12 & 0.103 & 0.062 & -2.894 \\
HD 107752 & [5] & 0.062 & 62 & 0.145 & 0.064 & -2.806 & 52 & 0.127 & 0.064 & -2.658 & 0.046 & 8 & 0.074 & 0.053 & -2.840 \\
HD 110184 & [9] & 0.082 & 52 & 0.150 & 0.084 & -2.731 & 41 & 0.109 & 0.084 & -2.570 & 0.059 & 7 & 0.099 & 0.07 & -2.668 \\
HD 115444 & [9] & 0.072 & 67 & 0.132 & 0.074 & -2.946 & 61 & 0.110 & 0.074 & -2.800 & 0.047 & 12 & 0.097 & 0.054 & -2.962 \\
HD 118055 & [10] & 0.067 & 16 & 0.093 & 0.071 & -2.020 & 5 & 0.067 & 0.073 & -1.910 & 0.066 & 9 & 0.131 & 0.079 & -1.888 \\
HD 122563 & [9] & 0.078 & 82 & 0.140 & 0.079 & -2.763 & 69 & 0.130 & 0.079 & -2.623 & 0.047 & 13 & 0.362 & 0.111 & -2.837 \\
HD 187111 & [10] & 0.058 & 14 & 0.114 & 0.066 & -1.791 & 4 & 0.057 & 0.065 & -1.661 & 0.074 & 10 & 0.19 & 0.095 & -1.722 \\
HD 204543 & [11] & 0.068 & 90 & 0.162 & 0.070 & -1.868 & 25 & 0.122 & 0.072 & -1.776 & 0.075 & 13 & 0.255 & 0.103 & -1.744 \\
HD 221170 & [12] & 0.065 & 125 & 0.112 & 0.066 & -2.084 & 40 & 0.112 & 0.068 & -1.988 & 0.056 & 21 & 0.151 & 0.065 & -2.103 \\
HD 25329 & [10] & 0.043 & 19 & 0.137 & 0.054 & -1.823 & 0 & $\dotsi$ & $\dotsi$ & $\dotsi$ & 0.079 & 1 & $\dotsi$ & 0.079 & -2.173 \\
HD 4306 & [9] & 0.067 & 65 & 0.144 & 0.070 & -2.819 & 60 & 0.133 & 0.069 & -2.709 & 0.044 & 10 & 0.114 & 0.057 & -2.812 \\
HD 6268 & [9] & 0.076 & 70 & 0.122 & 0.077 & -2.547 & 60 & 0.106 & 0.077 & -2.430 & 0.050 & 11 & 0.078 & 0.055 & -2.488 \\
HD 85773 & [5] & 0.079 & 52 & 0.100 & 0.081 & -2.603 & 37 & 0.090 & 0.081 & -2.462 & 0.059 & 7 & 0.075 & 0.066 & -2.533 \\
HD 8724 & [13] & 0.080 & 19 & 0.085 & 0.083 & -1.719 & 19 & 0.072 & 0.082 & -1.650 & 0.064 & 3 & 0.052 & 0.071 & -1.522 \\
HD 88609 & [9] & 0.084 & 62 & 0.266 & 0.091 & -3.096 & 56 & 0.259 & 0.091 & -2.919 & 0.040 & 11 & 0.094 & 0.049 & -3.035 \\
HE 1116-0634 & [14] & 0.057 & 74 & 0.140 & 0.059 & -3.468 & 51 & 0.088 & 0.058 & -3.172 & 0.052 & 10 & 0.083 & 0.058 & -3.417 \\
HE 1523-0901 & [1] & 0.068 & 55 & 0.216 & 0.074 & -2.594 & 38 & 0.196 & 0.075 & -2.462 & 0.048 & 7 & 0.21 & 0.093 & -2.602 \\
J0040+2729 & [15] & 0.071 & 99 & 0.120 & 0.073 & -2.669 & 58 & 0.100 & 0.073 & -2.522 & 0.052 & 12 & 0.103 & 0.06 & -2.724 \\
J2114-0616 & [15] & 0.065 & 68 & 0.107 & 0.067 & -2.414 & 40 & 0.088 & 0.067 & -2.296 & 0.052 & 10 & 0.12 & 0.065 & -2.312 \\
J2217+2104 & [16] & 0.083 & 49 & 0.223 & 0.089 & -4.003 & 40 & 0.200 & 0.089 & -3.702 & 0.088 & 2 & 0.212 & 0.173 & -4.064 \\
J2347+2851 & [15] & 0.066 & 93 & 0.109 & 0.067 & -2.315 & 56 & 0.106 & 0.068 & -2.215 & 0.054 & 12 & 0.089 & 0.06 & -2.457 \\
TYC 3407-1352-1 & [15] & 0.063 & 100 & 0.128 & 0.064 & -2.982 & 69 & 0.103 & 0.064 & -2.831 & 0.044 & 11 & 0.111 & 0.056 & -2.948 \\
TYC 3814-1598-1 & [4] & 0.078 & 32 & 0.039 & 0.079 & -2.875 & 15 & 0.060 & 0.080 & -2.707 & 0.051 & 9 & 0.081 & 0.058 & -2.819 \\
UCAC4 425-121652 & [1] & 0.069 & 61 & 0.197 & 0.074 & -2.754 & 41 & 0.157 & 0.073 & -2.617 & 0.058 & 8 & 0.152 & 0.079 & -2.577 \\
\enddata
\tablecomments{References for Fe equivalent widths: [1] \cite{2018ApJ...hansen}; [2] \cite{2018ApJ...cain}; [3] \cite{2013ApJ...ishigaki}; [4] \cite{2018ApJ...Holmbeck}; [5] \cite{2012ApJ...753...64I}; [6] \cite{Roederer_2014}; [7] \cite{2000AJ....120.1841F}; [8] \cite{2002ApJ...567.1166A}; [9] \cite{2004ApJ...Honda}; [10] \cite{2000AJ....120.1841F}; [11] \cite{2008ApJ...Aok}; [12] \cite{2005AA430..255Y}; [13] \cite{2017AA608A..89M}; [14] \cite{2011ApJ...Hollek}; [15] \cite{2022ApJ...Li}; [16] \cite{2018PASJ...70...94A} \\
Abbreviation: 1L = 1D/LTE; 1N = 1D/NLTE; I = Fe I lines; II = Fe II lines
}
\end{deluxetable*}
\end{longrotatetable}

\subsection{Oxygen Abundance} \label{ssec:O_abundance}
The oxygen abundance is derived from the \textit{H}-band first-overtone sequence vibro-rotational OH lines $(X^2\Pi,\;\Delta\nu=2)$. All of the spectral line data for the OH lines are adopted from \cite{2016JQSRT.168..142B}, which are compiled together with other spectral lines in the \textit{H}-band region in the \texttt{APOGEE} DR16 line list \citep{2021AJ....161..254S}. The abundance is derived through fitting with synthetic spectra based on 1D/LTE models to the observed IRD spectra. The input parameters derived from the previous step and also the carbon abundance from the literature are adopted. For LAMOST J2217+2104, which has $\log g<1$ and $[\text{Fe/H}]<-3$ that is not covered by the \texttt{MARCS} grid, the atmospheric model with $[\text{Fe/H}]=-3$ is adopted .

In some stars, the continuum around the line is heavily altered due to imperfections in the reduction process. {F}or the sake of simplicity, if the OH lines are too wide compared to other lines (as indicated by the derived macroturbulence velocity) or too weak/shallow comparable with noise level, the lines are excluded. To detect possible contamination/blending from other lines, theoretical spectra without oxygen are also synthesized, which then helps to select clean OH lines.

For reference, an analysis is also performed on Arcturus with the {two} high-resolution spectra of the \cite{1995PASP..107.1042H} atlas ($\mathcal{R}\sim100,000$) the summer and winter seasons. The adopted stellar parameters are $T_\text{eff}=4300\pm70$ K, $\log g =1.66\pm0.20$ dex, $\text{[Fe/H]}=-0.50\pm0.10$ dex, and $\xi_t=1.74\pm0.08\text{ km s}^{-1}$ \citep{2024A&A...684A..85V}, which agree well with those obtained by a previous study by \cite{2011ApJ...743..135R}. For higher precision, constant carbon and nitrogen abundances of $\text{[C/Fe]}=0.05\pm0.05$ and $\text{[N/Fe]}=0.45\pm0.11$, derived from {NIR} spectra by the same literature, are also adopted. The O abundance results on Arcturus and HD 122563 are discussed in Appendix \ref{ssec:o_arc_hd}.

\begin{figure*}[t!]
 \centering
 \includegraphics[width=\linewidth]{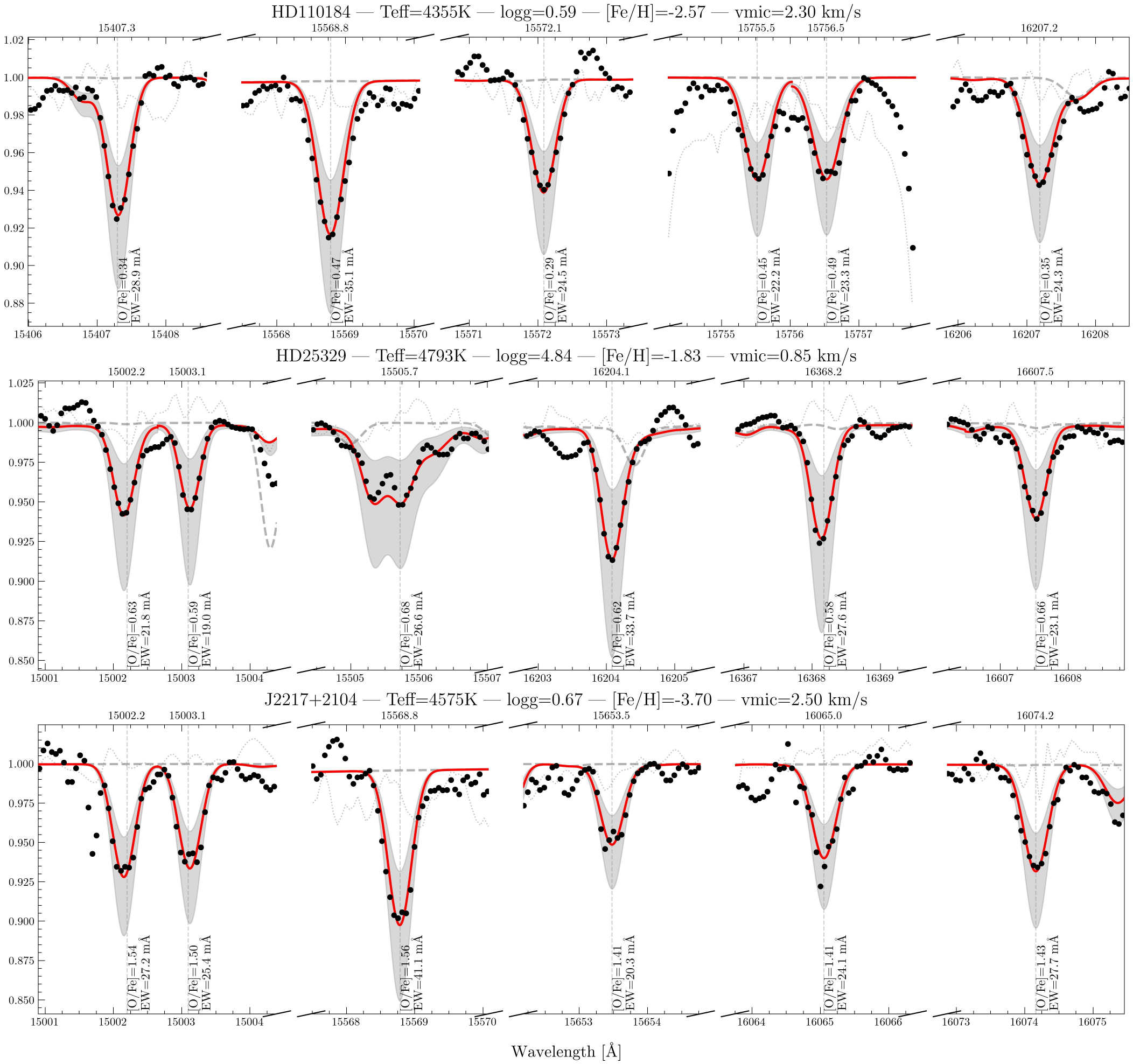}
 \caption{Subaru/IRD spectra (black dots) of several samples with the spectral region including several OH lines used in the analysis. The solid red line shows the best fit, with the grey shaded region representing $\pm0.2\text{ dex}$ O abundance changes to the lines. The dashed and dotted lines are the synthetic spectra without OH lines ($[\text{O/Fe}]=-20$) and the spectra of a rapidly rotating star {(indicating location of telluric lines, the continuum level fluctuated at $\sim1\%$)}, respectively.}
 \label{fig:OH_plot}
\end{figure*}

The average value and uncertainty of the {O} abundance from the OH lines is determined using the same approach as for the Fe abundance. All of the derived 1D/LTE O abundances from OH lines are given in Table \ref{tab:1D_OH_abund}{. T}he spectral fitting {for}several OH lines for some {examples}are presented in Figure \ref{fig:OH_plot}.

\begin{deluxetable}{lcccccc}
\centering
\digitalasset
\tabletypesize{\small}
\tablewidth{0pt}
\tablecaption{Derived 1D/LTE O abundances from OH lines\label{tab:1D_OH_abund}}
\tablehead{
\colhead{Designation} & \colhead{$\sigma_\text{atm}$} & \colhead{$N_\text{OH}$} & \colhead{$\sigma_\text{scat}$} & \colhead{$\sigma_\text{tot}$} & \colhead{$\langle\text{A(OH)}\rangle$} & \colhead{$\langle\text{[OH/Fe]}\rangle$}
}
\startdata
BD-02$^\circ$5957 & 0.162 & 18 & 0.101 & 0.163 & 6.620 & 0.768 \\
BD-07$^\circ$2674 & 0.145 & 14 & 0.082 & 0.147 & 6.995 & 1.068 \\
BD-14$^\circ$5890 & 0.121 & 11 & 0.171 & 0.132 & 7.608 & 0.909 \\
BD-15$^\circ$5781 & 0.135 & 9 & 0.129 & 0.141 & 7.525 & 1.096 \\
BD-18$^\circ$271 & 0.128 & 31 & 0.047 & 0.129 & 6.596 & 0.293 \\
BD-18$^\circ$5550 & 0.129 & 5 & 0.261 & 0.174 & 6.768 & 0.855 \\
BD-20$^\circ$6008 & 0.140 & 18 & 0.141 & 0.144 & 7.224 & 1.033 \\
BD+03$^\circ$2782 & 0.115 & 27 & 0.075 & 0.116 & 7.243 & 0.439 \\
BD+30$^\circ$2611 & 0.116 & 32 & 0.028 & 0.116 & 7.400 & 0.234 \\
CS29502-092 & 0.127 & 7 & 0.185 & 0.145 & 7.726 & 1.638 \\
CS30314-067 & 0.142 & 25 & 0.060 & 0.143 & 6.959 & 0.966 \\
HD 107752 & 0.138 & 12 & 0.103 & 0.141 & 7.342 & 1.231 \\
HD 110184 & 0.139 & 37 & 0.059 & 0.139 & 6.578 & 0.379 \\
HD 115444 & 0.141 & 17 & 0.077 & 0.142 & 7.288 & 1.318 \\
HD 118055 & 0.115 & 57 & 0.053 & 0.115 & 7.170 & 0.310 \\
HD 122563 & 0.146 & 18 & 0.097 & 0.147 & 7.020 & 0.873 \\
HD 187111 & 0.105 & 48 & 0.067 & 0.105 & 7.411 & 0.302 \\
HD 204543 & 0.109 & 21 & 0.073 & 0.110 & 7.492 & 0.499 \\
HD 221170 & 0.117 & 24 & 0.085 & 0.119 & 7.220 & 0.438 \\
HD 25329 & 0.107 & 52 & 0.050 & 0.108 & 7.547 & 0.600 \\
HD 4306 & 0.132 & 5 & 0.072 & 0.135 & 7.795 & 1.733 \\
HD 6268 & 0.143 & 3 & 0.095 & 0.153 & 7.258 & 0.918 \\
HD 85773 & 0.136 & 21 & 0.067 & 0.137 & 6.707 & 0.399 \\
HD 8724 & 0.102 & 30 & 0.078 & 0.103 & 7.506 & 0.386 \\
HD 88609 & 0.154 & 18 & 0.100 & 0.156 & 6.862 & 1.011 \\
HE 1116-0634 & 0.144 & 5 & 0.088 & 0.150 & 6.848 & 1.251 \\
HE 1523-0901 & 0.142 & 16 & 0.088 & 0.144 & 7.123 & 0.816 \\
J0040+2729 & 0.143 & 16 & 0.100 & 0.146 & 6.875 & 0.627 \\
J2114-0616 & 0.132 & 4 & 0.089 & 0.139 & 7.537 & 1.064 \\
J2217+2104 & 0.172 & 41 & 0.071 & 0.172 & 7.240 & 2.172 \\
J2347+2851 & 0.128 & 12 & 0.098 & 0.131 & 7.423 & 0.868 \\
TYC 3407-1352-1 & 0.145 & 6 & 0.053 & 0.147 & 7.033 & 1.094 \\
TYC 3814-1598-1 & 0.150 & 27 & 0.070 & 0.151 & 6.507 & 0.443 \\
UCAC4 425-121652 & 0.145 & 7 & 0.037 & 0.146 & 6.701 & 0.547 \\
UCAC4 515-137892 & 0.154 & 24 & 0.076 & 0.155 & 6.957 & 1.457 \\ 
\tableline
Arcturus Summer & 0.119 & 27 & 0.043 & 0.120 & 8.542 & 0.272 \\
Arcturus Winter & 0.116 & 23 & 0.039 & 0.116 & 8.553 & 0.283 \\
\enddata
\end{deluxetable}

\begin{deluxetable*}{lcccccccccc}
\centering
\digitalasset
\tablewidth{0pt}
\tablecaption{Derived O abundances from [OI] lines\label{tab:1D3D_OI_abund}}
\tablehead{
\colhead{Designation} & \colhead{$N_\text{obs}$} & \colhead{$\text{A(OI)}_\text{1L}$} & \colhead{$\text{[OI/Fe]}_\text{1L}$} & \colhead{$\Delta_{\text{1N}-\text{1L}}$} &  \colhead{$\Delta_{\text{3L}-\text{1L}}$} & \colhead{$\sigma_\text{atm}$} & \colhead{$\sigma_\text{scat}$} & \colhead{${\sigma_{\epsilon|W_\lambda}}^\star$}& \colhead{$\sigma_\text{tot}$} & \colhead{Note$^\dagger$}
}
\startdata
BD+03$^\circ$2782 & 1 & 7.400 & 0.607 & -0.006 & -0.006 & 0.055 & $\dotsi$ & 0.150 & 0.160 & out \\
BD+30$^\circ$2611 & 1 & 7.587 & 0.431 & -0.007 & -0.007 & 0.061 & $\dotsi$ & 0.122 & 0.137 & out \\
BD-07$^\circ$2674 & 1 & 6.655 & 0.737 & -0.010 & -0.010 & 0.054 & $\dotsi$ & 0.390 & 0.394 & out \\
BD-18$^\circ$271 & 2 & 6.883 & 0.590 & -0.005 & $\dotsi$ & 0.028 & 0.002 & 0.180 & 0.130 & $\dotsi$ \\
BD-18$^\circ$5550 & 1 & 6.752 & 0.812 & -0.010 & -0.010 & 0.056 & $\dotsi$ & 0.598 & 0.600 & in \\
CS30314-067 & 1 & 6.960 & 0.977 & -0.007 & -0.007 & 0.060 & $\dotsi$ & 0.225 & 0.234 & out \\
HD107752 & 2 & 6.755 & 0.653 & 0.013 & 0.013 & 0.049 & 0.015 & 0.459 & 0.331 & out \\
HD110184 & 1 & 6.761 & 0.571 & -0.004 & $\dotsi$ & 0.055 & $\dotsi$ & 0.087 & 0.103 & $\dotsi$ \\
HD115444 & 4 & 6.672 & 0.712 & 0.002 & 0.002 & 0.055 & 0.091 & 0.375 & 0.379 & out \\
HD118055 & 1 & 7.357 & 0.507 & -0.008 & $\dotsi$ & 0.059 & $\dotsi$ & 0.140 & 0.152 & $\dotsi$ \\
HD122563 & 8 & 6.793 & 0.656 & -0.007 & -0.007 & 0.034 & 0.062 & 0.323 & 0.167 & out \\
HD187111 & 1 & 7.629 & 0.530 & -0.009 & -0.009 & 0.057 & $\dotsi$ & 0.089 & 0.106 & out \\
HD204543 & 2 & 7.606 & 0.622 & -0.010 & -0.010 & 0.055 & 0.004 & 0.098 & 0.089 & out \\
HD221170 & 3 & 7.255 & 0.483 & -0.009 & -0.009 & 0.054 & 0.040 & 0.099 & 0.097 & out \\
HD4306 & 1 & 6.965 & 0.914 & -0.010 & -0.010 & 0.054 & $\dotsi$ & 0.398 & 0.402 & out \\
HD6268 & 3 & 7.050 & 0.720 & -0.010 & -0.010 & 0.054 & 0.056 & 0.296 & 0.183 & out \\
HD85773 & 2 & 6.842 & 0.544 & -0.006 & $\dotsi$ & 0.053 & 0.002 & 0.142 & 0.114 & $\dotsi$ \\
HD8724 & 1 & 7.664 & 0.555 & -0.010 & $\dotsi$ & 0.057 & $\dotsi$ & 0.126 & 0.138 & $\dotsi$ \\
HD88609 & 2 & 6.675 & 0.834 & -0.008 & -0.008 & 0.055 & 0.054 & 0.294 & 0.222 & out \\
J0040+2729 & 1 & 6.917 & 0.679 & -0.009 & -0.009 & 0.054 & $\dotsi$ & 0.408 & 0.412 & out \\
J2114-0616 & 1 & 7.256 & 0.792 & -0.010 & -0.010 & 0.054 & $\dotsi$ & 0.615 & 0.617 & out \\
J2217+2104 & 1 & 6.953 & 1.895 & -0.005 & $\dotsi$ & 0.096 & $\dotsi$ & 0.285 & 0.301 & $\dotsi$ \\
J2347+2851 & 1 & 7.276 & 0.731 & -0.010 & -0.010 & 0.054 & $\dotsi$ & 0.331 & 0.335 & out \\
\tableline
Arcturus & 3 & 8.676 & 0.406 & 0.000 & $\dotsi$ & 0.039 & 0.002 & $\sim0$ & 0.039 & $\dotsi$ \\
\enddata
\tablecomments{For 3D/LTE correction, $^\dagger$in = all input parameters are within the correction grid, out = some input parameters are outside the grid, closest-neighbor parameter is assumed. All of the abundances shown here are averages from multiple measurements if $N_\text{obs}>1$. \\$^\star$This is abundance uncertainty due to the EW uncertainty.}
\end{deluxetable*}

\begin{figure*}[t!]
 \centering
 \includegraphics[width=\linewidth]{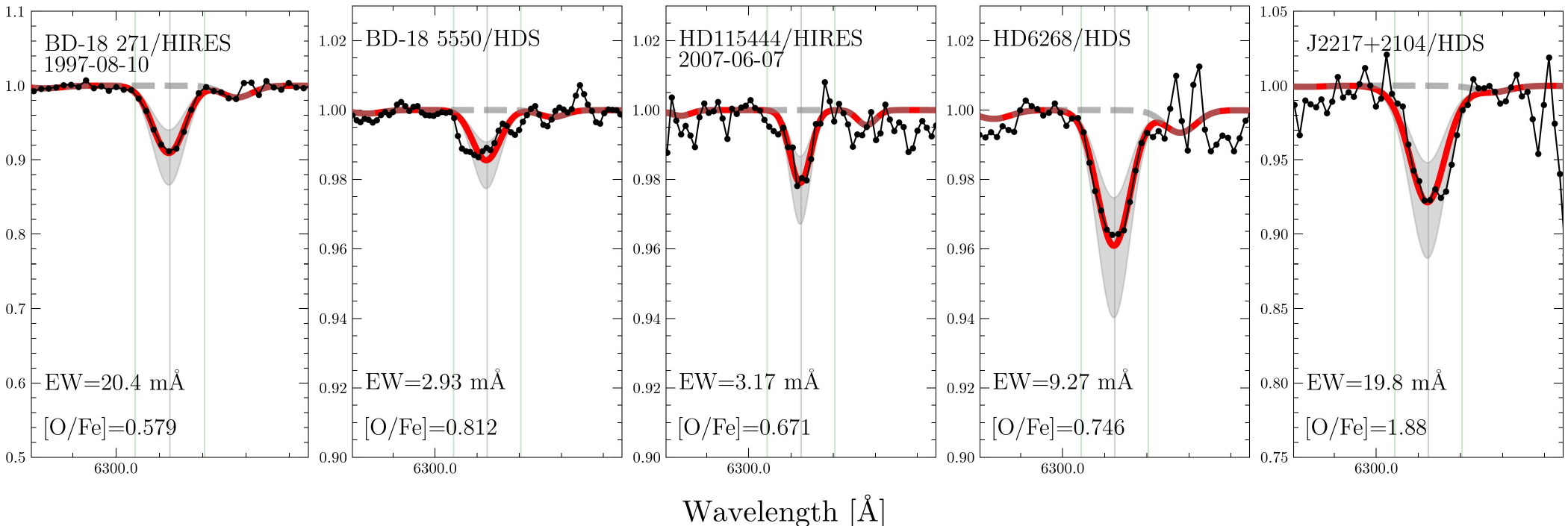}
 \caption{All archival optical spectra (black dots) of 24 samples from various instruments. The solid red line shows the best fit, with the grey shaded region representing $\pm0.2\text{ dex}$ O abundance changes to the lines. The dashed lines are synthetic spectra without [OI] lines ($[\text{O/Fe}]=-20$). Each x-axis minor tick mark represents a 0.25 \AA\ scale.}
 \label{fig:OI_plot}
\end{figure*}

To assess the reliability of the OH lines as an oxygen tracer, the oxygen abundance from the optical 6300 \AA\ [OI] line is also derived. The absorption lines (either tellurics or other lines) near the wing of the [OI] line are removed. This line overlaps with the Ni I line at $\lambda6300.34$ \AA. {The Ni blending effect is more pronounce for stars close to solar metallicity \citep{2021MNRAS.508.2236B}.} A recent study reveals that the Ni abundance may be as high as $\text{[Ni/Fe]}\sim+0.6$ dex in VMP stars if the 3D/NLTE correction is included \citep{2025MNRAS.538.3284S}. However, this line is very weak in this metallicity regime. Even if a very high $\text{[Ni/Fe]}=+1.0$ dex is assumed {in our analysis, we found that the blending contribution from Ni I line} on the [OI] equivalent width is barely noticeable. Therefore, a constant solar value of $\text{[Ni/Fe]}=0$ is simply assumed for all samples (see the dashed line in Figure \ref{fig:OI_plot}, which shows no absorption feature). For Arcturus, three archive spectra with high resolution and S/N are also analyzed. The same constant abundances for C and N as in the OH analysis are also adopted, along with a Ni abundance of $\text{[Ni/Fe]}=-0.14\pm0.07$ \citep{2024A&A...684A..85V}. 

 The average O abundance from [OI] line is taken for objects for which more than one spectrum is available. Since we derive {O abundances from only several spectra} (or one for some stars), the total error budget {might be} underestimated{. T}his is not the case on OH-based abundance with numerous lines available. {The uncertainty of O abundances due to spectral noise is estimated for the [O I] line, following} \cite{2006AN....327..862V} provide the equivalent width (EW) uncertainty for a weak line:
\begin{align}
\sigma_{W_\lambda}=\sqrt{2}\cdot\frac{(\Delta\lambda-W_\lambda)}{\text{S/N}},\label{eq:W_lambda}
\end{align}
where $\Delta\lambda\approx2\cdot\text{FWHM}\sim0.5$ \AA\ and $W_\lambda\sim6$ m\AA\ for the [OI] line. From this EW uncertainty, we run the program to compute corresponding O abundance uncertainty ($\sigma_{\epsilon|W_\lambda}$). Then, total error budget for [OI]-based abundance is determined by this equation:
\begin{align}
\sigma_\text{tot} = \sqrt{\sigma_\text{atm}^2 + \frac{\sigma_\text{rand}^2}{N_\text{obs}}}, \label{eq:sigma_tot_OI}
\end{align}
where $\sigma_\text{rand}$ is total random error defined as $\sigma_\text{rand}^2\sim\sigma^2_\text{scat}+\sigma^2_{\epsilon|W_\lambda}$, and $N_\text{obs}$ is the the number of spectra used to measure the [OI] line. The 1D/LTE O abundances obtained from the [OI] line for 23 stars with each corresponding error are tabulated in Table \ref{tab:1D3D_OI_abund}, and the spectral fitting results of [OI] line for some samples are presented in Figure \ref{fig:OI_plot}.

The sensitivity analysis is also performed for both OH and [OI] lines by changing adopted stellar parameters of $T_\text{eff},\;\log g,\;\xi_t,\;\text{[Fe/H]},\;\text{ and [C/Fe]}$ by $\pm100$ K, $\pm0.3$ dex, $\pm0.5$ km s$^{-1}$, $\pm0.3$ dex, and $\pm0.3$ dex, respectively.  The results of  sensitivity analysis for the OH and [OI] lines are discussed in Section \ref{ssec:sensi_Oxy}. 

To examine the NLTE effect and 3D atmospheric effect on the optical [OI] line, an additional analysis is performed as follows. First, for the NLTE analysis, the same spectral fitting is conducted using \texttt{TSFitPy} under the 1D/NLTE framework with the given O model atom \citep{2021MNRAS.508.2236B}. The resulting NLTE corrections $(\Delta_{\text{1N}-\text{1L}}=[\text{OI/Fe}]_\text{1D/NLTE}-[\text{OI/Fe}]_\text{1D/LTE})$ for all targets are provided in Table \ref{tab:1D3D_OI_abund}, typically between $-0.01$ and 0.0 dex, which are negligible compared to the uncertainties from the atmospheric parameters. 

For the 3D/LTE abundance correction of the [OI] line, a similar approach to \cite{2025MNRAS.538.3284S} is followed. The 3D LTE spectra are generated for all available model atmospheres for $T_\text{eff}$ = 4000, 4500, 5000 K; $\log g$ = 1.5, 2.0, 2.5, 3.0; [Fe/H] = -4, -3, -2, -1 dex; by changing the abundances of [O/Fe] from 0 to 2.5 in steps of 0.25 dex and [C/Fe] from -1.5 to 2 in steps of 0.5 dex. The profiles are integrated, {to derive} the corresponding 3D LTE EWs. Based on the given EWs, the corresponding 1D/LTE oxygen abundance is then computed, varying the microturbulence $\xi_t$ of 0.5, 1.5, 2.5 km s$^{-1}$. For lines with very weak EWs ($<1$ m\AA), the calculation is not performed.

Utilizing the grids, input parameters, and equivalent widths derived from spectral fitting for the 1D/LTE abundance analysis, the 3D/LTE correction $(\Delta_{\text{3L}-\text{1L}}=[\text{OI/Fe}]_{\text{3D/LTE}}-[\text{OI/Fe}]_{\text{1D/LTE}})$ is interpolated following the method from \citet{bergemann2014}. However, since most of the samples have $\log g < 1.5$ dex, which are not covered by the \texttt{M3DIS} models, samples with a $\log g$ difference of $< 0.5$ dex from the grid boundary are carefully selected and interpolated by adopting the parameters of the "closest-neighbour" models.

This analysis reveals that typical corrections within the grid are around 0.01 dex for the targets (see Figure \ref{fig:3Dcorr_OI} and Table \ref{tab:1D3D_OI_abund}). Similar to the NLTE correction, this indicates that the [OI] line in red giant stars is insensitive to both non-LTE and 3D atmospheric effects, consistent with previous studies on dwarfs and sub-giants \citep{2016MNRAS.455.3735A}.

\section{Results \& Discussions} \label{sec:Discussions}
\subsection{Sensitivity Analysis on Oxygen lines}\label{ssec:sensi_Oxy}

\begin{figure*}[htbp]
\centering
\includegraphics[width=\linewidth]{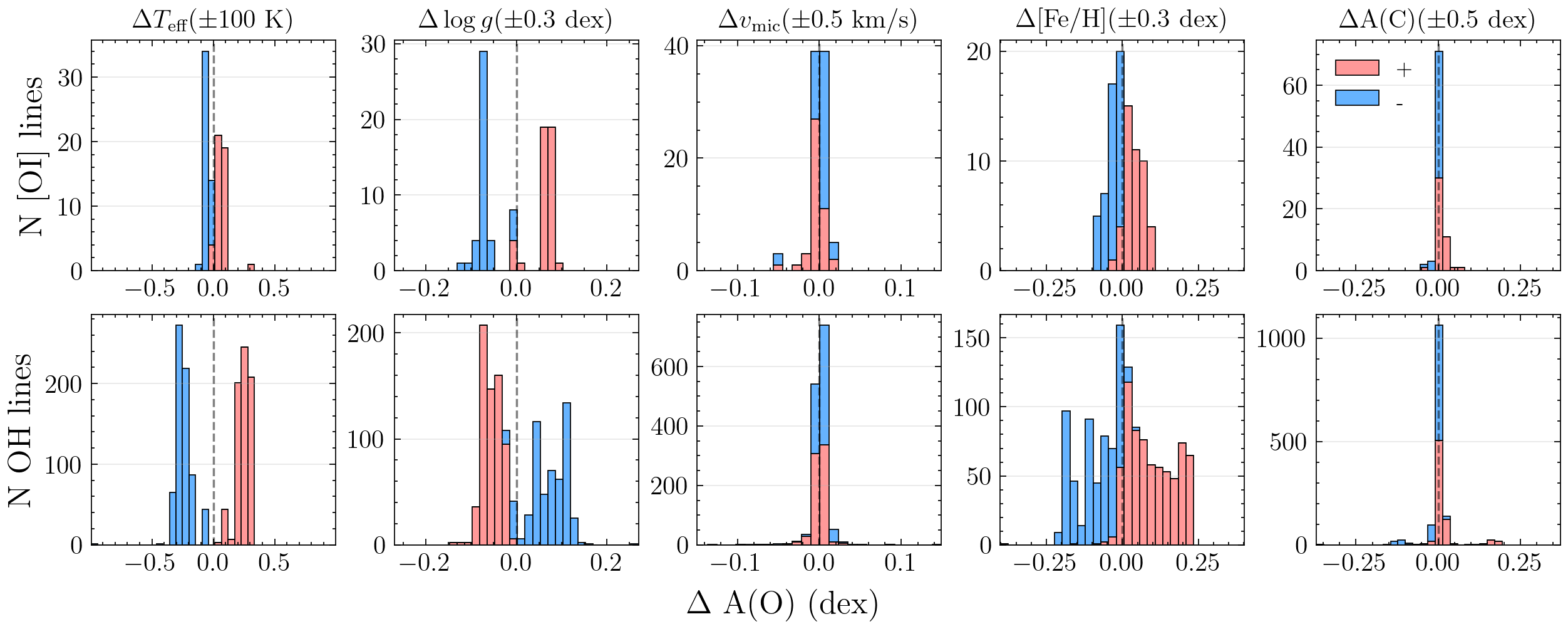}
\caption{The histogram distribution of [OI] line (first row) and OH lines (second row) abundance changes corresponding to each stellar parameter.}
\label{fig:sensi_hist}
\end{figure*}

Leveraging the large sample size and the availability of OH lines, a sensitivity analysis is conducted for all available lines across five input parameters, as described in the previous section. The details of the results of the sensitivity analysis are inspected. The abundance change $(\Delta\text{A(O)} = \text{A(O)}_f - \text{A(O)}_\text{init})$ for each line resulting from the perturbation of each input parameter is calculated. The distributions are shown in Figure \ref{fig:sensi_hist}. An initial inspection reveals that the atomic [OI] line exhibits significantly lower sensitivity to variations in stellar parameters $\Delta\text{A(O)} < 0.1$ dex in all cases.

\begin{figure*}[htbp]
\centering
\includegraphics[width=\linewidth]{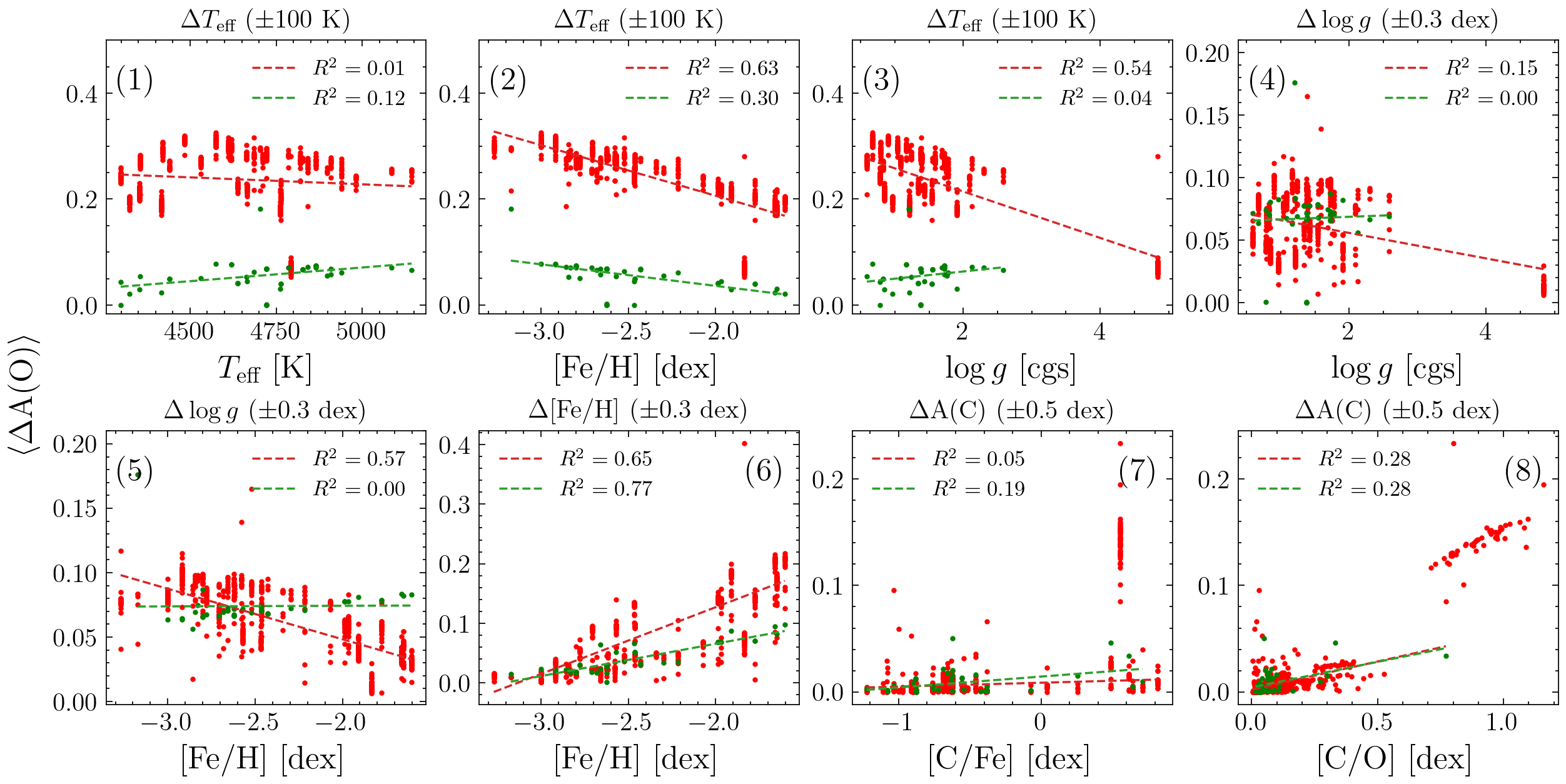}
\caption{Abundance change of OH lines (red points) and [OI] line (green points) due to changes of stellar parameters as a function of input parameters. See the text for further explanation.}
\label{fig:sensi_trend}
\end{figure*}

To evaluate the dependence of the sensitivity on parameters, the average absolute change in O abundance $(\langle\Delta\text{A(O)}\rangle=(|\Delta\text{A(O)}^+|+|\Delta\text{A(O)}^-|)/2)$ in response to parameter variations is plotted against the adopted stellar parameters (see Figure \ref{fig:sensi_trend}).{To} examine the correlations, the coefficient of determination $R^2$ is derived {by a simple linear regression}. Key findings are highlighted below.

The change in the adopted stellar temperature ($\pm100$ K) induces an abundance change of $|\Delta\text{A(O)}|\sim0.25$ dex for OH lines. No significant correlation with temperature is found (red data points in Panel 1 of Figure \ref{fig:sensi_trend}). In contrast, Panel 2 reveals a stronger dependence on metallicity with a negative slope: $|\Delta\text{A(O)}|\sim0.3$ dex at $\text{[Fe/H]}\sim-3.5$, decreasing to $|\Delta\text{A(O)}|\sim0.15$ dex at $\text{[Fe/H]}\sim-1.6$. Panel 3 indicates that the cool dwarf star HD 25329 ($\log g\sim4.8$) exhibits much weaker sensitivity than the other red giants ($\log g<3$). Overall, the [OI] 630 nm line (green data points in Figure \ref{fig:sensi_trend}) displays significantly lower sensitivity to effective temperature changes and only a slight dependence on metallicity.

This high sensitivity of molecular OH lines to temperature is an expected result, from the fact that OH molecules are minor population and it shifts dramatically toward atomic O and H with even slight temperature increases because its relatively small dissociation energy $(D_0\sim4.4\ \text{eV})$. In addition, molecular lines form in a region extending to shallower parts of the atmosphere {than} the forbidden line \citep{2015A&A...576A.128D,2016A&A...593A..48G}. Using the Saha equation, it {is} estimated that $n_\text{OH}\propto p_\text{OH}\propto T^{-3/2} \exp{\left[D_0/k_BT\right]}$, with $n_\text{OH}$ \& $p_\text{OH}$ are number density and partial pressure of OH molecules, respectively. In a typical red giant with $T_\text{eff}=4500\text{ K}$ ($D_0/k_B\sim5.1\times10^3$), a 100 K temperature increase reduce{s} the number of OH molecules by $\sim25\%$ or $-0.12$ dex, which is similar to what has been calculated by \cite{2016ApJ...819..103A}. Conversely, neutral oxygen atoms are far less sensitive to temperature, resulting in much smaller O abundance variations derived from the [OI] line. 

The higher sensitivity of OH lines to temperature in stars with lower metallicity is largely driven by the changing nature of the continuous opacity. In metal-poor atmospheres, the absence of low-ionization-potential metals causes the electron pressure ($P_e$) to be dominated by the ionization of hydrogen ($\chi = 13.6$ eV).
Because of this high ionization potential, the supply of free electrons--and, consequently, the H$^{-}$ continuous opacity--becomes extremely sensitive to temperature variations. Therefore, even small changes in the adopted temperature lead to large fluctuations in the opacity and temperature stratification, particularly in the shallow layers where OH lines form.

These findings align with previous studies utilizing {NIR} OH lines. \cite{2001ApJ...556..858M} reported an abundance change of $\Delta\text{A(O)}=+0.20$ dex for giants and $\Delta\text{A(O)}=+0.17$ dex for dwarfs given a shift of $\Delta T_\text{eff}=+100\ \text{K}$. Similar values ($\Delta\text{A(O)}=+0.22$ dex and $\Delta\text{A(O)}=+0.23$ dex) were reported by \cite{2002ApJ...Melendez} for HD 110184 and by \cite{2003ApJ...588.1072B} for HD 122563. Additionally, \cite{2013ApJ...765...16S} used a representative red giant model {with solar metallicity} $(T_\text{eff}/\log g/\text{[m/H]}=4000/1.3/0.0)$ and found $\partial\text{A(O)}/\partial T=+0.124\ \text{dex}/50\ \text{K}$, which is consistent with the measurements in this study. 

The sensitivities of OH and [OI] lines to changes in the adopted surface gravity are presented in Figure \ref{fig:sensi_trend} (Panels 4 and 5). Panel 4 shows no clear trend for OH lines regarding $\log g$ across the giant stars, though the scatter is large{. T}he dwarf star HD 25329 shows noticeably lower sensitivity than the giants. A clear trend emerges against metallicity in Panel 5. OH lines display a negative slope, where metal-rich stars generally show lower sensitivity to surface gravity. {In contrast, the [OI] line shows a flat trend or no dependency to the metallicity.}

The strength of a weak line is determined by the ratio of line and continuum opacities ($W\propto l_\nu/\kappa_\nu$). For neutral forbidden lines, the line opacity scales linearly with the total gas pressure and oxygen abundance $(l_{\nu,\text{[OI]}}\propto \text{A(O)} P_g)$. OH lines, formed by O+H collisions, follow the relation $l_{\nu,\text{OH}}\propto \text{A(O)}P_g^2$. In cool stars, the continuum is dominated by H$^-$ ions, scaling with gas and electron pressure $(\kappa_{\text{H}^-}\propto P_gP_e)$. In metal-rich stars, electrons come from ionized metals ($P_e\propto P_g$), implying $\kappa_{\text{H}^-}\propto P_g^2$. In metal-poor regimes, neutral hydrogen becomes the donor ($P_e\propto P_g^{1/2}$), implying $\kappa_{\text{H}^-}\propto P_g^{3/2}$. Consequently, OH lines in metal-poor conditions depend more strongly on surface gravity ($\delta W_\text{OH}\propto \delta\text{A(O)}\;\delta P_g^{1/2}$) than in metal-rich conditions $(\delta W_\text{OH}\sim \delta\text{A(O)}\;\delta P_g^0)$. Conversely, the [OI] line in metal-poor conditions shows less dependency on $\log g$ $(\delta W_\text{[OI]}\propto \delta\text{A(O)}\;\delta P_g^{-1/2})$ {than in} metal-rich conditions $(\delta W_\text{[OI]}\propto\delta\text{A(O)}\;\delta P_g^{-1})$. This framework explains the trends observed in Panel 5.

Furthermore, OH and [OI] respond oppositely to pressure. OH line strength correlates positively with gas pressure (or surface gravity), whereas the [OI] line exhibits an inverse correlation. This behavior explains the {opposite} abundance responses observed in the second column panels from the left of Figure \ref{fig:sensi_hist}.

Figure \ref{fig:sensi_trend} (Panel 6) illustrates the sensitivity to a typical iron abundance uncertainty $(\sim0.3\text{ dex})$. Both tracers exhibit positive correlations, indicating higher sensitivity in the metal-rich regime, with OH showing twice the sensitivity of [OI], while remaining practically insensitive in extremely metal-poor stars $([\text{Fe/H}]\lesssim-2.5)$. In a very metal-poor 1D atmosphere model, changing the Fe abundance by $\sim0.3$ dex does not change the temperature stratification as significantly as under more metal-rich conditions, which explains the O abundance insensitivity to the changing adopted metallicity. Since the OH molecular line is more sensitive to temperature changes, higher dependence is expected than that for the neutral [OI] line.

Finally, the effect of changing the carbon abundance on the O abundance derived from OH and [OI] lines is addressed. As mentioned in Section \ref{ssec:vmic_cabund}, carbon and oxygen are coupled through the formation of stable CO molecules. Increasing the carbon abundance promotes CO formation, thereby reducing the number of OH molecules. In contrast, [OI] lines are less sensitive to changes in carbon abundance because the majority of oxygen atoms remain in the neutral state in typical red giants under LTE conditions. In our results, as shown in Figure \ref{fig:sensi_hist} first column from the right and Figure \ref{fig:sensi_trend} Panel 7 \& 8, the derived O abundance shows only a negligible change $(\langle\Delta\text{A(O)}\rangle<0.05\text{ dex})$ for both [OI] and OH lines when the carbon abundance is changed by $\pm0.5$ dex. An outlier is observed for OH lines with $[\text{C/Fe}]\sim+0.7$, which correspond to a VMP star, HD25329. This cool dwarf star exhibits significantly higher sensitivity to carbon abundance than the rest of the sample. {Because of i}ts higher surface gravity, combined with a relatively high metallicity and [C/O] ratio, {a substantial fraction of O species is locked in stable CO molecules.}

For other samples of red giant stars reveals no clear trend in O abundance sensitivity against $[\text{C/Fe}]$ (Figure \ref{fig:sensi_trend} Panel 7), since almost all of the targets (including C-enhanced metal-poor stars, CEMPs, $[\text{C/Fe}]>0.5$) {still} have a low $[\text{C/O}]$. This becomes apparent when O sensitivity trends are plotted against the $[\text{C/O}]$ ratio (Figure \ref{fig:sensi_trend} Panel 8), which reveals a slight{ly} positive trend for both OH and [OI], with a higher $R^2$ indicating better correlation. As the stars approach high metallicity, the $[\text{C/O}]$ ratio approaches 0, or the abundances of C and O are comparable; thus, increasing C removes a noticeable amount of atomic O available for OH formation. This weakens the OH lines, requiring a larger increase in O abundance {than very metal-poor stars}. 

In addition to the low [C/O] of the sample, {NIR} OH lines appear relatively independent of carbon abundance. This is consistent with {the result of} \cite{2018MNRAS.475.3369C}{(hereafter \citetalias{2018MNRAS.475.3369C})}, who performed simultaneous 3D-LTE analysis of CH G-band{,} near-UV OH lines, {and NIR} OH lines in HD 122563. They noted that near-IR lines form in deeper atmospheric layers compared to their near-UV counterparts (see \cite{2015A&A...576A.128D} for near-IR and \cite{2017A&A...599A.128P} for near-UV), making the derived O abundance less sensitive to CO formation.

\subsection{Oxygen Abundance Discrepancy between near-IR OH lines and optical [OI] line}\label{ssec:oxygen_discrepancy}

The fundamental question of this work—echoing many previous investigations into metal-poor stars—is: \textit{Are OH lines reliable enough as oxygen tracers in the extremely metal-poor regime?} As discussed in the above sections, the [OI] $\lambda6300.3$ \AA\ line is insensitive to both non-LTE and 3D granulation effects. This stability arises because the most populated oxygen species is insensitive to over-ionization by NLTE effects and line formation occurs deep enough in the atmosphere to be shielded from 3D cooling effects \citep[see][and references therein]{2015A&A...576A.128D,2016MNRAS.455.3735A,2019A&A...622L...4A, 2019A&A...630A.104A}. However, at very low metallicities, this line is excessively weak and unusable. {Moreover, this line lies near atmospheric O$_2$ lines, requiring high-resolution spectra to resolve each line. Even with high resolution, some blending remains possible, making it unusable.}

\begin{figure*}[htbp]
\centering
\includegraphics[width=\linewidth]{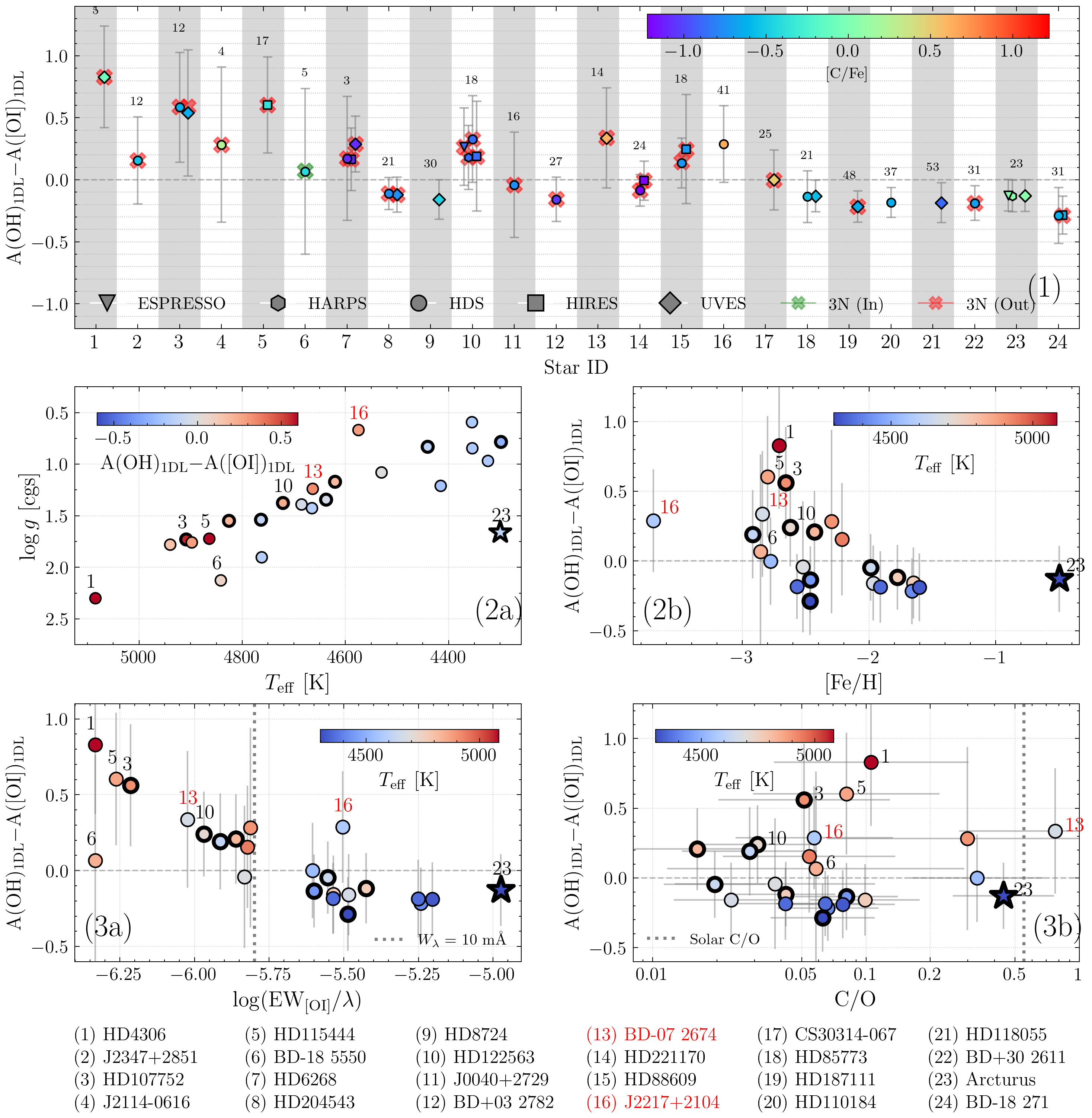}
\caption{Oxygen abundance differences between 1D/LTE \textit{H}-band OH lines and optical [OI] against several parameters. \textbf{Panel 1}: Abundance differences are plotted for all 24 targets (including Arcturus), ordered by decreasing effective temperature from left to right. Each marker shape represents the average abundance difference derived with different instruments. Cross markers represent abundance differences after 3D/NLTE correction $(\Delta_{\text{3N}-\text{1L}} \approx \Delta_{\text{3L}-\text{1L}} + \Delta_{\text{1N}-\text{1L}})$ for [OI], which are practically negligible. Error bars represent the total uncertainty of both OH and [OI] abundances. Each data point is color-coded by [C/Fe], and the number of OH lines used in the analysis is annotated in the plots. \textbf{Panel 2a}: $T_\text{eff}$ vs $\log g$ plot of the 24 samples, color-coded by the average OH and [OI] abundance difference. \textbf{Panel 2b-3b}: Average OH and [OI] abundance differences on the Y-axis versus adopted [Fe/H], forbidden line strength, and absolute C/O ratio, respectively on X-axes. All data points in these panels are color-coded by effective temperature. The X-axis on Panel 1 and annotated numbers in Panels 2a-3b represent the Star ID, which is given in the bottom legend. Red text represents CEMP stars ([C/Fe]$>+0.7$ dex).}
\label{fig:difference_O}
\end{figure*}

To examine the reliability of oxygen abundances obtained from OH lines and estimate the possible systematics, the abundances derived from \textit{H}-band OH lines and the optical [OI] line are compared. First, the O abundances obtained from both OH and [OI] lines are compared in the absolute scale: $A(\text{O})=[\text{O/Fe}]+[\text{Fe/H}]+A(\text{O})_\odot$, for which $A(\text{O})_\odot=8.77$ dex is adopted \citep{2021MNRAS.508.2236B}. For stars for which more than one studies using optical spectra are available, the average abundance from the same instrument is taken to identify any potential systematics between instruments. The O abundance difference between measurements using OH lines and the [OI] line $(\Delta_{\text{OH}-\text{[OI]}}\equiv A(\text{OH})_\text{1DL}-A(\text{[OI]})_\text{1DL})$ for each instrument is presented in Panel 1 of Figure \ref{fig:difference_O}. 

Before addressing the general trends across the parameters, the peculiar stars within the dataset are examined. Out of the 24 stars for which the O abundance is obtained from both OH and [OI] lines, three stars {,} HD 4306 (ID:1), HD 107752 (ID:3), and HD 115444 (ID:5), show a large discrepancy of $\Delta_{\text{OH}-\text{[OI]}}>+0.5$ dex.

    In these three outliers, the [OI] line is remarkably weak ($W_\lambda\sim3$ m\AA). Given the S/N levels, the resulting uncertainties in equivalent width ($\sigma_{W_\lambda}$) lead to significant oxygen abundance errors: $\sigma_{\epsilon|W_\lambda}\sim\pm0.56\text{ dex}$ for HD 4306, $\sigma_{\epsilon|W_\lambda}\sim\pm0.64\text{ dex}$ for HD 107752, and $\sigma_{\epsilon|W_\lambda}\sim\pm0.70\text{ dex}$ for HD 115444. These uncertainties dominate the total error budget ($\sigma_\epsilon=(\sigma_\text{atm}^2+\sigma_{\epsilon|W_\lambda}^2)^{1/2}>0.6$ dex), where $\sigma_\text{atm}\sim0.05\text{ dex}$ (see Table \ref{tab:1D3D_OI_abund}), giving the abundance from [OI] line less reliable for these stars.

Regarding the OH lines in these three targets, HD 4306 is one of the hottest stars showing the weakest OH lines in the sample. Of the 5 derived OH lines, three exhibit clear absorption features. An analysis excluding the two weaker lines, however, does not significantly change the average abundance. The EW for these three lines ranges from $13-21$ m\AA\ with an S/N $\sim75$, corresponding to $\sigma_{W_\lambda}\sim9$ m\AA\ or $\sigma_{\epsilon|W_\lambda}\approx\pm0.25$ dex. 

A larger number of OH lines is detected in HD 107752 and HD 115444 (12 and 17 OH lines, respectively) with similar quality and O abundance consistency measured from individual OH lines. While the uncertainty due to atmospheric parameters for each OH line is notable ($\sigma_\text{atm}\sim0.2\text{ dex}$), the numerous OH lines available make the random error contribution ($\sigma_\text{rand}^2/N_\text{line}=(\sigma_\text{scat}^2+\sigma_{\epsilon|W_\lambda}^2)/N_\text{line}$) to the total error of the OH-based abundance significantly smaller than that of the single [OI] lines (with $W_\lambda<10\text{ m\AA}$).

For the remaining stars in the sample, the abundance difference is generally less than 0.1 dex. An exception is HD 122563, which shows a discrepancy of up to 0.2 dex that is attributable to the uncertainty due to the weakness of the {[OI]} line (see Appendix \ref{ssec:o_arc_hd}). The higher $\Delta_{\text{OH}-\text{[OI]}}$ found in the CEMP star LAMOST J2217+2104 (ID:16) can be attributed to its much lower metallicity, making it more susceptible to 3D effects compared to other samples with similar temperatures. Similarly, the larger discrepancy in the CEMP star BD$-07^\circ2674$ (ID:13) is likely due to lower quality of the optical spectra, resulting in a {larger} [OI] abundance uncertainty ($\sigma_\text{tot}\sim0.4$ dex). If the current [OI] EW is underestimated by $\sim3$ m\AA, {the actual} $A(\text{[OI]})$ {could be 0.2 dex higher}, bringing it into agreement with the OH abundance.

Whereas warm red giants show higher OH-based O abundance than the [OI]-based ones {in general}, a warm red giant BD$-18^\circ5550$ (ID:6) exhibits a significantly smaller OH-[OI] difference. In this context, BD$-18^\circ5550$ deviates from other stars with similar [OI] line strengths and temperatures. It is worth noting that a large deviation is observed among the five {OH} lines measured for this star{, which }{u}nlike other stars where low S/N spectra allow the exclusion of poor-quality OH lines, this star has excellent S/N \textit{H}-band spectra free from telluric contamination.

This star was also studied by \cite{2015A&A...576A.128D} (DO15). It is noted that the OH lines used differ between the present study and \citetalias{2015A&A...576A.128D} due to different telluric {lines} locations. Although the temperature, surface gravity, and metallicity adopted here are higher by $\sim100$ K, 0.6 dex, and 0.2 dex, respectively, than \citetalias{2015A&A...576A.128D}, the average 1D/LTE OH-based O abundance remains similar ($A(\text{OH})\sim6.8$ dex). This is an expected result from the sensitivity analysis for a change of $\Delta\text{A(O)}=\Delta\epsilon_{T_\text{eff}|\text{O}}+\Delta\epsilon_{\log g|\text{O}}+\Delta\epsilon_{\text{[Fe/H]}|\text{O}}=(+0.3\text{ dex})+(-0.2\text{ dex})+(0.0\text{ dex})\approx+0.1$ dex, which explains the similarity. 

The strongest OH line in BD$-18^\circ5550$ is $\lambda15422.4$ \AA\ that yields an abundance twice {higher than}the weakest line at $\lambda16052.8$ \AA. Unlike other samples, this star exhibits a very steep slope in $A(\text{OH})$ versus excitation potential $\chi_\text{exc}$, where the $\lambda15422.4$ \AA\ also has the smallest $\chi_\text{exc}$. Excluding this strongest line would lower the OH abundance by $\sim0.1$ dex, further decreasing $\Delta_{\text{OH}-\text{[OI]}}$. 

Another notable discrepancy between our study and \citetalias{2015A&A...576A.128D} is in the [OI]-based abundance. \citetalias{2015A&A...576A.128D} reported a significantly lower [OI]-based abundance $(A(\text{[OI]})_\text{DO15}=6.13\text{ dex})$ compared to the value derived here $(A(\text{[OI]})_\text{our}=6.75\text{ dex})$. \citetalias{2015A&A...576A.128D} adopted [OI]-based abundance from \cite{2004A&A...416.1117C} (CA04). \citetalias{2004A&A...416.1117C} reported a significantly smaller $\text{EW}_{\text{CA04}}=1.5$ m\AA, which is half of ours with  $\text{EW}_{\text{our}}=2.93$ m\AA. A comparison of spectra of \citetalias{2004A&A...416.1117C} and ours indicates that the [OI] feature in their spectra is evidently much weaker and broader than in our spectra. The difference of the [OI]-based abundances between our study and \citetalias{2015A&A...576A.128D} could be attributed to the difference in the [OI] features in the spectra used for the analysis.

For the remaining stars show that show difference smaller than 0,25 dex, we investigate the dependence of the difference on stellar parameters and C/O abundance ratio.

Panel 2a of Figure \ref{fig:difference_O} displays temperature versus surface gravity for the 24 samples, color-coded by abundance difference. Observationally, the trends clarify that most samples with $\Delta_{\text{OH}-\text{[OI]}}<0$ are cool red giants. 

Panel 2b, shows the abundance differences between the two indicators as a function of metallicity and color-coded by effective temperature. 
At higher metallicity ([Fe/H]$ > -2.0$), both cool ($T_\text{eff} \lesssim 4600$ K) and warm red giants show a similar $\Delta_{\text{OH}-\text{[OI]}} \sim -0.2$ dex. In contrast, at [Fe/H]$ < -2.0$, the separation becomes more pronounced between the two groups, with larger differences observed for higher temperatures and lower [Fe/H].

A clear correlation between the [OI] line strength and abundance differences is found (as seen in Figure \ref{fig:difference_O}, Panel 3a). This correlation results from an \textit{indirect} effect of temperature and metallicity on {$\Delta_\text{OH-[OI]}$ through} the [OI] line strength. As the [OI] lines weaken due to higher temperatures and/or lower metallicity (EW $<10$ m\AA){the } $\Delta_{\text{OH}-\text{[OI]}}$ increases linearly. In contrast, cool red giants generally have stronger [OI] lines and exhibit an almost constant $\Delta_{\text{OH}-\text{[OI]}} \sim -0.2$ dex.

A limited number of previous studies have investigated comparisons of O abundance from OH and [OI] lines. The pioneering work was by \cite{1996AJ....111..946B} on the metal-poor dwarf HD 103095 that reported consistent abundances between \textit{H}-band OH and [OI]. \cite{2003ApJ...588.1072B} derived oxygen abundance in HD 122563, finding a higher abundance from OH than that from [OI], which they attributed primarily to uncertainties in the adopted $\log g$. Another theoretical studies on the 3D effect on the OH molecule in LTE condition indicate negative 3D corrections ($\Delta_{\text{3L}-\text{1L}}\sim-0.25$ dex) for {NIR} OH \citep{2015A&A...576A.128D,2017A&A...599A.128P,2018MNRAS.475.3369C}. 

However, the results presented here exhibit higher and lower O abundances from OH than those from [OI] for warm ($T_\text{eff}\gtrsim4600$ K) and cool stars, respectively. Previous works generally find abundances, whether from UV or IR OH lines, to be higher than those from the forbidden lines. In 3D simulations of atmospheres for warmer stars, the upper atmosphere diverges significantly from 1D models due to convective overshoot compared to cooler stars. Hot gas parcels rise into low-pressure regions, expanding and cooling adiabatically—a process excluded from 1D models that assume radiative equilibrium \citep{1999A&A...346L..17A,2007A&A...469..687C}. Since molecular lines typically form in these shallower, cooler layers in warmer stars, 3D/LTE simulations predict stronger OH lines than 1D/LTE models, suggesting that 1D/LTE approaches generally overestimate the abundances from molecular lines. 

The only study showing the opposite trend {(where the 1D/LTE [OI]-based abundance is higher than the 1D/LTE OH-based one)} is \citetalias{2018MNRAS.475.3369C}{, which is} discussed in Appendix \ref{ssec:o_arc_hd}{. This discrepancy can be} attributed to {the} $T_\text{eff}$ and $\log g$ offsets {adopted in \citetalias{2018MNRAS.475.3369C}}. {Using updated values for these parameters would increase} the {1D/LTE OH-based} abundance , which {then} might align with {the [OI]-based abundance} if 3D/LTE corrections are applied.


The opposite trend is, however, observed here. OH-based abundances are lower than [OI]-based one, in many of the samples. Most of these are cool red giants, a regime not well-covered in previous studies. This discrepancy is statistically significant, supported by numerous detected OH lines yielding consistent abundances and relatively strong [OI] lines, both of which reduce uncertainty compared to warmer stars (see Panel 3a on Figure \ref{fig:difference_O}). Attributing this difference to parameter errors, as in the case of HD 122563, is unlikely. To resolve the discrepancy by changing the adopted $T_\text{eff}$, an increase of at least $\sim100$ K would be needed, given that a 100 K increase shifts the total abundance difference by $\Delta\epsilon_\text{OH-[OI]}=\Delta\epsilon_{\text{OH}|\text{Teff}}-\Delta\epsilon_{\text{[OI]}|\text{Teff}}\approx(+0.25)-(+0.05)=+0.20$ dex. Since the temperatures adopted in the present work are already $\sim100-200$ K higher than in many previous studies and possess high precision ($\sigma_T<50$ K between colors), this option is considered improbable. It is also unlikely that the result can be explained by adjustments of surface gravity given the high-precision parallax used in this work.

A plausible interpretation involves the interplay between the true temperature stratification predicted by 3D modeling and non-LTE effects on IR OH lines. \textit{H}-band OH lines form slightly deeper in the photosphere than UV OH lines, which extend towards the upper atmosphere as demonstrated in  \cite{2015A&A...576A.128D,2017A&A...599A.128P}. Consequently, these lines exhibit different sensitivities to 3D effects. In cooler red giants ($T_\text{eff}\sim4500$ K, $\log g\sim1.5$ dex), particularly towards metal-rich atmospheres, the $\langle\text{3D}\rangle$ temperature profile can be similar to or even 100-200 K hotter than the 1D prediction \citep{2013A&A...557A..26M}. At higher metallicities, increased molecular opacity absorbs radiation more efficiently, causing radiative heating that raises local temperatures \citep{2011A&A...529A.158H}. Furthermore, the constant mixing-length parameter assumed in 1D $\texttt{MARCS}$ models ($\alpha_\text{MLT}=1.5$) \citep{2008AAMARCS} has a strong dependency on stellar parameters \citep{2014MNRAS.445.4366T} and fails to capture the efficient convective overshooting seen in real 3D atmospheres \citep{2013A&A...557A..26M,2024A&A...688A..52E}. This large sample, covering the transition of very metal-poor red giant stars towards solar metallicity, provides observational evidence supporting these theoretical predictions.

Literature on non-LTE corrections for OH lines is sparse due to the complexity of modeling vibrational and rotational transition levels. \cite{2001A&A...372..601A} utilized a simple two-level calculation for the OH $\lambda3139$ \AA\ line in dwarf stars (with metallicity from solar to [Fe/H] = -3), finding that LTE predicts stronger lines and thus underestimates abundance by $\simeq0.2$ dex compared to non-LTE. This occurs because the source function ($S_\nu$) deviates from the Planckian distribution ($B_\nu$) such that $S_\nu>B_\nu$, {enhancing} radiative scattering and photon pumping. This over-ionization shifts the population away from {thermal} equilibrium, weakening the line—a mechanism common for near-UV transitions \citep{2024ARA&A..62..475L} and low excitation energy lines \citep[{e.g., } ][]{2012MNRAS.427...27B}. On the other hand, {NIR} OH lines possess distinct transition signatures. {The}rovibrational transitions {within the} $X^2\Pi${state} are controlled by hydrogen collisions. Since the energy gaps are relatively small, collisions are frequent enough to maintain the population closer to thermalized (LTE) conditions \citep{1973ApJ...181.1039T,1975MNRAS.170..447H}, unlike the radiatively coupled electronic transitions of near-UV lines ($A^2\Sigma^+-X^2\Pi$) \citep{1975MNRAS.170..447H}. This suggests that non-LTE effects on {NIR} OH are likely small {, though} non-negligible.

Panel 3b of Figure \ref{fig:difference_O} shows the observational abundance difference as a function of the absolute C/O ratio $(\text{C/O}=N(\text{C})/N(\text{O})=10^{\log\epsilon(\text{C})-\log\epsilon(\text{O})})$; no strong correlation is observed. A previous theoretical study highlighted that the C/O ratio strongly influences the 3D correction $(\Delta_\text{3L-1L}=\Delta_\text{3D/LTE}-\Delta_\text{1D/LTE})$ of both CH and OH molecular bands, with an anticorrelated behavior that becomes more pronounced in metal-poor atmospheres \citep{2016A&A...593A..48G}. In oxygen-rich stars $(\text{C/O}\ll1)$, most carbon forms CO molecules, weakening CH and reducing $\Delta_\text{3L-1L}$ for CH lines. Since atomic O is more abundant under these conditions, 3D effects lead to more OH molecules forming in cooler pockets, strengthening OH lines and increasing $\Delta_\text{3L-1L}$ for OH lines as C/O decreases \citep[see also Figure D.1 in][]{2017A&A...599A.128P}. 

If it is assumed the [OI]-based O abundance approximates the 3D OH-based abundance, it would be expected that $\Delta_{\text{OH}-\text{[OI]}}$ increases as C/O decreases. This trend is not observed here: the carbon abundance appears to have a negligible effect on $\Delta_{\text{OH}-\text{[OI]}}$, especially for stars with low [C/O] ratio.  It should be noted, however, that the C abundance (from the CH band) is not rederived in the present work. The adopted temperatures of the samples are generally higher than in many previous studies, which likely results in higher C (or CH) abundances and thus higher C/O ratios. Recalibrating the C abundance may reveal the expected trends described above.

In summary, assuming that [OI] is insensitive to both 3D and non-LTE effects, and its abundance should reflect the "true" O abundance that determined by 3D/NLTE analysis of OH lines, the 3D/NLTE effect on OH is estimated by $\Delta_{\text{OH}-\text{[OI]}}\sim A(\text{OH})_{\text{1D/LTE}}-A(\text{OH})_{\text{3D/NLTE}}=-\Delta_\text{3N}$. Approximating the 3D/NLTE correction as $\Delta_\text{3N,app}=A_\text{3N,app}-A_\text{1L}\approx\Delta_{\text{3L}-\text{1L}}+\Delta_{\text{1N}-\text{1L}}$ \citep{2024ARA&A..62..475L}, the total contribution depends on the competing size of each factor. It is assumed that non-LTE corrections for {NIR} OH have less or similarly positive values compared to the estimation by \cite{2001A&A...372..601A} for UV OH lines. Then in cooler red giants with  slightly higher metallicity, $\Delta_{\text{3L}-\text{1L}}>0$ due to radiative heating as discussed earlier. Combined with the non-LTE effect, lines are weaken{ed} even more, and an even higher O abundance is required to match observed line strengths, which hence explains the lower 1D/LTE OH-based abundance. Conversely, towards higher temperatures and lower metallicities, while non-LTE corrections might become slightly more positive, radiative heating diminishes in favor of adiabatic cooling. This cooling strengthens the lines, potentially overcoming the non-LTE effect and resulting in a total $\Delta_{\text{OH}-\text{[OI]}}>0$ (or $\Delta_\text{3N,app}<0$). 

It must be cautioned that the above linear approximation could underestimate the true 3D/NLTE abundance, as demonstrated for Fe I \citep{2017A&A...607A..75N} and Li I \citep{2021MNRAS.500.2159W}. In full 3D/NLTE conditions, lines form deeper due to over-ionization/dissociation  from stronger UV photons, enhancing the NLTE effect compared to 1D/NLTE \citep{2025MNRAS.538.3284S}, thereby yielding a more positive $\Delta_\text{3N}$. However, if this mechanism were dominant, it would be difficult to explain the positive $\Delta_{\text{OH}-\text{[OI]}}$ (or negative $\Delta_\text{3N}$) observed in warmer red giants. To approximate the real 3D/NLTE abundance more accurately, adopting 1D/NLTE abundances may be preferable to 3D/LTE \citep{2024ARA&A..62..475L}. 

Another potential factor contributing to lower O abundance from OH lines is the efficient photodissociation of OH molecules by UV photons. While \citetalias{2018MNRAS.475.3369C} calculated OH photodissociation rates and found it to be negligible compared to other formation and destruction processes, detailed calculations including more transitional levels might reach different results. \cite{2023A&A...670A..25P} for other molecules with low dissociation energies like OH. 

Recent high-resolution radiative transfer modeling of irradiated exoplanet atmospheres by \cite{2026ApJ...997..203B} offers insights into the 1D/non-LTE behavior of OH. Using the ultra-hot Jupiter KELT-20b, they demonstrated that non-LTE effects on vibrational transition of OH could enhance the total OH photodissociation rate by up to two orders of magnitude compared to LTE predictions. Future studies dedicated to metal-poor stars are essential to reliably estimate 1D (or ideally 3D) non-LTE corrections for OH lines.


\subsection{[O/Fe] vs. [Fe/H] diagram}
\begin{figure*}[htbp!]
\centering
\includegraphics[trim={0.8cm 0 0.8cm 0},width=0.95\linewidth]{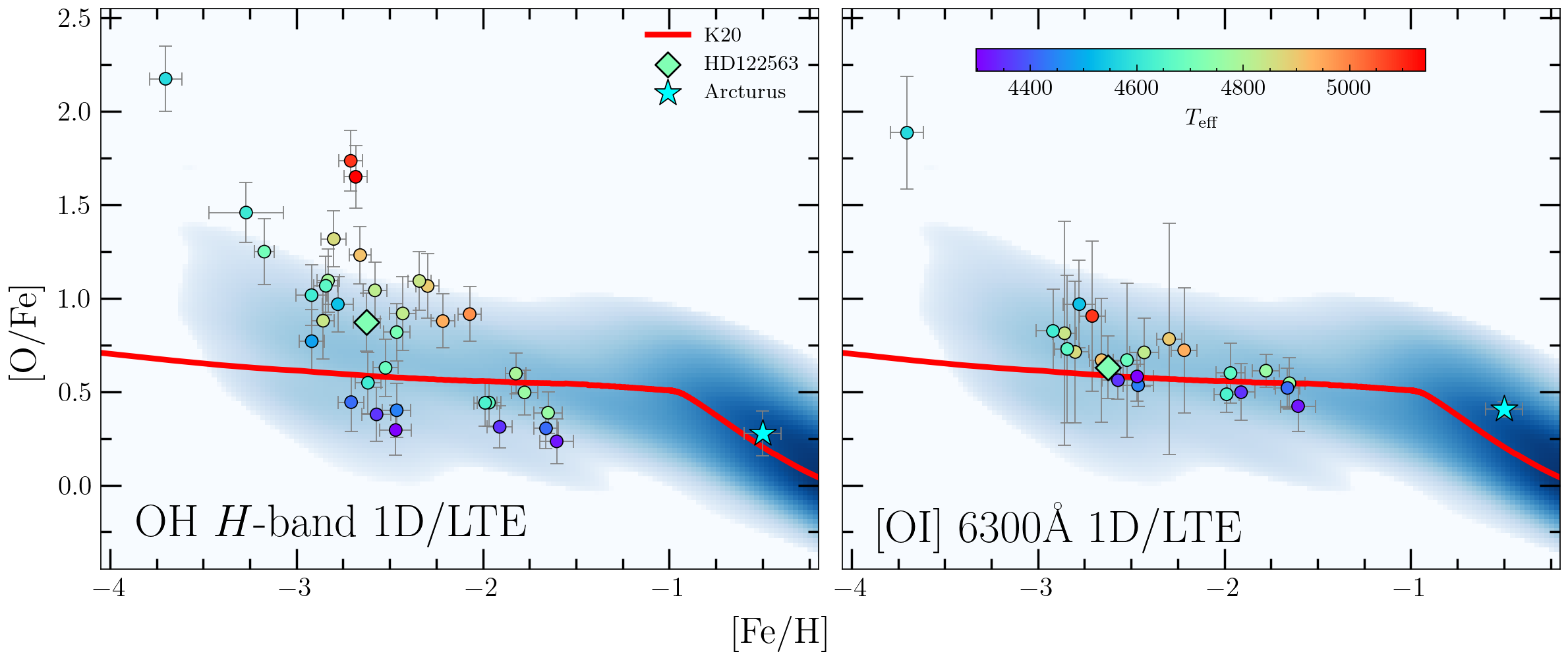}
\caption{The adopted [O/Fe] versus adopted [Fe/H] plot for near-IR OH (\textbf{left panel}) and 6300.3 \AA\ [OI] line (\textbf{right panel}), color-coded by effective temperature. Both oxygen abundances shown here are derived in 1D/LTE. The error bars represent the total uncertainty $(\sigma_\text{tot})$. The solid red line represents the average solar-neighborhood chemical evolution model for oxygen taken from \cite{2020ApJ...900..179K}. Blue color density background represent star number density that Oxygen abundance have been derived which are taken from \texttt{SAGA} database \citep{2008PASJ...60.1159S}.}
\label{fig:gce_oxygen}
\end{figure*}
The [O/Fe] ratios derived by analysis for OH IR and optical [OI] lines are presented in the left and right panels of Figure \ref{fig:gce_oxygen}, respectively. The background color contour presents the number densities of derived oxygen abundances compiled by the \texttt{SAGA} Database \citep{2008PASJ...60.1159S}. The result from OH covers [Fe/H] $>-3$. For the stars with [Fe/H] $<-3$, two notable differences distinguish these tracers: (1) the abundance trend and scatter, and (2) their total uncertainties. Specifically, OH-based [O/Fe] exhibits an increasing trend with a steeper slope towards lower metallicity {than} the [OI]-based measurements. Furthermore, [O/Fe] derived from the forbidden line displays significantly smaller scatter, aligning more closely with predictions from chemical evolution models \citep{2020ApJ...900..179K}.

A comparison of abundance uncertainties indicates that, thanks to the numerous lines available for measurement, random errors (including line-to-line scatter and EW uncertainty) of the OH-based abundances are reduced to levels comparable to or even smaller than systematic errors (dominated by atmospheric parameter uncertainties $\sigma_\text{atm}\sim0.15$ dex). In contrast, most of the 24 samples have only a single [OI] line measurement, causing random errors to exceed systematic errors by more than a factor of three, whereas the 3D/NLTE effect would be much smaller than that for OH lines. Hence, the large scatter found in the O abundances from OH lines is not due to the random errors, but primarily caused by the variation of the 3D/NLTE effects depending on stellar parameters. 

To investigate the abundance trend by taking advantage of the small random errors in OH-based [O/Fe] and the small systematic errors of [OI]-based measurements, an empirical calibration is applied to the OH-based O abundance in order to correct for its 3D/NLTE abundance. It is important to emphasize that this correction is not equivalent to the full 3D/NLTE analysis, as discussed in detail in the previous section. However, utilizing the 24 high-quality samples, a linear relationship can be estimated between the abundance discrepancy $\Delta_{\text{OH}-\text{[OI]}}$ and key stellar parameters ($T_\text{eff}$, $\log g$, $\text{[Fe/H]}$, and $\text{[C/Fe]}$). After evaluating various potential relations, the following linear equation is found to best describe the sample ($4300\leq T_\text{eff}/\text{K}\leq5000$, $0.5\leq\log g\leq2.5$, $-1.5\leq[\text{Fe/H}]\leq-3.0$, and $-1.0\leq[\text{C/Fe}]\leq0.7$), including the uncertainty in $\Delta_{\text{OH}-\text{[OI]}}$:
\begin{align}
\Delta^\text{corr}
_{\text{OH}-\text{[OI]}}=&(-2.23\pm2.45)+(1.59\pm3.08)\left(\frac{T_\text{eff}}{4500\text{ K}}\right)\notag\\
&+(3.08\pm5.13)\times10^{-1}\left(\frac{\log g}{1.5\text{ dex}}\right)\notag\\
&+(4.57\pm4.34)\times10^{-1}\left(\frac{\text{[Fe/H]}_\text{1N}}{-2.5\text{ dex}}\right)\notag\\
&+(9.80\pm14.16)\times10^{-2}\text{[C/Fe]}.\label{eq:correction}
\end{align}
It is noted that the sample size is insufficient to include nonlinear terms (e.g., $T_\text{eff}^2$, $T_\text{eff}\cdot\log g$). Furthermore, despite including one EMP target (LAMOST J2217+2104) in the calibration, applying this empirical correction to stars with $\text{[Fe/H]}<-3$ or to subgiants/dwarfs with $\log g>2.5$ is not recommended. The corrected OH-based O abundance with its uncertainy are obtained simply via:
\begin{align}
A(\text{OH})_\text{corr}&=A(\text{OH})_\text{1L}-\Delta^\text{corr}_{\text{OH}-\text{[OI]}}\\
\sigma_{A(\text{OH})_\text{corr}}&=\sqrt{\sigma_{A(\text{OH})_\text{1L}}^2+\sigma_{\Delta^\text{corr}}^2}
\end{align}

\begin{figure*}[htbp]
\centering
\includegraphics[width=\linewidth]{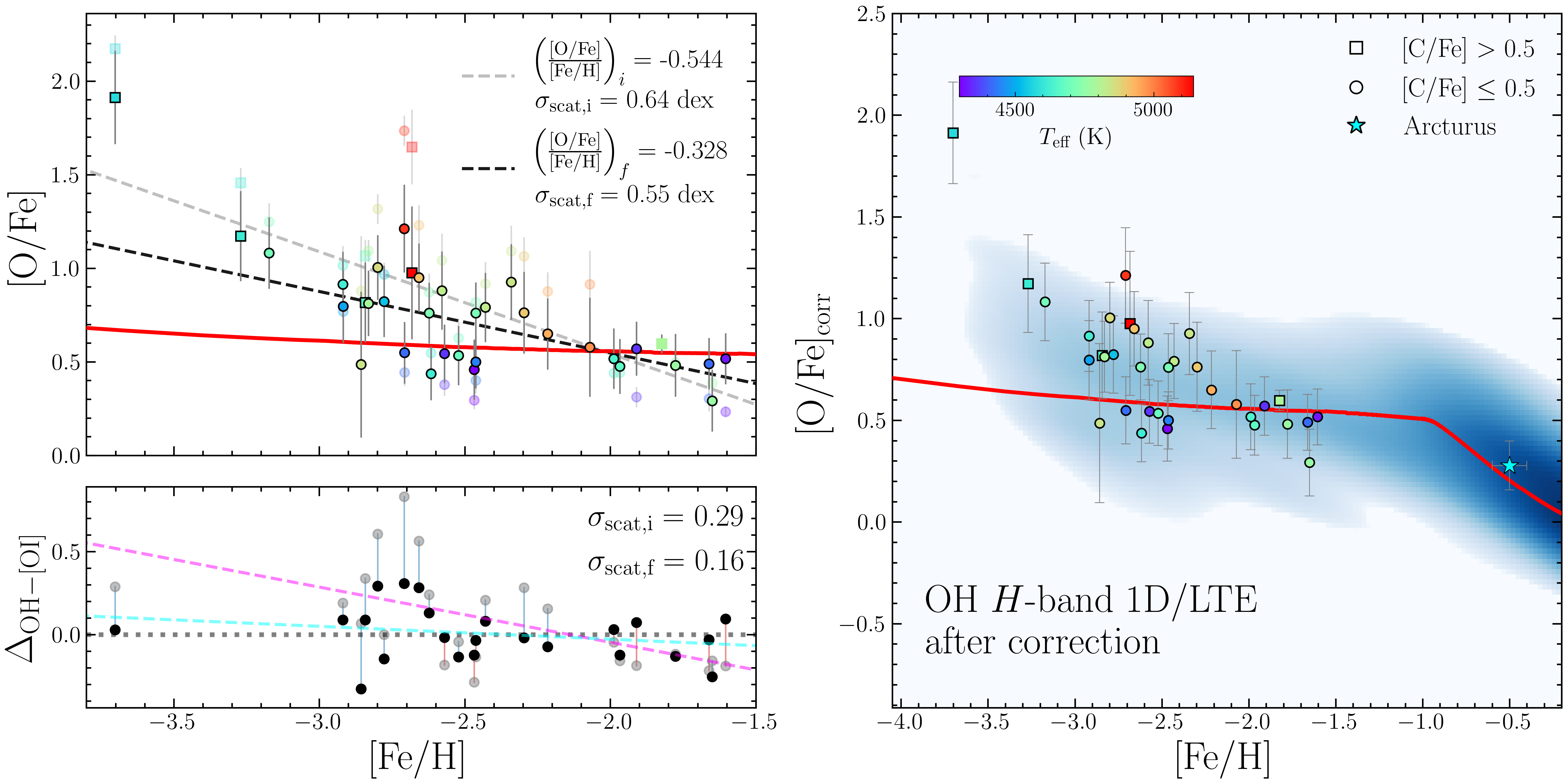}
\caption{\textbf{Upper left panel:} Corrected OH-based [O/Fe] ratios for the complete sample of 35 stars, overlaid with a linear regression fit (gray dashed line). The solid blue line represents the Galactic chemical evolution of oxygen \citep[K20:][]{2020ApJ...900..179K}. For comparison, the uncorrected [O/Fe] ratios are shown as fainter data points, with their corresponding linear fit indicated by a faint gray dashed line. \textbf{Bottom left panel:} The abundance difference between OH and [OI] ($\Delta_{\text{OH}-\text{[OI]}}$). Fainter symbols represent values before correction, while the solid lines indicate the shift for each star (red for increasing, blue for decreasing). The linear trends before and after correction are plotted as magenta and cyan dashed lines, respectively, highlighting a significant reduction in the discrepancy. Note that the scatter ($\sigma$) reported in this figure is calculated with respect to the corresponding linear trend fit. \textbf{Right panel:} Same description as Figure \ref{fig:gce_oxygen}, but $[\text{O/Fe}]_\text{corr}$ and $\sigma_{\text{corr}}$ are plotted with circle and square-shaped points represent stars with normal Carbon and Carbon-enhanced respectively. }
\label{fig:OFe_corrected}
\end{figure*}

The corrected [O/Fe] ratios with their uncertainties for 34 samples (excluding HD25329) are derived using a Monte Carlo simulation, which accounts for the uncertainties in each stellar parameter and coefficient. The results are tabulated in Table \ref{tab:corrected_OH_abund} and presented in the [O/Fe]--[Fe/H] diagram, in Figure \ref{fig:OFe_corrected}.

\begin{deluxetable}{lcccc}
\centering
\tablewidth{0pt}
\tabletypesize{\small}
\tablecaption{Corrected OH-based O Abundances\label{tab:corrected_OH_abund}}
\tablehead{
\colhead{Designation} & \colhead{$\Delta^\text{corr}_{\text{OH}-\text{[OI]}}$} & \colhead{$\sigma_{\Delta}$} & \colhead{$\langle\text{[OH/Fe]}\rangle_{\text{corr}}$} & \colhead{$\sigma_{\text{corr}}$}
}
\startdata
BD-02$^\circ$5957 & -0.026 & 0.164 & 0.796 & 0.194 \\
BD-07$^\circ$2674 & 0.248 & 0.196 & 0.818 & 0.214 \\
BD-14$^\circ$5890 & 0.337 & 0.195 & 0.578 & 0.265 \\
BD-15$^\circ$5781 & 0.165 & 0.148 & 0.926 & 0.202 \\
BD-18$^\circ$271 & -0.164 & 0.153 & 0.459 & 0.160 \\
BD-18$^\circ$5550 & 0.394 & 0.259 & 0.485 & 0.390 \\
BD-20$^\circ$6008 & 0.161 & 0.150 & 0.881 & 0.209 \\
BD+03$^\circ$2782 & -0.034 & 0.125 & 0.476 & 0.147 \\
BD+30$^\circ$2611 & -0.282 & 0.134 & 0.516 & 0.137 \\
CS29502-092 & 0.672 & 0.293 & 0.976 & 0.354 \\
CS30314-067 & 0.145 & 0.179 & 0.823 & 0.190 \\
HD 107752 & 0.280 & 0.148 & 0.950 & 0.183 \\
HD 110184 & -0.165 & 0.145 & 0.544 & 0.157 \\
HD 115444 & 0.312 & 0.154 & 1.004 & 0.173 \\
HD 118055 & -0.259 & 0.133 & 0.570 & 0.144 \\
HD 122563 & 0.110 & 0.127 & 0.761 & 0.162 \\
HD 187111 & -0.186 & 0.120 & 0.490 & 0.137 \\
HD 204543 & 0.014 & 0.150 & 0.481 & 0.168 \\
HD 221170 & -0.077 & 0.136 & 0.517 & 0.161 \\
HD 4306 & 0.522 & 0.220 & 1.212 & 0.234 \\
HD 6268 & 0.127 & 0.143 & 0.791 & 0.184 \\
HD 85773 & -0.099 & 0.123 & 0.499 & 0.141 \\
HD 8724 & 0.095 & 0.144 & 0.292 & 0.164 \\
HD 88609 & 0.102 & 0.141 & 0.914 & 0.175 \\
HE 1116-0634 & 0.167 & 0.164 & 1.081 & 0.191 \\
HE 1523-0901 & 0.057 & 0.137 & 0.761 & 0.164 \\
J0040+2729 & 0.093 & 0.120 & 0.534 & 0.158 \\
J2114-0616 & 0.302 & 0.193 & 0.762 & 0.219 \\
J2217+2104 & 0.260 & 0.239 & 1.913 & 0.250 \\
J2347+2851 & 0.228 & 0.161 & 0.649 & 0.191 \\
TYC 3407-1352-1 & 0.282 & 0.165 & 0.812 & 0.175 \\
TYC 3814-1598-1 & -0.105 & 0.148 & 0.549 & 0.165 \\
UCAC4 425-121652 & 0.110 & 0.136 & 0.437 & 0.142 \\
UCAC4 515-137892 & 0.285 & 0.227 & 1.172 & 0.240 \\
\enddata
\end{deluxetable}

After deriving the correction for all 34 samples (excluding the dwarf star HD 25329 and Arcturus), the large uncertainty of the [OI]-based abundances propagate into the large polynomial coefficients uncertainties in Equation \ref{eq:correction}, resulting in a typical $\sigma_\Delta^\text{corr} \sim 0.20$ dex. This leads to a larger abundance uncertainty than that before correction for each star, as seen in the right panel of Figure \ref{fig:OFe_corrected}, compared to the left panel of Figure \ref{fig:gce_oxygen}. In contrast, both the slope of the [O/Fe] versus [Fe/H] trend and the scatter ($\sigma_\text{scat}$) decrease noticeably in the upper left panel of Figure \ref{fig:OFe_corrected}, bringing the OH-based results into closer agreement with the [OI]-based abundances. The lower left panel demonstrates that the linear relation effectively reduces the discrepancy between OH and [OI] abundances without overfitting the data.

We note that, for BD$-18^\circ5550$, the correction exacerbates the discrepancy, driving the OH abundance even lower and increasing the abundance uncertainty.

Comparing the corrected [O/Fe] trends to previous large-sample studies in this metallicity regime, agreement is found between the OH results and 3D/NLTE OI triplet-based abundances, although the latter still show higher values by $\sim0.1$ dex than those from [OI] \citep{2019A&A...630A.104A}. An additional factor not detailed in this study is the 3D/NLTE effect on Fe lines . The adopted 1D/NLTE [Fe I/H] abundances are generally 0.15--0.3 dex higher (see Table \ref{tab:Fe_abund} and Figure \ref{fig:nlteFeI}) than the 1D/LTE values. While this has a negligible effect on $A(\text{O})$ for VMP stars, it shifts [O/Fe] lower by a similar amount. Previous studies indicate that 3D/NLTE corrections for Fe can be as high as +0.3 dex for V/EMP stars \citep{2016MNRAS.463.1518A,2022A&A...668A..68A}, though potentially lower for cooler VMP red giants like HD 122563 (+0.2 dex in \cite{2016MNRAS.455.3735A}), which is comparable to the 1D/NLTE correction for the same star ($\Delta_{\text{1N}-\text{1L}}\sim+0.14$ dex). Consequently, a complete 3D/NLTE analysis of both oxygen and iron abundances in V/EMP stars are likely to reveal more or less different trends and bring the [O/Fe] ratios from all oxygen tracers into closer agreement in the future.

\section{Summary \& Conclusion} \label{sec:conclusion}

In this study, we explore oxygen abundance derived from NIR OH lines in the $H$-band ($1.5-1.7\ \mu\text{m}$) using high-quality Subaru/IRD spectra for 35 very and extremely metal-poor stars (34 red giants and 1 dwarf) with metallicities between $\text{[Fe/H]} = -4.0$ and $-1.5$.

We uniformly redetermine the fundamental stellar parameters ($T_\text{eff}$, $\log g$, $\xi_t$) using high-precision \textit{Gaia} DR3-\textit{2MASS} colors and \textit{Gaia} DR3 parallax, and re-derive the 1D/NLTE Fe abundances with literature Fe equivalent width. The oxygen abundance{s} are {determined} through spectral synthesis method for NIR OH lines for 35 stars and optical [OI] 6300Å line for 24 stars with \texttt{Turbospectrum} code and 1D \text{MARCS} atmospheric models. NIR OH lines are very sensitive to temperature, shifting by $\sim\pm0.25$ dex for 100 K change. In contrast, the $\text{[OI]}$ line remain{s} remarkably stable against changes in all these parameters. We also confirmed through 3D/LTE and 1D/NLTE tests that corrections for the $\text{[OI]}$ line in our red giant stars are negligible ($\lesssim 0.01$ dex), proving 1D/LTE [OI]-based O abundance as an accurate baseline. Additionally, most of our stars have low carbon-to-oxygen ratios, both tracers are  mostly unaffected by the assumed carbon abundance.

The 1D/LTE oxygen abundances from the two tracers exhibit a clear temperature-dependent abundance difference ($\Delta_{\text{OH}-\text{[OI]}}$). Whereas warmer red giants ($T_\text{eff} \gtrsim 4600$ K) show higher OH-based abundances as expected from 3D effects, for the cool red giants in our sample, the OH-based abundances are {systematically} lower than the $\text{[OI]}$ measurements by 0.05 to 0.25 dex. We suggest this reversed trend in our cooler red giants is caused by 3D radiative heating in the deeper atmospheric layers{,} where NIR OH lines form{,} that weakens the molecular lines. When combined with non-LTE effects driven by UV photons, it forces the 1D/LTE analysis to underestimate the OH-based abundance to match the observed line strengths.

The non-corrected 1D/LTE OH-based abundances exhibit a steep increase toward lower metallicities with a large scatter partially due to the 3D/NLTE effect depending on effective temperature. To use the advantage that there are large number of OH lines, we calibrate the abundance from OH lines as they match those from [OI] line by employing an empirical calibration, deriving the abundance difference as the function of $T_\text{eff}$, $\log g$, $\text{[Fe/H]}$, and $\text{[C/Fe]}${. We} succesfully bring the corrected  1D/LTE OH abundances closer to the reliable $\text{[OI]}$-based abundance.

By applying this calibration, the corrected OH-based $\text{[O/Fe]}$ trend show{s} much less scatter and a flatter trend across the metal-poor regimes. This adjusted trend brings our observed oxygen abundances into an agreement with the Galactic chemical evolution models for oxygen. While future work will need full 3D/NLTE modeling for both O and Fe, the present work demonstrates that abundances derived from {NIR} OH lines with careful calibration are reliable and useful to explore the oxygen abundance distribution in the oldest stars.
\begin{acknowledgments}
{Based on data collected at Subaru Telescope, which is operated by the National Astronomical Observatory of Japan.} The authors are honored and grateful for the opportunity of observing the Universe from Maunakea, which has the cultural, historical and natural significance within the Hawaiian community. {The archive data of Subaru Telescope and obtained from the SMOKA and the \href{https://jvo.nao.ac.jp/portal/}{JVO portal} \footnote{https://jvo.nao.ac.jp/portal/}, which are operated by the Astronomy Data Center and National Astronomical Observatory of Japan.} 
Some of the archival spectra used in this study are based on observations made with ESO Telescopes at the La Silla Paranal Observatory under programme ID 111.251N.001; 165.N-0276; 0104.D-0059; 090.B-0605; 080.D-0347; 0103.D-0118; 71.B-0529; 68.D-0546; 073.D-059; 273.D-5032; 0103.D-0118. 
{This research has made use of the Keck Observatory Archive (KOA), which is operated by the W. M. Keck Observatory and the NASA Exoplanet Science Institute (NExScI), under contract with the National Aeronautics and Space Administration.} The Observatory was made possible by the generous financial support of the W.M. Keck Foundation. {We thank the anonymous referees for their constructive comments and suggestions. BDSB thanks Vinicius Placco for insightful discussions regarding $T_\text{eff}$ derivations using \textit{Gaia} colors, and Richard Hoppe for suggestions on interpreting 3D atmospheric effects.} BDSB acknowledges fundings from The Graduate University for Advanced Studies, SOKENDAI that also partially supported this research. WA and MI are  supported by JSPS KAKENHI grant No. 21H04499 and 25K01046. NS acknowledges funding from the European Research Council (ERC) under the European Union’s Horizon 2020 research and innovation programme (Grant agreement No. 949173). This publication also makes use of data products from the Two Micron All Sky Survey, which is a joint project of the University of Massachusetts and the Infrared Processing and Analysis Center/California Institute of Technology, funded by the National Aeronautics and Space Administration and the National Science Foundation.

\end{acknowledgments}





%
\facilities{Subaru (IRD, HDS), Keck:I (HIRES), VLT:Kueyen (UVES), ESO:3.6m (HARPS), VLT (ESPRESSO)}

\software{\texttt{Astropy} \citep{2013AA558A..33A,2018AJ....156..123A,2022ApJ...935..167A}, \texttt{TSFitPy}\citep{2023AAGerberTSFitPy}, \texttt{jupyter} \citep{2016ppap.book...87K}, \texttt{iSpec} \citep{2014AAispec}, \texttt{IRAF} \citep{1986SPIE..627..733T}
}


\clearpage      
\appendix  
\twocolumngrid
     

\section{Oxygen Abundance on Benchmark Stars: Arcturus and HD 122563}\label{ssec:o_arc_hd}
\begin{figure*}[hbpt!]
    \centering
    \includegraphics[width=0.87\linewidth]{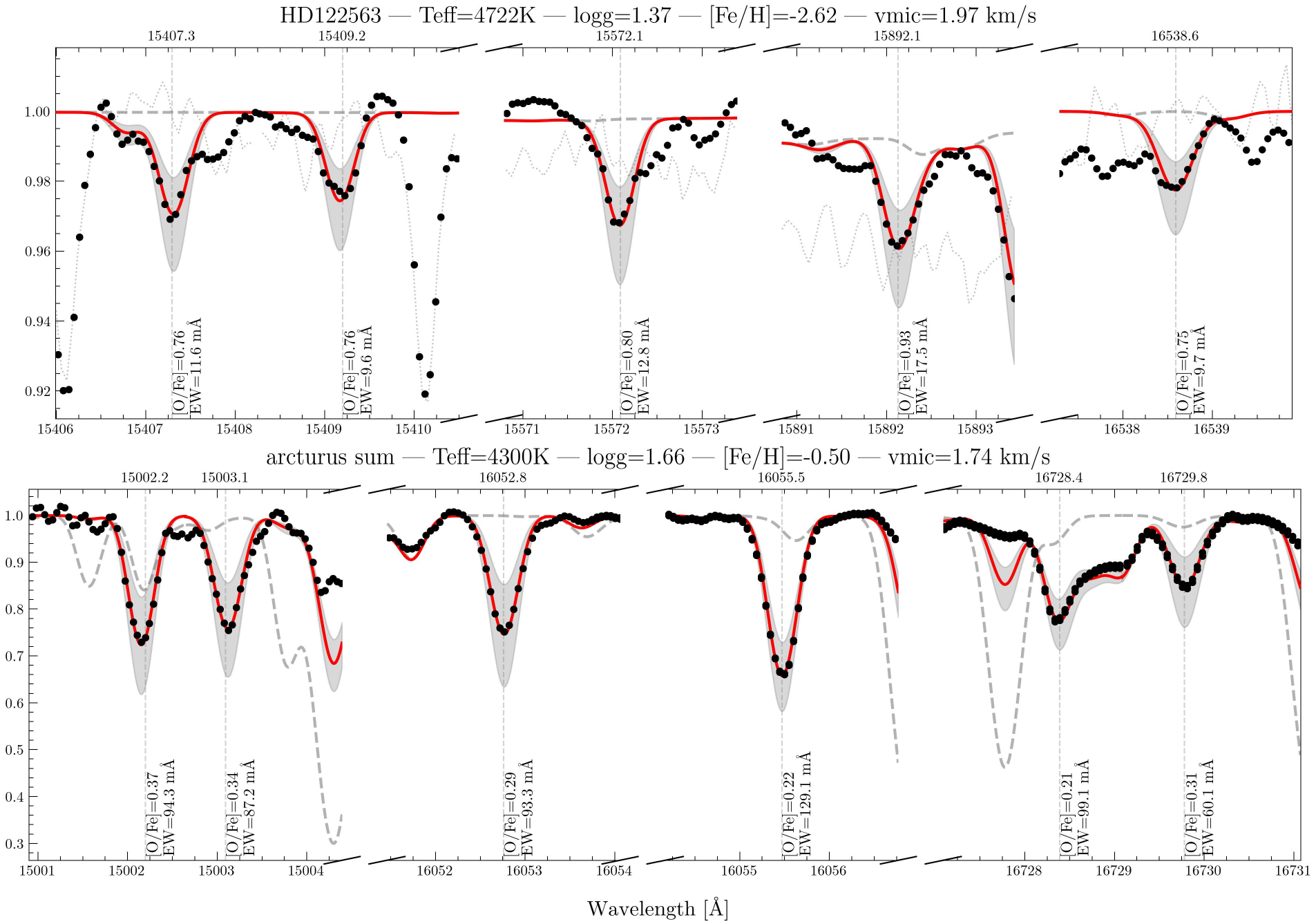}
    \caption{Same description as Figure \ref{fig:OH_plot}, but for Arcturus and HD 122563.}
    \label{fig:OH_arc_HD}
\end{figure*}

\begin{figure*}[hbpt!]
    \centering
    \includegraphics[width=0.87\linewidth]{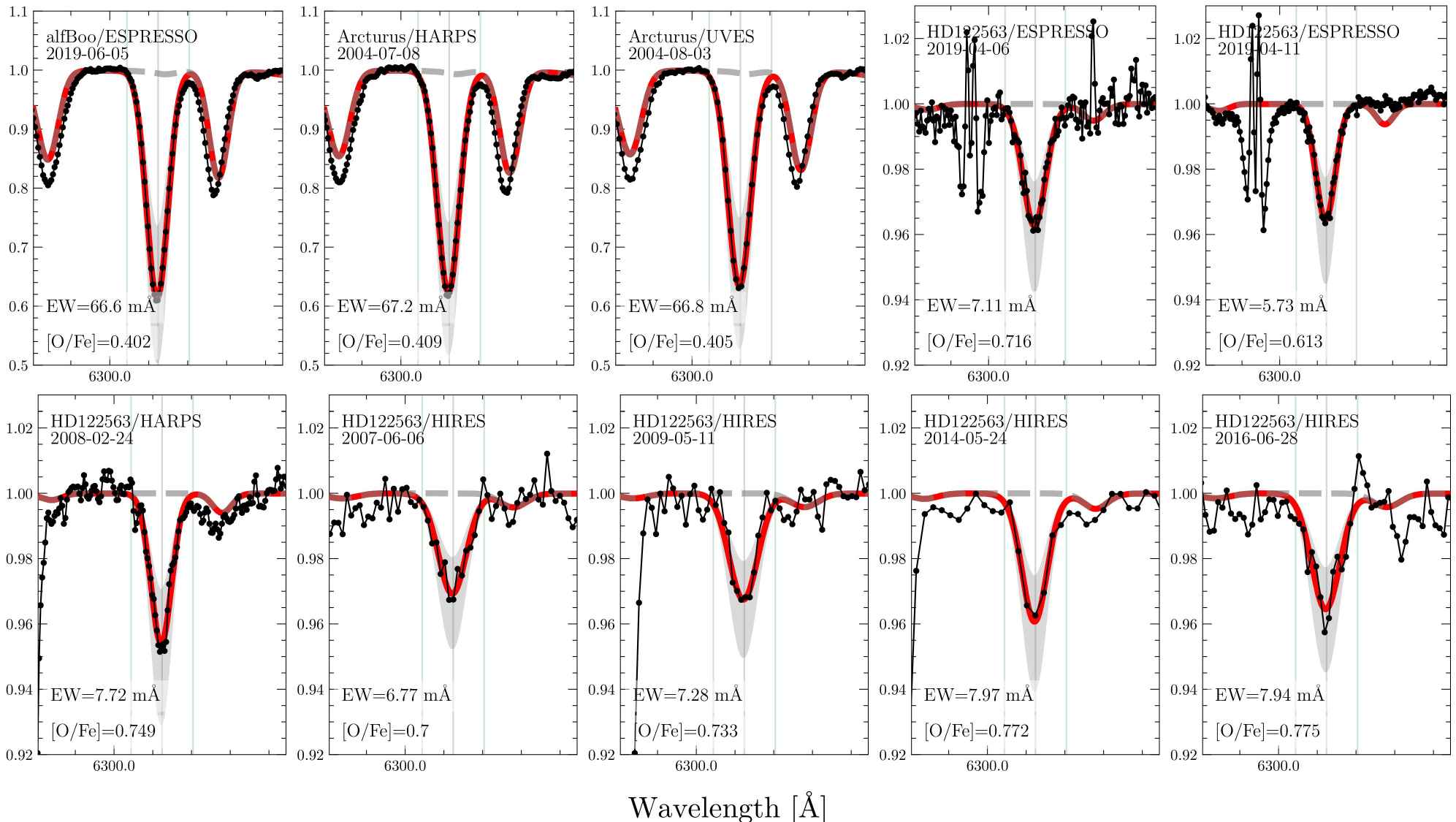}
    \caption{Same description as Figure \ref{fig:OI_plot}, but for Arcturus and HD 122563.}
    \label{fig:OI_arc_HD}
\end{figure*}

In this section, we discuss our derived O abundance on benchmark stars: Arcturus and HD 122563, in comparison to many previous studies. The example of spectral fitting around OH lines and [OI] line for each target are shown in Figure \ref{fig:OH_arc_HD} \& \ref{fig:OI_arc_HD}, respectively.


Arcturus is a well-known very bright star $(V\sim-0.05\text{ mag})$ with sub-solar metallicity, for which very high-resolution and excellent S/N $H$-band spectra are available. HD 122563 is a bright $(V\sim6.19\text{ mag})$, very metal-poor star that has been investigated in various literature. 

\begin{figure*}[htbp!]
\centering
\includegraphics[trim={0.8cm 0.8cm 0 0},width=0.72\linewidth]{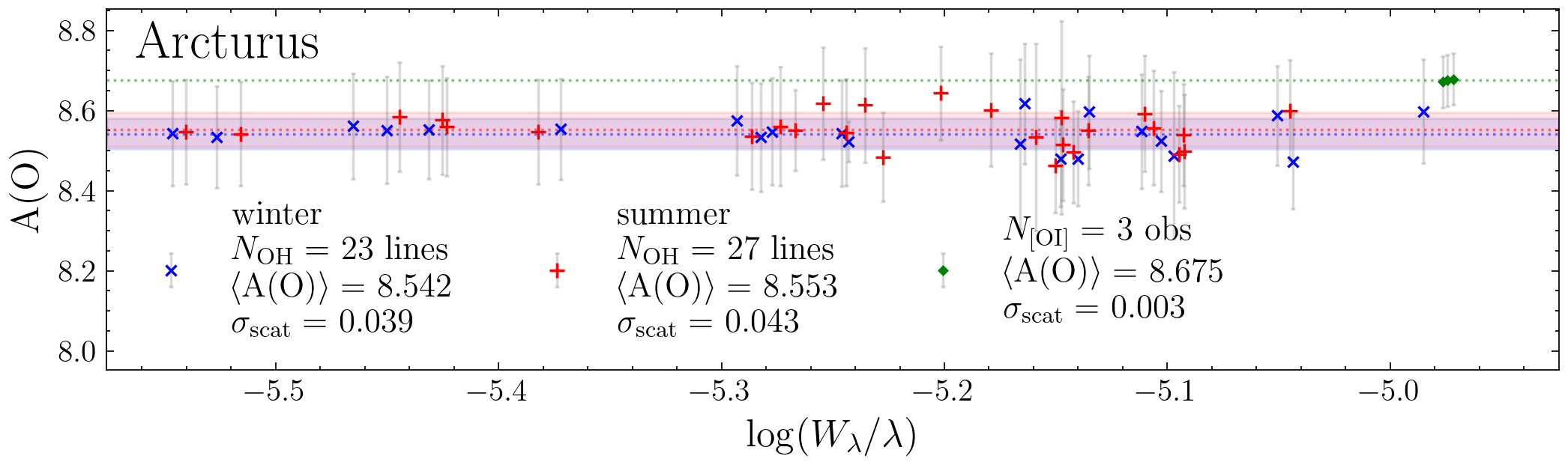}
\includegraphics[trim={0.8cm 0 0 0},width=0.72\linewidth]{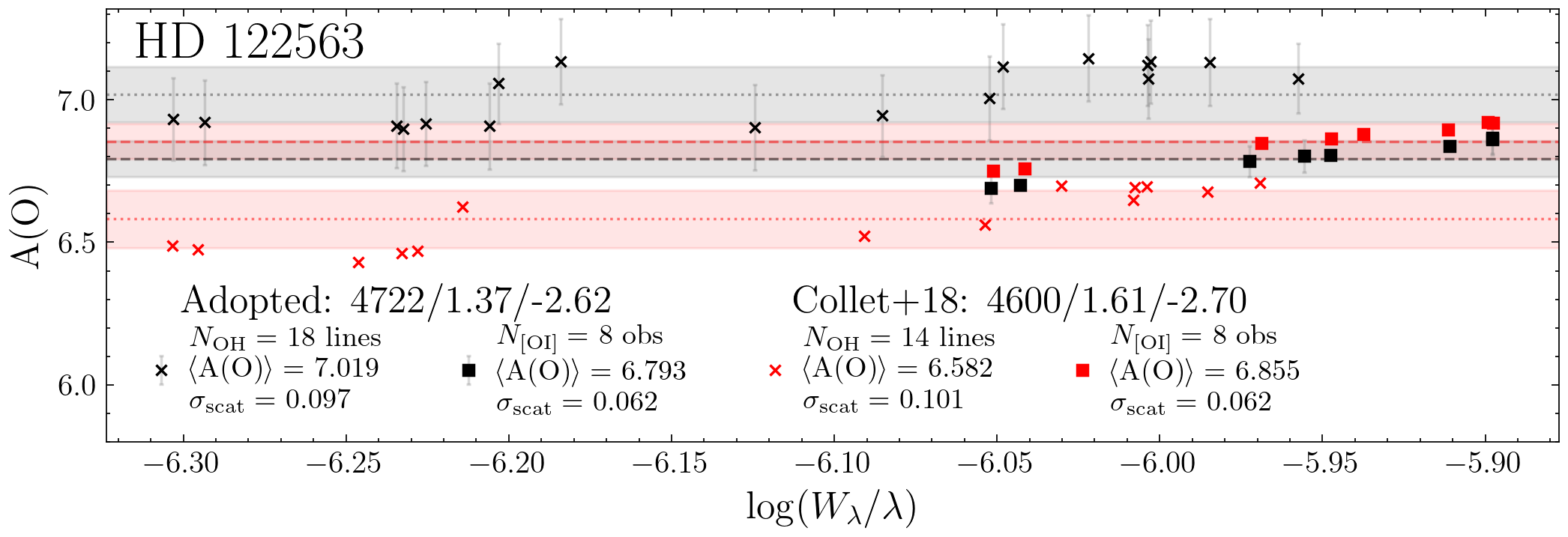}
\caption{Plot of O abundance (A(O)) derived from both OH and [OI] lines as a function of line strength for Arcturus (\textbf{upper panel}) and HD 122563 (\textbf{lower panel}). The error bar in each data point represents the uncertainty due to the input atmospheric parameters: $\sigma_\text{atm}$. The mean abundance for each tracer is indicated as a dotted horizontal line. For the OH lines (cross), each data point represents a different line. For the [OI] line (square), measurement of different spectra are represented as different data points. Note for HD 122563, the O abundance from OH and [OI] lines are derived using two different set of stellar parameters: our adopted parameters \& parameters from \citetalias{2018MNRAS.475.3369C} which are indicated as black and red datapoints, respectively.}
\label{fig:arc_hd_line_strength}
\end{figure*}

For Arcturus, two high-resolution spectra observed in the summer and winter seasons \citep{1995PASP..107.1042H} are analyzed. The O abundance from each line with its uncertainty is shown in the upper panel of Figure \ref{fig:arc_hd_line_strength}. In the winter and summer spectra, 23 and 27 OH lines are obtained, respectively, yielding a consistent O abundance with a line-to-line scatter of $\sigma_\text{scat}<0.05$ dex. The mean abundance difference between the summer and winter spectra is $\Delta\text{A(O)}<0.01$ dex. For the optical spectra, three archival spectra from different epochs are analyzed, yielding identical abundances with a scatter of $<0.01$ dex.

\begin{figure*}[htbp!]
\centering
\includegraphics[width=0.8\linewidth]{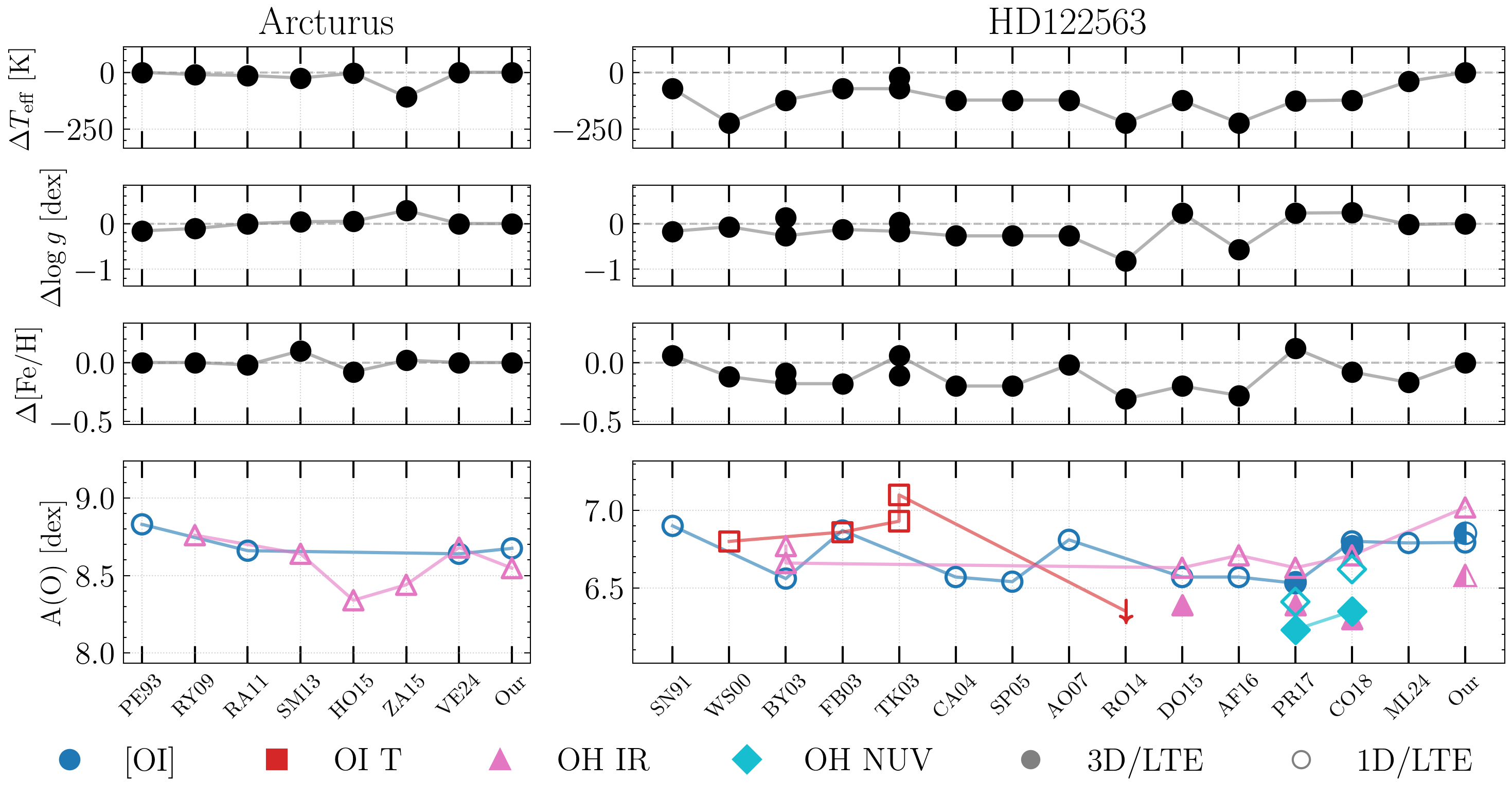}
\caption{The literature and derived stellar parameters and O abundances for Arcturus and HD 122563. For $T_\text{eff},\;\log g$, and [Fe/H], the Y-axis shows the differences between each literature source and the adopted values. Half-filled data points represent the O abundance derived using stellar parameters from \cite{2018MNRAS.475.3369C}. The lower legends are given for the O abundance panels only. (PE93: \cite{1993ApJ...404..333P}; RY09: \cite{2009A&A...496..701R}; RA11: \cite{2011ApJ...743..135R}; SM13: \cite{2013ApJ...765...16S}; HO15: \cite{2015AJ....150..148H}; ZA15: \citep{2015AJ....149..181Z}; VE24: \cite{2024A&A...684A..85V}; SN91: \cite{1991AJ....102.2001S}; WS00: \cite{2000ApJ...530..783W}; BY03: \cite{2003ApJ...588.1072B}; FB03: \cite{2003ApJ...fulbright}; TK03: \cite{2003AA402..343T}; CA04: \cite{2004A&A...416.1117C}; SP05: \cite{2005A&A...430..655S}; AO07: \cite{2007ApJ...660..747A}; RO14: \cite{Roederer_2014}; DO15: \cite{2015A&A...576A.128D}; AF16: \cite{2016ApJ...819..103A}; PR17: \cite{2017A&A...599A.128P}; CO18: \cite{2018MNRAS.475.3369C}; ML24: \cite{2024A&A...691A.220F})}
\label{fig:comp_arc_hd}
\end{figure*}

All available literature on 1D/LTE {O} abundances in Arcturus is also compiled, shown in the left panels of Figure \ref{fig:comp_arc_hd}. No study performing 3D abundance analysis is found for this star. In this work, no attempt is made to rederive stellar parameters for Arcturus. The most recent study for the optical and {NIR} spectra of Arcturus is \cite{2024A&A...684A..85V}. Despite using the same input parameters and analyzing the same Atlas spectra, the O abundance from {NIR} OH by the present work is lower by 0.1 dex than that of \cite{2024A&A...684A..85V}. This difference could be mainly due to the different lines used in the analysis and systematic errors in deriving the abundance using different spectral synthesis codes (\texttt{MOOG} vs \texttt{TS}) and atmosphere models (\texttt{Kurucz} vs \texttt{MARCS}). For other studies, all O abundance differences derived from {NIR} OH lines in the literature can be explained by slight differences in input parameters. In contrast, the O abundance from the [OI] line is consistent with previous studies \citep{2024A&A...684A..85V,2011ApJ...743..135R} within 0.05 dex.

For HD 122563, the O abundance from OH and [OI] lines is derived using the same {NIR} and optical spectra for two different sets of atmospheric parameters: the adopted parameters and those from \citetalias{2018MNRAS.475.3369C}. The major difference is found in the effective temperature adopted from the IRFM method \citep{2014MNRAS.439.2060C}, which is 122 K lower than the one adopted here. In addition, our surface gravity is lower than \citetalias{2018MNRAS.475.3369C} by 0.23 dex due to the different parallax used (\citetalias{2018MNRAS.475.3369C} uses the \textit{Hipparcos} parallax: $\varpi_\text{Hip}=4.22\pm0.35$ mas, while this study uses \textit{Gaia} DR3: $\varpi_\text{GDR3}=3.099\pm0.033$ mas). The surface gravity of the present work is closer to that estimated by the 3D/NLTE analysis of Fe lines \citep{2016MNRAS.455.3735A}. The OH-based O abundance exhibits a line-to-line scatter of $\sim0.1$ dex, which is typical among those in the sample. No clear dependence of abundance on line strength is found (see Figure \ref{fig:arc_hd_line_strength}).

As seen in the lower panel of Figure \ref{fig:arc_hd_line_strength}, there is a large difference in the O abundance derived from OH lines between the results using our adopted stellar parameters ($\text{A(O)}_\text{OH,our}=7.019$ dex) and the one using \citetalias{2018MNRAS.475.3369C}'s stellar parameters ($\text{A(O)}_\text{OH,CO18}=6.582$ dex) by $\text{A(O)}_\text{our}-\text{A(O)}_\text{CO18}\sim+0.44$ dex. Again, this difference can primarily be explained by the different input parameters, using Table \ref{tab:sensi_tab_HD_arc} as the reference, an increased of temperature by $\Delta T_\text{eff}=T_\text{eff,our}-T_\text{eff,CO18}=+122$ K contributes to an increasing abundance of $\Delta\text{A(O)}\approx${+0.34} dex. Also, decreasing the surface gravity of $\Delta \log g=\log g_\text{our}-\log g_\text{CO18}=-0.23$ dex will contribute to the abundance change by $\sim${+0.08} dex. Since the sensitivity to metallicity is negligible, the total \textit{expected} abundance change due to difference of stellar parameters is $\sim${+0.42} dex, which explains the result. 

Differences in the selected OH lines, atmospheric models, and spectral synthesis codes possibly explain the $\sim0.13$ dex offset between our OH-based O abundances and those of \citetalias{2018MNRAS.475.3369C}. Even when adopting their exact stellar parameters and C \& N abundances, our IRD spectra yield a 1D/LTE abundance of $\text{A(O)}_\text{OH}=6.58\pm0.10$ dex (Figure \ref{fig:arc_hd_line_strength}, lower panel), compared to their $6.71\pm0.09$ dex derived from NIR spectra used in \citetalias{2015A&A...576A.128D} study.

The O abundance obtained from the [OI] line by this analysis is slightly lower $(\Delta\text{A(O)}\sim-0.06\text{ dex})$ than \citetalias{2018MNRAS.475.3369C}, whereas no significant offset is expected ($\Delta\text{A(O)}_\text{expected}\sim+0.007\text{ dex}$) from the difference in stellar parameters. Indeed, this {small difference is} possibly originate{d} from the internal error of the fitting process. In addition, a noticeable scatter of $\sigma_\text{scat}\sim0.06$ dexor a $0.2$ dex difference from the lowest to the highest [OI] line abundance{,} is observed using 8 spectra observed with several instruments and at different epochs. This abundance scatter could be explained by the {measurement errors for this} weak-line . All high-resolution spectra for HD 122563 have a typical $\text{S/N}\sim200$. Using Equation \ref{eq:W_lambda}, the expected EW uncertainty is $\sigma_{W_\lambda}\approx2.5$ m\AA, which is already $\sim42\%$ of the EW. It is calculated that the respective abundance change will be $\sigma_{\epsilon|W_\lambda}\approx^{+0.17}_{-0.27}$ dex, which hence explains the observed line scatter.

The stellar parameters and O abundance are compared between this work and previous studies, as shown in the right panels of Figure \ref{fig:comp_arc_hd}. The adopted effective temperature is the highest {among the} studies {for this star}. The most recent studies on HD 122563, \cite{2025AJ....169..172M} and \cite{2024A&A...691A.220F}, derive similar temperatures and surface gravities using the same method of $T_\text{eff}$ determination as the one used here. \cite{2024A&A...691A.220F} find a similar [OI] abundance to this study, while \cite{2025AJ....169..172M} do not derive the O abundance for HD 122563. The average O abundance obtained by previous studies using {NIR} OH is $A(\text{O})\approx6.65$ dex, which is 0.35 dex lower than the value obtained here. Most studies adopt a $T_\text{eff}$ that is 150 K lower than the one used here, explaining this abundance difference. \citetalias{2018MNRAS.475.3369C} find that the O abundance from the forbidden line is slightly higher than those from OH lines. The discrepancy is larger if a 3D correction is applied. However, with the newly adopted $T_\text{eff}$ and $\log g$, the opposite is found. If a 3D correction similar to \citetalias{2018MNRAS.475.3369C} is assumed, the {NIR} OH abundance could be lower, bringing it closer to the [OI] abundance.

\begin{deluxetable}{lcccc}
\centering
\digitalasset
\tablewidth{0pt}
\tablecaption{1D/LTE abundance sensitivities on Arcturus and HD 122563\label{tab:sensi_tab_HD_arc}}
\tablehead{
\colhead{\multirow{3}{*}{Parameters}} & \multicolumn{4}{c}{$\langle\Delta\text{A(O)}\rangle$} \\
\colhead{} & \multicolumn{2}{c}{Arcturus} & \multicolumn{2}{c}{HD 122563} \\
\colhead{} & \colhead{OH IR} & \colhead{[OI]} & \colhead{OH IR} & \colhead{[OI]}
}
\startdata
$\Delta T_\text{eff}$ (+100 K) & +0.158 & +0.013 & +0.276 & +0.046 \\
$\Delta T_\text{eff}$ (-100 K) & -0.141 & -0.015 & -0.288 & -0.042 \\
$\Delta \log g$ (+0.3 dex) & -0.012 & +0.069 & -0.071 & +0.042 \\
$\Delta \log g$ (-0.3 dex) & +0.006 & -0.067 & +0.105 & -0.041 \\
$\Delta \xi_t$ (+0.5 $\text{km s}^{-1}$) & -0.010 & -0.030 & +0.000 & -0.002 \\
$\Delta \xi_t$ (-0.5 $\text{km s}^{-1}$) & +0.018 & +0.036 & +0.004 & +0.002 \\
$\Delta \text{[Fe/H]}$ (+0.3 dex) & +0.188 & +0.134 & +0.034 & +0.022 \\
$\Delta \text{[Fe/H]}$ (-0.3 dex) & -0.287 & -0.116 & -0.014 & -0.014 \\
$\Delta\log\epsilon_\text{C}$ (+0.3 dex) & -0.011 & +0.095 & +0.001 & +0.002 \\
$\Delta\log\epsilon_\text{C}$ (-0.3 dex) & -0.002 & -0.056 & +0.000 & +0.000 \\
$\Delta\log\epsilon_\text{N}$ (+0.3 dex) & -0.002 & +0.000 & $\dotsi$ & $\dotsi$ \\
$\Delta\log\epsilon_\text{N}$ (-0.3 dex) & +0.000 & +0.000 & $\dotsi$ & $\dotsi$ \\
$\Delta\log\epsilon_\text{Ni}$ (+0.3 dex) & $\dotsi$ & -0.002 & $\dotsi$ & $\dotsi$ \\
$\Delta\log\epsilon_\text{Ni}$ (-0.3 dex) & $\dotsi$ & +0.002 & $\dotsi$ & $\dotsi$ \\
\enddata
\end{deluxetable}

For Arcturus, a small discrepancy ($\sim0.11\text{ dex}$) is found between tracers. The forbidden line unexpectedly gives a higher abundance than OH lines. Many previous studies also show that OH lines provide a lower abundance compared to [OI] lines, raising a tension in explaining this discrepancy. According to Table \ref{tab:sensi_tab_HD_arc}, increasing the adopted temperature by $\sim75$ K results in a final abundance of $A(\text{O})=8.67$ dex for both tracers. One possibility is that previous studies underestimate the effective temperature. \cite{2011ApJ...743..135R} adopt a slightly lower temperature ($T_\text{eff}=4286\pm30\text{ K}$) that is derived from SED fits of optical to mid-IR spectra. The authors also mentioned that the 1D/LTE ionization balance of Fe yields a much higher temperature of 4380 K. Adopting this value would achieve agreement between the OH and [OI] abundances, but it is not preferable due to known non-LTE affecting Fe I lines. Measuring temperature with extraordinary precision is challenging (except for the Sun). Although the O abundance from OH lines is sensitive to $T_\text{eff}$, the $\sim0.1$ dex difference in O abundance in Arcturus cannot simply be attributed to temperature differences. We find a similar discrepancy, where the OH-based abundance is lower than that from [OI], in most of our sample of cool red giants, as discussed in Section \ref{ssec:oxygen_discrepancy}.




\section{3D/LTE correction grid for [OI] 630nm line}
We present our derived 3D/LTE correction of [OI] 6300Å line which cover $0.0\leq[\text{O/Fe}]\leq2.5$. Our stellar parameters grid covers metal-poor to very metal-poor Red Giant stars with $4000\leq T_\text{eff}/\text{K}\leq5000$, $1.5\leq\log g\leq3.0$, $-4.0\leq[\text{Fe/H}]\leq-1.0$, and $0.5\leq\xi_t/(\text{km s}^{-1})\leq2.5$. In addition, in this work, we also account for {C} abundance {in} the [OI] 3D/LTE abundance correction which cover{s} $-1.5\leq[\text{C/Fe}]\leq2.0$. 

\begin{figure*}[htbp!]
 \centering
 \includegraphics[trim={0 23mm 0 0},width=0.95\linewidth]{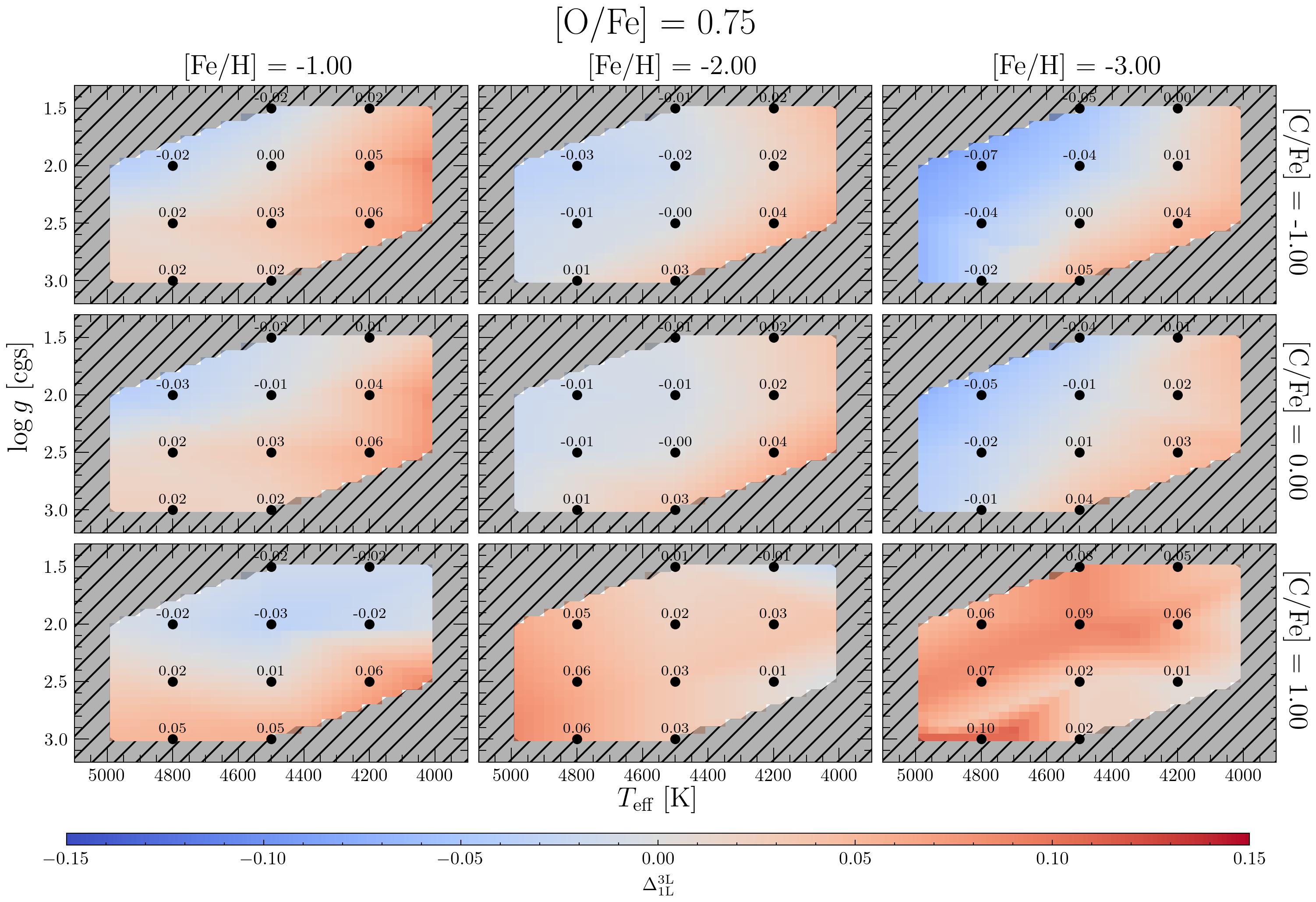}
 \includegraphics[width=0.95\linewidth]{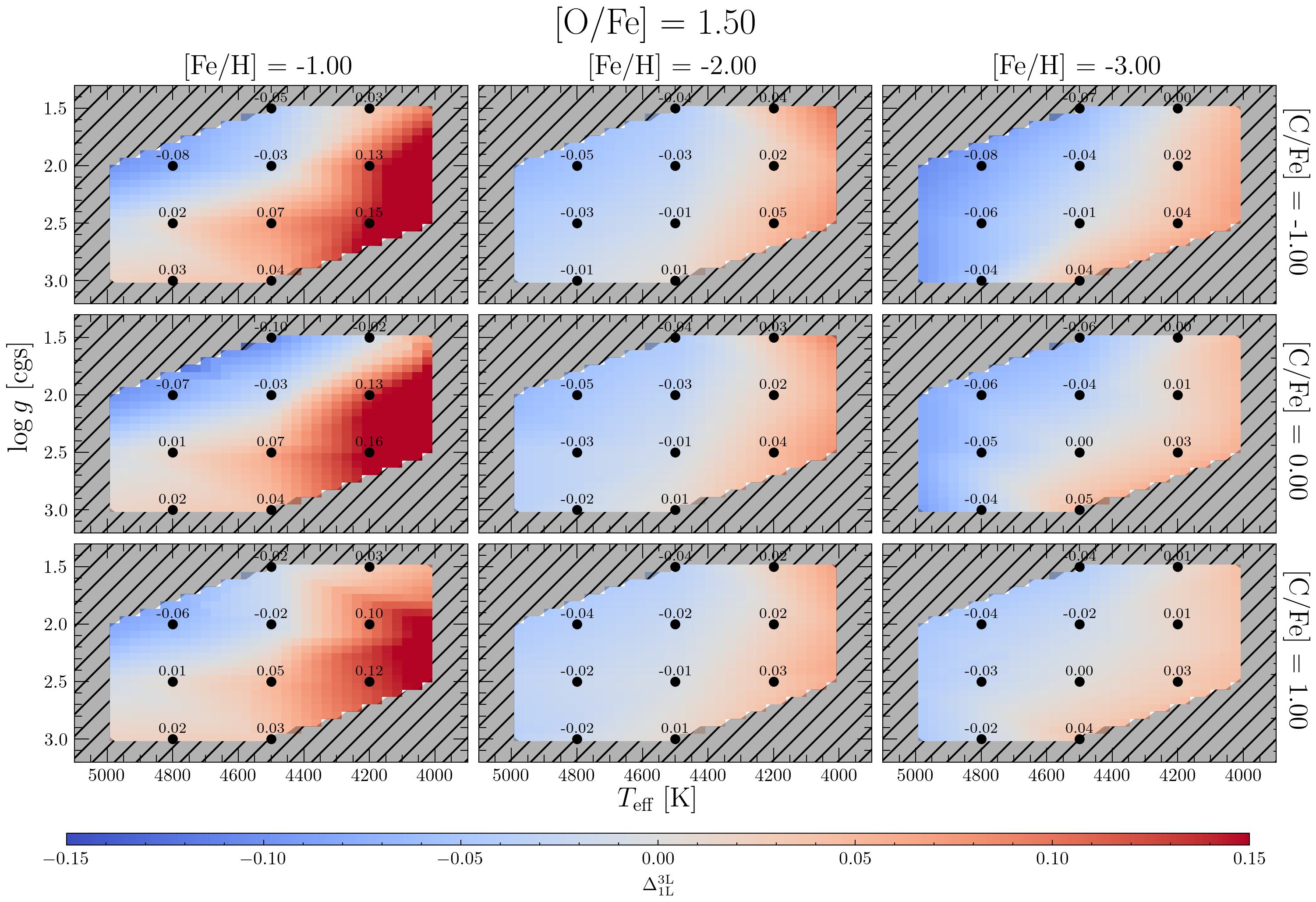}
 \caption{Example of the 3D/LTE correction $(\Delta_\text{1L}^\text{3L}\equiv \text{A(O)}_\text{3D/LTE}-\text{A(O)}_\text{1D/LTE})$ derived in this works (show as color-coded map) for [O/Fe]=0.75 dex (upper panel) and 1.50 dex (lower dex) for various Fe and C abundances. In each subplot, X and Y-axes are effective temperature and surface gravity respectively.}
 \label{fig:3Dcorr_OI}
\end{figure*}

The correctional grid used in this study could be found in Table \ref{tab:3Dcorrtable}, in which contains stellar parameters, the equivalent width computed from 3D/LTE spectra, 3D/LTE [O/Fe], and the 3D/LTE correction. Also, Kiel diagram of 3D/LTE correction at different metallicities (\textit{columns}) and Carbon abundances (\textit{rows}) for [O/Fe]=0.75 and 1.50 dex, are shown in Figure \ref{fig:3Dcorr_OI}.

\begin{deluxetable*}{ccccccccc}
\centerwidetable
\digitalasset
\tablewidth{0pt}
\tablecaption{The 3D/LTE O abundance correction grid for [OI] line\label{tab:3Dcorrtable}}
\tablehead{
\colhead{$T_\text{eff}$} & \colhead{$\log g$} & \colhead{[Fe/H]} & \colhead{[C/Fe]} & \colhead{$W_\lambda$}   & \colhead{$\xi_t$} & \colhead{$\text{[O/Fe]}_\text{3D/LTE}$} & \colhead{$\text{[O/Fe]}_\text{1D/LTE}$} & \colhead{$\Delta^\text{3L}_\text{1L}$} \\
\colhead{(K)} & \colhead{(dex)} & \colhead{(dex)} & \colhead{(dex)} & \colhead{(m\AA)}   & \colhead{($\text{km s}^{-1}$)} & \colhead{(dex)} & \colhead{(dex)} & \colhead{(dex)}
}
\startdata
4000 & 1.5  & -4.0 & -1.5       & 2.09 & 2.5                  & 1.00       & 0.890      & 0.110    \\
4000 & 1.5  & -4.0 & -1.5       & 2.09 & 1.5                  & 1.00       & 0.893      & 0.107    \\
4000 & 1.5  & -4.0 & -1.5       & 2.09 & 0.5                  & 1.00       & 0.895      & 0.105    \\
4000 & 1.5  & -4.0 & -1.0       & 2.08 & 1.5                  & 1.00       & 0.892      & 0.108    \\
4000 & 1.5  & -4.0 & -1.0       & 2.08 & 2.5                  & 1.00       & 0.894      & 0.106    \\
4000 & 1.5  & -4.0 & -1.0       & 2.08 & 0.5                  & 1.00       & 0.895      & 0.105    \\
4000 & 1.5  & -4.0 & -0.5       & 2.06 & 2.5                  & 1.00       & 0.891      & 0.109    \\
4000 & 1.5  & -4.0 & -0.5       & 2.06 & 1.5                  & 1.00       & 0.895      & 0.105    \\
4000 & 1.5  & -4.0 & -0.5       & 2.06 & 0.5                  & 1.00       & 0.898      & 0.102    \\
4000 & 1.5  & -4.0 & -1.5       & 3.67 & 2.5                  & 1.25       & 1.137      & 0.113    \\
4000 & 1.5  & -4.0 & -1.5       & 3.67 & 0.5                  & 1.25       & 1.156      & 0.094    \\
4000 & 1.5  & -4.0 & -1.5       & 3.67 & 1.5                  & 1.25       & 1.156      & 0.094    \\
4000 & 1.5  & -4.0 & -1.0       & 3.66 & 2.5                  & 1.25       & 1.138      & 0.112    \\
$\vdots$ & $\vdots$  & $\vdots$ & $\vdots$       & $\vdots$ & $\vdots$                  & $\vdots$       & $\vdots$      & $\vdots$  \\
\enddata
\end{deluxetable*}

\section{Atomic and Molecular Linelist}
The Fe, OH, and [OI] lines used in this study, with their equivalent width, line parameters, and abundances are given in Table \ref{tab:linebyline}.

\begin{deluxetable*}{lcccccccccc}
\centerwidetable
\digitalasset
\tablewidth{0pt}
\tablecaption{Line-by-line parameters and derived abundances\label{tab:linebyline}}
\tablehead{
\colhead{\multirow{2}{*}{Designation}} & \colhead{\multirow{2}{*}{Species}} & \colhead{EW} & \colhead{Wavelength} & \colhead{$\chi_\text{exc}$} & \colhead{$\log gf$} & \colhead{$\log\epsilon(X)$} & \colhead{$\log\epsilon(X)$} & \colhead{$\sigma_\text{atm}$} & \colhead{\multirow{2}{*}{Note}} \\
\colhead{} &  & \colhead{(m\AA)} & \colhead{(\AA)} & \colhead{(eV)} & \colhead{} & \colhead{1D/LTE (dex)} & \colhead{1D/NLTE(dex)} & \colhead{(dex)} & \colhead{} 
}
\startdata
Arcturus   & {[}OI{]} & 66.543   & 6300.31   & 0.000    & -9.715   & 8.672    & $\dotsi$ & 0.064    & ESPRESSO   \\
Arcturus   & {[}OI{]} & 67.240   & 6300.31   & 0.000    & -9.715   & 8.679    & $\dotsi$ & 0.064    & HARPS      \\
Arcturus   & {[}OI{]} & 66.808   & 6300.31   & 0.000    & -9.715   & 8.675    & $\dotsi$ & 0.064    & UVES       \\
Arcturus   & OH       & 94.3037  & 15002.153 & -5.653   & 0.134    & 8.643    & $\dotsi$ & 0.116    & Hinkle sum \\
Arcturus   & OH       & 87.1772  & 15003.113 & -4.656   & 2.579    & 8.614    & $\dotsi$ & 0.143    & Hinkle sum \\
Arcturus   & OH       & 108.9753 & 15278.525 & -5.453   & 0.205    & 8.568    & $\dotsi$ & 1.065    & Hinkle sum \\
\multicolumn{1}{c}{$\vdots$}   & $\vdots$ & $\vdots$ & $\vdots$  & $\vdots$ & $\vdots$ & $\vdots$ & $\vdots$ & $\vdots$ & $\vdots$   \\
Arcturus   & OH       & 114.2753 & 15278.525 & -5.453   & 0.205    & 8.595    & $\dotsi$ & 0.542    & Hinkle win \\
Arcturus   & OH       & 118.1738 & 15281.055 & -5.453   & 0.205    & 8.548    & $\dotsi$ & 0.143    & Hinkle win \\
Arcturus   & OH       & 123.162  & 15407.294 & -5.435   & 0.255    & 8.488    & $\dotsi$ & 0.208    & Hinkle win \\
\multicolumn{1}{c}{$\vdots$}   & $\vdots$ & $\vdots$ & $\vdots$  & $\vdots$ & $\vdots$ & $\vdots$ & $\vdots$ & $\vdots$ & $\vdots$   \\
BD-02 5957 & Fe I     & 56.75    & 4032.627  & 1.485    & -2.377   & 4.311    & 4.472    & 0.018    & literature \\
BD-02 5957 & Fe I     & 15.70    & 4080.209  & 3.283    & -1.220   & 4.529    & $\dotsi$ & 0.013    & literature \\
BD-02 5957 & Fe I     & 25.40    & 4098.175  & 3.241    & -0.879   & 4.438    & $\dotsi$ & 0.021    & literature \\
BD-02 5957 & Fe I     & 30.49    & 4114.445  & 2.832    & -1.303   & 4.517    & 4.668    & 0.017    & literature \\
BD-02 5957 & Fe I     & 53.48    & 4134.677  & 2.832    & -0.649   & 4.288    & 4.508    & 0.034    & literature \\
\multicolumn{1}{c}{$\vdots$}   & $\vdots$ & $\vdots$ & $\vdots$  & $\vdots$ & $\vdots$ & $\vdots$ & $\vdots$ & $\vdots$ & $\vdots$  \\
\enddata
\end{deluxetable*}

\newpage

\bibliography{sample7}{}
\bibliographystyle{aasjournal}
\end{document}